\pdfoutput=1
\documentclass[a4paper,12pt]{article}

\usepackage{amsfonts}
\usepackage{mathrsfs}
\usepackage{amsmath}
\usepackage{amssymb}
\usepackage{framed} 
\usepackage{subfigure}
\usepackage[medium]{titlesec}
\usepackage{bm}
\usepackage{cite}

\usepackage[normalem]{ulem}
\usepackage{extarrows}
\usepackage{slashed}
\usepackage{isodateo}
\usepackage{graphicx}
\usepackage{xcolor}
\usepackage[bookmarksnumbered=true,bookmarksopen=true]{hyperref}
 \hypersetup{colorlinks,%
             linkcolor=[rgb]{0,0.3,0.6}, %
             citecolor=[rgb]{0,0.3,0.6}, %
             urlcolor=[rgb]{0,0.3,0.6}}
\usepackage[hmargin=.7in,vmargin=1.1in]{geometry}
\usepackage{indentfirst}
\usepackage{booktabs}
\usepackage{soul}
\usepackage{bbm}

\newcommand{\FR}[2]{\displaystyle\frac{\,{#1}\,}{#2}}
\newcommand{\fr}[2]{\mbox{$\frac{\,{#1}\,}{#2}$}}
\newcommand{\n}{\nonumber}

\graphicspath{{fig/}}

\def\bge{\begin{equation}}
\def\ede{\end{equation}}
\def\bga{\begin{aligned}}
\def\eda{\end{aligned}}
\def\bgb{\begin{bmatrix}}
\def\edb{\end{bmatrix}}
\def\bgp{\begin{pmatrix}}
\def\edp{\end{pmatrix}}
\def\bgm{\begin{matrix}}
\def\edm{\end{matrix}}
\def\bgs{\begin{subequations}}
\def\eds{\end{subequations}}
\newcommand{\order}[1]{\mathcal{O}({#1})}
\def\di{{\mathrm{d}}}

\def\mb{\mathbf}

\def\pd{\partial}
\def\ld{{\mathscr{L}}}

\def\la{\langle}\def\ra{\rangle}

\def\to{\rightarrow}

\def\ii{\mathrm{i}}

\def\de{\delta}

\def\lam{\lambda}

\def\si{\sigma}

\def\aa{\mathsf{a}}
\def\bb{\mathsf{b}}
\def\cc{\mathsf{c}}

\usepackage{mdframed}

\newmdenv[skipabove=0pt,%
          skipbelow=5pt,%
          leftmargin=0pt,%
          rightmargin=0pt,%
          innertopmargin=-5pt,%
          innerbottommargin=7pt,%
          innerleftmargin=2pt,%
          innerrightmargin=2pt,%
          splittopskip=0pt,%
          splitbottomskip=0pt,%
          linewidth=0pt,%
          nobreak=true]%
          {keyeqn2}

\newmdenv[backgroundcolor=gray!15,%
          skipabove=0pt,%
          skipbelow=5pt,%
          leftmargin=0pt,%
          rightmargin=0pt,%
          innertopmargin=-5pt,%
          innerbottommargin=7pt,%
          innerleftmargin=2pt,%
          innerrightmargin=2pt,%
          splittopskip=0pt,%
          splitbottomskip=0pt,%
          linewidth=0pt,%
          nobreak=true]%
          {keyeqn}

\makeatletter
\@ifpackageloaded{hyperref}%
  {\newcommand{\mylabel}[2]
    {\protected@write\@auxout{}{\string\newlabel{#1}{{#2}{\thepage}%
      {\@currentlabelname}{\@currentHref}{}}}}}%
  {\newcommand{\mylabel}[2]
    {\protected@write\@auxout{}{\string\newlabel{#1}{{#2}{\thepage}}}}}
\makeatother

\usepackage{titlesec}          
\titleformat{\section}
{\normalfont\fontsize{15}{20}\bfseries}{\thesection}{1em}{}

\newcommand{\wt}[1]{\mkern 2mu \widetilde{\mkern -2mu #1 \mkern -2mu}\mkern 2mu}
\newcommand{\wh}[1]{\mkern 2mu \widehat{\mkern-2mu#1\mkern-2mu}\mkern 2mu}

\newcommand{\fnemail}[1]{\footnote{Email: \href{mailto:#1}{\nolinkurl{#1}}}}

\begin{document}

\title{\Large\textbf{Inflation Correlators with Multiple Massive Exchanges\\[2mm]}}

\author{Zhong-Zhi Xianyu$^{\,a}$\fnemail{zxianyu@tsinghua.edu.cn}~~~~~ and ~~~~~Jiaju Zang$^{\,b}$\fnemail{zangjj20@mails.tsinghua.edu.cn}\\[5mm]
\normalsize{\emph{${}^{a}$Department of Physics, Tsinghua University, Beijing 100084, China}}\\
\normalsize{\emph{${}^{b}$Zhili College, Tsinghua University, Beijing 100084, China}}
}
\date{}
\maketitle

\vspace{20mm}

\begin{abstract}
\vspace{10mm}

The most general tree-level boundary correlation functions of quantum fields in inflationary spacetime involve multiple exchanges of massive states in the bulk, which are technically difficult to compute due to the multi-layer nested time integrals in the Schwinger-Keldysh formalism. On the other hand, correlators with multiple massive exchanges are well motivated in cosmological collider physics, with the original quasi-single-field inflation model as a notable example. In this work, with the partial Mellin-Barnes representation, we derive a simple rule, called family-tree decomposition, for directly writing down analytical answers for arbitrary nested time integrals in terms of multi-variable hypergeometric series. We present the derivation of this rule together with many explicit examples. This result allows us to obtain analytical expressions for general tree-level inflation correlators with multiple massive exchanges. As an example, we present the full analytical results for a range of tree correlators with two massive exchanges. 
 
\end{abstract}

\newpage
\tableofcontents

\newpage
\section{Introduction}

Recent years have witnessed increasing interests in the theoretical study of cosmological correlation functions of large-scale fluctuations, which are believed to be sourced by quantum fluctuations of spacetime and matter fields during cosmic inflation \cite{Achucarro:2022qrl}. By observing the correlation functions of the large-scale structure, we can access quantum field theory in inflationary spacetime. This connection has far-reaching consequences to both early-universe cosmology and fundamental particle physics. It has been emphasized that heavy particles produced during cosmic inflation could leave characteristic and oscillatory signals in certain soft limits of correlation functions. Many recent studies have exploited this Cosmological Collider (CC) signal to study particle physics at the inflation scale \cite{Chen:2009we,Chen:2009zp,Baumann:2011nk,Chen:2012ge,Pi:2012gf,Noumi:2012vr,Gong:2013sma,Arkani-Hamed:2015bza,Chen:2015lza,Chen:2016nrs,Chen:2016uwp,Chen:2016hrz,Lee:2016vti,An:2017hlx,An:2017rwo,Iyer:2017qzw,Kumar:2017ecc,Chen:2017ryl,Tong:2018tqf,Chen:2018sce,Chen:2018xck,Chen:2018cgg,Chua:2018dqh,Wu:2018lmx,Saito:2018omt,Li:2019ves,Lu:2019tjj,Liu:2019fag,Hook:2019zxa,Hook:2019vcn,Kumar:2018jxz,Kumar:2019ebj,Alexander:2019vtb,Wang:2019gbi,Wang:2019gok,Wang:2020uic,Li:2020xwr,Wang:2020ioa,Fan:2020xgh,Aoki:2020zbj,Bodas:2020yho,Maru:2021ezc,Lu:2021gso,Sou:2021juh,Lu:2021wxu,Pinol:2021aun,Cui:2021iie,Tong:2022cdz,Reece:2022soh,Chen:2022vzh,Niu:2022quw,Niu:2022fki,Aoki:2023tjm,Chen:2023txq,Tong:2023krn,Jazayeri:2023xcj,Jazayeri:2023kji,Meerburg:2016zdz,MoradinezhadDizgah:2017szk,MoradinezhadDizgah:2018ssw,Kogai:2020vzz}. 
At the same time, there are considerable works devoting to the analytical or numerical study of correlation functions of quantum field theory in inflationary spacetime, or \emph{inflation correlators} for short \cite{Arkani-Hamed:2018kmz,Baumann:2019oyu,Baumann:2020dch,Pajer:2020wnj,Hillman:2021bnk,Baumann:2021fxj,Hogervorst:2021uvp,Pimentel:2022fsc,Jazayeri:2022kjy,Wang:2022eop,Baumann:2022jpr,Goodhew:2020hob,Goodhew:2021oqg,Melville:2021lst,Meltzer:2021zin,DiPietro:2021sjt,Tong:2021wai,Salcedo:2022aal,Agui-Salcedo:2023wlq,Sleight:2019hfp,Sleight:2019mgd,Sleight:2020obc,Sleight:2021plv,Jazayeri:2021fvk,Premkumar:2021mlz,Qin:2022lva,Qin:2022fbv,Xianyu:2022jwk,Qin:2023ejc,Qin:2023bjk,Qin:2023nhv,Loparco:2023rug,Maldacena:2011nz,Baumann:2017jvh,Bonifacio:2022vwa,Lee:2022fgr,Cabass:2022rhr,Cabass:2022oap,Lee:2023jby,Arkani-Hamed:2017fdk,Arkani-Hamed:2018bjr,Gomez:2021qfd,Gomez:2021ujt,Wang:2021qez,Werth:2023pfl,Stefanyszyn:2023qov}. 
These studies have revealed many interesting structures of inflation correlators or wavefunctions, which deepen our understanding of quantum field theory in de Sitter spacetime. On the other hand, explicit analytical results are indispensable for a precise understanding of CC signals and for comparing theoretical predictions of CC models with observational data.

Many explicit analytical results have been obtained in recent years for inflation correlators relating to CC physics \cite{Arkani-Hamed:2018kmz,Baumann:2019oyu,Pimentel:2022fsc,Jazayeri:2022kjy,Qin:2022lva,Qin:2022fbv,Xianyu:2022jwk,Qin:2023ejc,Qin:2023bjk,Qin:2023nhv}. Most of these results are for the exchange of a single massive particle in the bulk of dS, with a few exceptions at loop orders. However, previous works on CC model building have shown that correlators with multiple exchanges of massive particles could be phenomenologically important. Already in the early studies of quasi-single-field inflation, it was noticed that the correlator with cubic self interaction of a bulk massive scalar can greatly enhance the size of the correlation function. In such models, tree-level graphs exchanging more than one massive scalar make dominant contributions to the 3-point correlator \cite{Chen:2009zp,Chen:2017ryl,Chen:2018sce}. However, due to the technical complications, explicit analytical results for inflation correlators with more than one bulk massive field are still beyond our reach at the tree level. 

It may come as a surprise to flat-space field theorists that scalar tree graphs are hard to compute. Indeed, setting aside the issues of tensor and flavor structures, the complexity of a scalar Feynman graph in flat spacetime largely increases with the number of loops $L$: Each loop gives rise to a loop momentum integral, and carrying out these loop integrals are not trivial. However, so long as we stay at the tree level ($L=0$), Feynman graphs are simply products of propagators and vertices and are typically rational functions of external momenta. So, increasing the number of vertices and propagators does not generate any difficulty per se. 

Things are a little different in inflationary spacetime: Here we normally have full spatial translation and rotation symmetries, but the time translation is usually broken. Accordingly, we Fourier transform only spatial dependence of a function to momentum space, and leave the time dependence untransformed. In this hybrid ``time-momentum'' representation, we get additional time integrals at all interaction vertices in the  Schwinger-Keldysh (SK) formalism \cite{Schwinger:1960qe,Feynman:1963fq,Keldysh:1964ud,Weinberg:2005vy,Chen:2017ryl}. As a result, the complexity of graphs in inflation increases in two directions: either with the number of loops, or with the number of vertices. 

Partly for this reason, full analytical computation of tree correlators with multiple massive exchanges remains challenging: In a tree graph, the number of bulk vertices is always equal to the number of bulk propagators plus 1. Thus, a tree graph with $I$ internal legs requires time integrals of $(I+1)$ layers. Worse still, each bulk propagator $D_{\aa\bb}(k;\tau_1,\tau_2)$ in SK formalism comes in four types, depending on the four choices of SK indices $\aa,\bb=\pm$ at the two endpoints. The two propagators with same-sign indices $D_{\pm\pm}(k;\tau_1,\tau_2)$ involve expressions that depend on the time ordering, which make the $(I+1)$-layer time integral nested. That is, the integration limit of one layer could depend on the integration variable of the next layer. So, the integration quickly becomes intractable with increasing number of bulk lines or vertices. 

One may wish to bypass the difficulty of bulk time integrals by taking a boundary approach. For instance, one can try to derive differential equations satisfied by the correlators starting from simple bootstrapping inputs \cite{Arkani-Hamed:2018kmz,Baumann:2019oyu,Baumann:2020dch,Jazayeri:2022kjy,Pimentel:2022fsc,Qin:2022fbv,Qin:2023ejc}. As explored in many previous works, this approach turns out quite successful for single massive exchanges, where the ``bootstrap equations'' are usually a simple set of second-order ordinary differential equations and usually have well-known analytical solutions.  However, when one goes to the two massive exchanges, the resulting differential equations become much more complicated, and it seems rather nontrivial to directly solve such equations  \cite{Pimentel:2022fsc}.

One can also try other methods such as a full Mellin-space approach, where one still works in the bulk, but rewrites correlators in Mellin space \cite{Sleight:2019hfp,Sleight:2019mgd,Sleight:2020obc,Sleight:2021plv}. Then, the time ordering of the same-sign propagators $D_{\aa\bb}$ becomes an overall cosecant factor that nests two Mellin variables. While this is enormously simpler than the time-momentum representation, eventually we need to transform the Mellin-space correlators back to a normal time-momentum representation and push the time variables to the boundary: The future boundary is where the observables are naturally defined, and the momentum space is where the cosmological data are presented and analyzed. However, the nested Mellin variables make the inverse Mellin transform nontrivial. Thus, in a sense, in the full Mellin-space approach, we are moving the difficulty of nested time integral to the difficulty of nested Mellin integral.

There are other studies considering inflation correlators or wavefunction coefficients with multiple massive bulk lines. Rather than full analytical computations, most works focused on general properties of such amplitudes, such as the analyticity, unitarity, causality, cutting rules, etc. There is a special case where one does achieve full results for tree graphs with arbitrary number of bulk lines, namely when the bulk field's mass is tuned to the conformal value $m^2=2H^2$ and all couplings are dimensionless. In such cases, the amplitudes reduce to the flat-space results, and one can find nice recursion relations to directly build arbitrary tree amplitudes or even loop integrand \cite{Arkani-Hamed:2017fdk,Arkani-Hamed:2018bjr}. However, this result only applies to very special class of theories which are not of direct interest to CC physics. One might want to restore general mass and couplings by integrating the conformal-scalar amplitudes with appropriate weighting functions. However, the complication here is that we encounter fractions of nested energy variables which are hard to integrate. 

As we see, no matter what representation we take, there is always a nested part of the amplitude that makes the computation difficult. There is a physical reason behind it: The nested time integrals are from the time ordering of the bulk propagator, and the time-ordered bulk propagator is a solution to the field equation with a local $\de$-source. Thus, the nested part of the amplitude is closely related to the EFT limit where several or all nested vertices are pinched to a single bulk vertex. Very schematically, we can express this fact with the position-space Feynman propagator $D(x,y)$, which is a solution to the sourced equation of motion $(\square_x-m^2)D(x,y)=\ii\de(x-y)$. Then, we can make an EFT expansion of $D(x,y)\sim \fr{\ii}{\square_x-m^2}\de(x-y)$. The leading order term is simple, which is just the contact graph with $D(x,y)\sim -\ii\de(x-y)/m^2$. However, there are higher order terms coming from acting on powers of $\square_x/m^2$ on $\de(x-y)$, which produce a series of momentum ratios when transformed to the momentum-space representation. Technically, as we shall see, such series are typically multi-variable hypergeometric series which in general do not reduce to any well known functions. So, one just has no way to get around with this result; the complication has to show up somewhere. The best we can do is to find a way to write down the analytical result as a convergent hypergeometric series for some kinematic configurations, and then try to find ways to do analytical continuation for other configurations. This is the goal we are going to pursue in this work. Below, we introduce the main results of this work before detailed expositions in subsequent sections.

\paragraph{Summary of main results.}
In this work, we tackle the problem of analytically computing tree-level inflation correlators with arbitrary number of massive exchanges, via a standard bulk calculation in the SK formalism. The main technical tool is the partial Mellin-Barnes (PMB) representation proposed in \cite{Qin:2022lva,Qin:2022fbv}. The basic idea is very simple: One takes the Mellin-Barnes representation for all factorized bulk propagators, but leaves all the bulk-to-boundary propagators in the original time-momentum representation. Also, one leaves all the time-ordering Heaviside $\theta$-functions untransformed. In this way, one takes the advantage of Mellin-Barnes (MB) representation that it resolves complicated bulk mode functions into simple powers, but still retains the explicit time-domain representation for external modes. As has been shown in several previous works, the PMB representation is suitable for analyzing a range of problems related to inflation correlators, including explicit results at tree and loop levels \cite{Qin:2022lva,Qin:2022fbv}, and the analytical properties and on-shell factorizations for arbitrary loop correlators \cite{Qin:2023bjk,Qin:2023nhv}. The general procedure of using PMB representation to compute an arbitrary tree-level inflation correlator is detailed in Sec.\ \ref{sec_tree}.

As mentioned above, the time orderings are not removed in the PMB representation. So, we still need to deal with them. We solve this problem in Sec.\ \ref{sec_timeint}. As we will see, the PMB representation greatly simplifies the integrand of nested time integrals. As a result, the most general nested time integral we have to compute takes the following form:
\begin{align}
  \mathbb{T}_{q_1\cdots q_V}(E_1,\cdots,E_V)=\int\prod_{\ell=1}^V\Big[\di\tau_{\ell}(-\tau_\ell)^{q_\ell-1}e^{\ii E_{\ell}\tau_\ell}\Big]\prod_{i,j}\theta(\tau_i-\tau_j),
\end{align}
where we have time integrals at $V$ vertices, nested arbitrarily by the Heaviside $\theta$ functions from the $I$ internal lines. While this integral is still somewhat complicated, it is already in a form that allows us to directly write down the analytical answer. The way to make progress is to recognize that every bulk propagator has a time ordering in a fully nested integral, and we are free to flip the direction of time orderings using a simple relation of the Heaviside $\theta$-function, so that any nested integral can be recast into a partially ordered form. To explain the partial ordering, we adopt this convenient terminology: Whenever we have a factor $\theta(\tau_i-\tau_j)$, we call $\tau_j$ to be $\tau_i$'s mother and $\tau_i$ to be $\tau_j$'s daughter. Then, a partially ordered graph simply means that every time variable in the graph can have any number of daughters but must have only one mother, except the earliest member, who is motherless. In plain words, a partially ordered graph can be thought of as a maternal family tree.

After rewriting a given nested integral into a partially ordered form, we get new terms with less layers of nested integrals, which can be further rewritten into partially ordered form with additional terms generated. This procedure can be carried out recursively, until all nested integrals are partially ordered. This procedure has a very similar structure with the conventional cluster decomposition in statistical mechanics or quantum field theory. We will call it family-tree decomposition. Then, each of the partially ordered nested integrals is a family tree, which we also call ``family'' or ``family integral'' for short. A family is denoted by $\mathcal{C}_{q_1\cdots q_N}(E_1,\cdots,E_N)$. The details of this family-tree decomposition will be presented in Sec.\ \ref{sec_family}. In practice, this family-tree decomposition takes a very simple form. An example of family-decomposing a graph with 5-layer nested integral is shown in Fig.\ \ref{fig_fivesite}. 

As we shall see, the partial order structure allows us to find a simple one-line formula for general family integral $\mathcal{C}_{q_1\cdots q_N}(E_1,\cdots,E_N)$. Working in the configurations where $E_1\gg E_i$ with $i=2,\cdots,N$, we find:
\begin{align}  
\label{eq_CqIntro}
   {\mathcal{C}}_{q_1\cdots q_N}(\wh{E}_1,E_2,\cdots,E_N)=\FR{1}{(\ii E_1)^{q_{1\cdots N}}}\sum_{n_2,\cdots,n_N=0}^\infty \Gamma(q_{1\cdots N}+n_{2\cdots N})\prod_{j=2}^N\FR{(-E_j/E_1)^{n_j}}{(\wt{q}_j+\wt{n}_j)n_j!}.
\end{align}
Here the hatted energy $\wh{E}_1$ denotes the maximal energy, which sits at the vertex with the earliest time. On the right hand side, we have $N-1$ layers of summations corresponding to the $N-1$ descendents of $E_1$-site. We use shorthands such as $q_{1\cdots N}\equiv q_1+\cdots+q_{N}$. The quantity $\wt q_j$ means to take the sum of all $q_i$ where either $i=j$ or $i$ is a descendent of $j$. $\wt n_j$ is similarly defined. Explicit application of this formula to the 5-layer graph in Fig.\ \ref{fig_fivesite} is given in (\ref{eq_C5site1})-(\ref{eq_C5site4}). We give many examples and also a general proof of the formula (\ref{eq_CqIntro}) in Sec.\ \ref{sec_example} and Sec.\ \ref{sec_general}.

One important point is that the maximal energy variable can be chosen at will: To take the analytical continuation of (\ref{eq_CqIntro}) to kinematic regions where $E_1$ is no longer maximal, all we need to do is to rearrange the original integral into a different partial order such that the new maximal energy sits at the earliest time. Thus, our method provides a practical way to do analytical continuation of multi-variable hypergeometric series  (\ref{eq_CqIntro}) beyond its convergence region. An example of this analytical continuation is given for the example of 5-layer integral in Fig.\ \ref{fig_fivesite2}. As will be shown in Sec.\ \ref{sec_alter}, one can exploit the flexibility of MB representation to rewrite (\ref{eq_CqIntro}) as Taylor series of the sum of several or all energy variables, which further extends the domain of validity of our expressions. 

With the formula for general time integrals at hand, the computation of the tree-level inflation correlators becomes a matter of collecting appropriate Mellin poles in the PMB representation. As a demonstration of this procedure, we compute the general tree-level graphs with two bulk massive exchanges in Sec.\ \ref{sec_2mass} and present the full analytical result of this type of correlators for the first time. In Sec.\ \ref{sec_4point}, we show how to take folded limits of these results, by computing a tree-level 4-point graph with two massive exchanges. We conclude the paper with further discussions in Sec.\ \ref{sec_concl}. Useful mathematical formulae on Mellin-Barnes representations and hypergeometric functions are collected in App.\ \ref{app_math}, and some intermediate steps of computing graphs with two massive exchanges are collected in App.\ \ref{app_detail}.

\paragraph{Notation and convention.} We work in the Poincaré patch of the dS spacetime with inflation coordinates $(\tau,\bm x)$ where $\tau\in (-\infty,0)$ is the conformal time, and $\bm x\in\mathbb{R}^3$ is the comoving coordinate. In this coordinate system, the spacetime metric is $\di s^2 = a^2(\tau)(-\di\tau^2+\di\bm x^2)$, where $a(\tau)=-1/(H\tau)$ is the scale factor, and $H$ is the inflation Hubble parameter. We set $H=1$ throughout this work for simplicity. 
We use bold letters such as $\bm k$ to denote 3-momenta and the corresponding italic letter $k \equiv |\bm k|$ to denote its magnitude, which is also called an energy. The energies are often denoted by $E_i$, and the energy ratios such as $\varrho_{ij}\equiv E_i/E_j$ are often used. We follow the diagrammatic methods reviewed in \cite{Chen:2017ryl} to compute inflation correlators in SK formalism.
We often use shorthand for sums of several indexed quantities. Examples include:
\begin{align}
  &k_{12}\equiv k_1+k_2, 
  &&E_{ij}\equiv E_i+E_j, 
  &&s_{123}\equiv s_1+s_2+s_3, 
  &&q_{1\cdots N}\equiv q_1+\cdots+q_N. 
\end{align}
Finally, the Mellin integral measures are very often abbreviated in the following way:
\bge
  \int_{s_1,s_2\cdots}\equiv\int_{-\ii\infty}^{+\ii\infty}\FR{\di s_1}{2\pi\ii}\FR{\di s_2}{2\pi\ii}\cdots.
\ede

\section{Tree Graphs with Partial Mellin-Barnes}
\label{sec_tree}

In this section, we review the method of PMB representation for a general tree-level inflation correlator with arbitrary massive exchanges. Our starting point is a general $B$-point connected equal-time correlation function of a bulk field $\varphi$ in the late time limit:
\bge
  \lim_{\tau\to 0}\big\la \Omega \big| \varphi_{\bm k_1}(\tau)\cdots \varphi_{\bm k_B}(\tau)\big| \Omega \big\ra_\text{C}=(2\pi)^3\de^{(3)}(\bm k_1+\cdots+\bm k_B)\mathcal{T}(\bm k_1,\cdots,\bm k_B).
\ede
As shown above, the correlation function is defined as an equal-time expectation value of the product of $B$ operators $\varphi_{\bm k_i}$ in 3-momentum space, over a state $|\Omega\ra$ which is taken asymptotic to the Bunch-Davies vacuum state in the early time limit $\tau\to-\infty$. We assume that the bulk theory of $\varphi$ is a weakly coupled local quantum field theory. Therefore, after stripping off the momentum-conserving $\de$-function, the amplitude on the right hand side $\mathcal{T}(\bm k_1,\cdots,\bm k_B)$ can be represented as an expansion of connected graphs $\mathcal{G}(\bm k_1,\cdots,\bm k_B)$ with increasing number of loops. Thus, the leading contribution is from the tree graphs, which are the focus of this work. 

We do not specify the type of the field $\varphi$, but we do assume that it has a simple mode function $\varphi(k,\tau)$. More explicitly, if we expand the mode $\varphi_{\bm k}$ in terms of canonically normalized creation and annihilation operators $a_{\bm k}$ and $a_{-\bm k}^\dag$, we get mode function $\varphi(k,\tau)$ as the coefficient: 
\begin{align}
  \varphi_{\bm k}(\tau)=\varphi(k,\tau)a_{\bm k}+\varphi^*(k,\tau)a_{-\bm k}^\dag.
\end{align}
We suppress helicity indices if there are any. We assume that all the time dependence of the mode function $\varphi(k,\tau)$ can be expressed as an exponential factor $e^{-\ii k\tau}$ times a polynomial of $-k\tau$. This covers essentially all cases relevant to cosmological collider phenomenology where the mode function survives the late-time limit, including the massless spin-0 inflaton field and the massless spin-2 graviton. For instance, the mode function for the inflaton is given by:
\bge
\label{eq_mlmode}
  \varphi(k,\tau)=\FR{1}{\sqrt{2k^3}}(1+\ii k\tau)e^{-\ii k\tau}.
\ede
Our assumption also covers the case where the mode does not survive the late-time limit but is of theoretical interest, such as a conformal scalar $\phi_c$ with mass $m_c=\sqrt2$ in $3+1$ dimensions, whose mode function is:
\bge
  \phi_c(k,\tau)=\FR{\tau}{\sqrt{2k}}e^{-\ii k\tau}.
\ede 
The bulk fields appearing in the tree graphs of $\mathcal{T}(\bm k_1,\cdots,\bm k_B)$ can be rather arbitrary. In general, they can have arbitrary mass and spin. They can also have dS-boost-breaking dispersion relations, and thus can have nonzero (helical) chemical potential or non-unit sound speed. They can also have rather arbitrary couplings among themselves and to the boundary field $\varphi$. In particular, these couplings can break dS boosts and even the dilatation symmetry. However, we do assume that these couplings are well behaved in the infrared so that the diagrammatic expansion remains perturbative in the late-time limit.

However, for definiteness, we shall take a fixed type of bulk field, namely a scalar field in the principal series (i.e., with mass $m>3/2$), in all the following discussions. Generalization to other cases should be straightforward. For a massive scalar with $m>3/2$, it is convenient to introduce a \emph{mass parameter} $\wt\nu\equiv\sqrt{m^2-9/4}$. Then, according to the SK formalism \cite{Chen:2017ryl}, we can construct four bulk propagator $D_{\aa\bb}^{(\wt\nu)}(k;\tau_1,\tau_2)$ with $\aa,\bb=\pm$ for such a field. More explicitly:
\begin{align}
\label{eq_Dmp}
  D_{-+}^{(\wt\nu)}(k;\tau_1,\tau_2)
  =&~\FR{\pi}{4}e^{-\pi\wt\nu}(\tau_1\tau_2)^{3/2}\mathrm{H}_{\ii\wt\nu}^{(1)}(-k\tau_1)\mathrm{H}_{-\ii\wt\nu}^{(2)}(-k\tau_2),\\
\label{eq_Dpm}
  D_{+-}^{(\wt\nu)}(k;\tau_1,\tau_2)
  =&~\Big[D_{-+}^{(\wt\nu)}(k;\tau_1,\tau_2)\Big]^*, \\
\label{eq_Dpmpm}
  D_{\pm\pm}^{(\wt\nu)}(k;\tau_1,\tau_2)=&~D_{\mp\pm}^{(\wt\nu)}(k;\tau_1,\tau_2)\theta(\tau_1-\tau_2)+D_{\pm\mp}^{(\wt\nu)}(k;\tau_1,\tau_2)\theta(\tau_2-\tau_1).
\end{align}
Then, a general tree graph consisting of massive scalar bulk propagators and massless/conformal scalar bulk-to-boundary propagators can be computed by an integral of the following form:
\begin{align}
\label{eq_SKint}
  \mathcal{I}=\sum_{\aa_1,\cdots,\aa_V=\pm}\int\prod_{\ell=1}^V\Big[\di\tau_\ell\,\ii\aa_\ell(-\tau_\ell)^{p_\ell}e^{\ii\aa_\ell E_\ell\tau_\ell}\Big]\prod_{i=1}^I D_{\aa_{i1}\aa_{i2}}(K_i,\tau_{i1},\tau_{i2}).
\end{align}
Here we assume that there are $V$ vertices and $I$ bulk propagators in the graph. For each vertex, we have an integral over the conformal time variable $\tau_\ell$ ($\ell=1,\cdots,V$). Also, we introduce a factor of $\ii\aa_\ell$ as required by the diagrammatic rule \cite{Chen:2017ryl}, and a factor of $(-\tau_\ell)^{p_\ell}$ to account for various types of couplings as well as power factors in the external mode function (such as the $\ii k\tau$ term in the massless mode function (\ref{eq_mlmode})). The exponential factor $e^{\ii a_\ell E_\ell\tau_\ell}$ comes from the external mode function, and $E_\ell$ represents the sum of magnitudes of 3-momenta of all \emph{external} modes. Following the terminology in the literature, we call it the \emph{energy} at the Vertex $\ell$. However, we note that $E_\ell$ is not the total energy at Vertex $\ell$ since we do not include energies of bulk lines. For each bulk line, we have a bulk propagator $D_{\aa_{i1}\aa_{i2}}(K_i,\tau_{i1},\tau_{i2})$ with momentum $\bm K_i$, which is completely determined by external momenta via the 3-momentum conservation at each vertex. The two time variables $\tau_{i1},\tau_{i2}$ as well as the two SK variables $\aa_{i1},\aa_{i2}$ should be identified with the corresponding time and SK variables at the two vertices to which the bulk propagator attach. 

The computation of the integral (\ref{eq_SKint}) is complicated by the products of Hankel functions, as well as the time-ordering $\theta$-functions in the bulk propagators. To tackle these problems, we use the MB representation for all the \emph{bulk} propagators, but leave all the bulk-to-boundary propagators untransformed. This is the so-called PMB representation \cite{Qin:2022lva,Qin:2022fbv}. The MB representations for the two opposite-sign bulk propagators (\ref{eq_Dmp}) and (\ref{eq_Dpm}) are given by \cite{Qin:2022lva,Qin:2022fbv}:
\begin{align}
\label{eq_DScalarMB1}
    D_{\pm\mp}^{(\wt\nu)}(k;\tau_1,\tau_2) =&~ \FR{1}{4\pi}
    \int_{-\ii\infty}^{+\ii\infty}
    \FR{\di s}{2\pi\ii}\FR{\di \bar s}{2\pi\ii}\,
    e^{\mp\ii\pi(s-\bar s)}\Big(\FR{k}2\Big)^{-2(s+\bar s)}
    (-\tau_1)^{-2s+3/2}(-\tau_2)^{-2\bar s+3/2}\n\\
    &\times \Gamma\Big[s-\FR{\ii\wt\nu}2,s+\FR{\ii\wt\nu}2,\bar s-\FR{\ii\wt\nu}2,\bar s+\FR{\ii\wt\nu}2\Big],
\end{align} 
This follows directly from the MB representation of the Hankel function (\ref{eq_HankelMB}), which we collect in App.\ \ref{app_math}. In particular, the Mellin variable $s$ is associated with time $\tau_1$ and $\bar s$ is associated with $\tau_2$. The same-sign propagators $D_{\pm\pm}$ are obtained by substituting in the above expression into (\ref{eq_Dpmpm}). We note that the time-ordering $\theta$-functions are left untransformed. 

After taking the above PMB representation, the original SK integral (\ref{eq_SKint}) becomes:  
\begin{align}
\label{eq_SKintMB}
  \mathcal{I}
  =&\int_{-\ii\infty}^{+\ii\infty}\prod_{i=1}^I\bigg\{\FR{1}{4\pi}\FR{\di s_i}{2\pi\ii}\FR{\di\bar s_i}{2\pi\ii}\Big(\FR{K}2\Big)^{-2s_{i\bar i}}
  \Gamma\Big[s_i-\FR{\ii\wt\nu}2,s_i+\FR{\ii\wt\nu}2,\bar s_i-\FR{\ii\wt\nu}2,\bar s_i+\FR{\ii\wt\nu}2\Big]\bigg\}\n\\
  &\times\bigg\{\sum_{\aa_1,\cdots,\aa_V=\pm}\int_{-\infty}^0\prod_{\ell=1}^V\Big[\di\tau_\ell\,\ii\aa_\ell(-\tau_\ell)^{p_\ell-2\sum_\ell s}e^{\ii\aa_\ell E_\ell\tau_\ell}\Big]\mathcal{N}_{\aa_1\cdots\aa_V}\Big(\tau_1,\cdots,\tau_V;\{s,\bar s\}\Big)\bigg\}.
\end{align}
Here we have switched the order of the time integral and the Mellin integral, assuming all integrals are well convergent. With this representation, we see that all SK-index-dependent part goes into the time integral, namely the second line of the above expression. In this time integral, we have used a shorthand $(-\tau_\ell)^{p_\ell-2\sum_\ell s}$ where $\sum_\ell s$ denotes the sum of all Mellin variables associated to $\tau_\ell$, and the Mellin variables in this summation can be either barred or unbarred. An important fact we shall use below is that the Mellin variables always appear with negative signs in this exponent. Also, we have introduce a function $\mathcal{N}_{\aa_1\cdots\aa_V}(\tau_1,\cdots,\tau_V;\{s,\bar s\})$ to represent all combinations of time-ordering $\theta$-functions, as well as the SK-index-dependent phase factor $e^{\mp\ii\pi(s-\bar s)}$ in (\ref{eq_DScalarMB1}).

The reason we introduce the PMB representation is that the time integral now only involves exponentials and powers in its integrand, as shown in the second line of (\ref{eq_SKintMB}). This is significantly simpler than the original time integral, which involves Hankel functions. While this simplification is powerful enough for a single layer time integral, the computation of time-ordered integrals remain nontrivial. In previous works using PMB representation, only the two-layer nested integral was explicitly computed \cite{Qin:2022fbv}:
\begin{align}
  \int_{-\infty}^0(-\tau_1)^{q_1-1}(-\tau_2)^{q_2-1}e^{\ii E_1\tau_1+\ii E_2\tau_2}\theta(\tau_2-\tau_1)=\FR{1}{(\ii E_1)^{\ii q_{12}}}\,{}_2\mathcal{F}_1\left[\bgm q_2,q_{12}\\ q_2+1\edm\middle|-\FR{E_2}{E_1}\right],
\end{align}
where ${}_2\mathcal{F}_1$ is the dressed hypergeometric function, defined in App.\ \ref{app_math}. For computing inflation correlators with single massive exchange, this result is enough. However, if we wish to go beyond the single massive exchange and consider the most general tree graphs, it is necessary to tackle the problem of computing time integrals of exponentials and powers with arbitrary layers and arbitrary time orderings. We will systematically solve this problem in the next section. 

From (\ref{eq_SKintMB}), we see that, if the time integral in the second line can be done, then it only remains to finish the Mellin integrals. This is typically done by closing the Mellin contour and collecting the residues of all enclosed poles. So, we need knowledge about the pole structure of the Mellin integrand. Although the answer to the time integral was not explicitly known in the previous studies, it was proved in \cite{Qin:2023bjk} that such time integrals, however nested, only contribute \emph{right poles} of the Mellin integrand. That is, their poles only appear on the right side of the integral contour that goes from $-\ii\infty$ to $+\ii\infty$. As a result, all \emph{left poles} are contributed by the $\Gamma$-factors from the bulk propagators, shown in the first line of (\ref{eq_SKintMB}). These are all the poles of the Mellin integrand for a tree graph. 

Another important observation is that, if we sum the \emph{arguments} of all $\Gamma$-factors in the first line of (\ref{eq_SKintMB}), we get:
\bge
  +2\sum_{i=1}^I (s_{i}+\bar s_{i})+\cdots.
\ede
That is, all Mellin variables are summed together, with an overall coefficient $+2$. Here ``$\cdots$'' denotes $s$-independent terms, which are irrelevant to our current argument, and happen to be 0 in this particular case.  On the other hand, as we shall see from the explicit results in the next section, the right poles contributed by the time integrals are also from $\Gamma$-factors of the form $\Gamma[\cdots - 2\sum s]$. If we sum over the arguments of all right poles, we will get:
\bge
\label{eq_RightS}
  -2\sum_{i=1}^I (s_{i}+\bar s_{i})+\cdots,
\ede
which is exactly the $s$ terms from the left-pole $\Gamma$-arguments with an additional sign. In this sense, we say that the Mellin variables in the integrand are \emph{balanced}. In such a balanced situation, the convergence of the Mellin integral is determined by the power factors such as $(K_i/2)^{-2s_{i\bar i}}$ in (\ref{eq_SKintMB}). Typically, one can first work in the kinematic region where the internal momenta $K_i$ are small (compared to relevant external energies), so that the Mellin integrals will be convergent if we pick up all the \emph{left poles}, which are all from the bulk propagator $\Gamma$-factors in (\ref{eq_DScalarMB1}). Their poles and residues are well understood. So, if we can finish the time integral, then we only need to collect all left poles from the first line of (\ref{eq_SKintMB}). The result will be a series expansion in $K_i$. So, this result will be valid at least when the bulk momenta $K_i$ are not too large.\footnote{\label{fn_internal}There is a degenerate situation where a vertex is \emph{internal}, in the sense that it is not attached to any bulk-to-boundary propagator. In this case, the energy variable $E=0$ at this vertex. Then, if we compute a factorized time integral for this vertex alone, we get a $\de$ function for Mellin variables. Consequently, one should integrate out one Mellin variable using this $\de$ function instead of picking up poles. We will come back to this point at the end of Sec.\ \ref{sec_timeint}.} In the opposite limit, when $K_i$ becomes large compared to the relevant energy variables, we can instead close the Mellin contour from the right side and pick up all the right poles. In this way, we get analytical continuation of the result from small $K_i$ region to large $K_i$ region. This will cover most of the parameter space of interest. The narrow intermediate region will be difficult to be expressed by a series solution. Analytically, one needs to take the analytical continuation of the series solutions for those intermediate regions, which is a separate mathematical problem. Practically, however, we can use numerical interpolation to bridge the gap between different regions. This strategy has been shown to be workable in previous studies \cite{Xianyu:2022jwk}. So, barring possible issues of analytical continuation for special configurations, we can say that, the problem of analytical computation of arbitrary tree-level inflation correlators is solved, if we can compute the arbitrary nested time integral. We will solve the latter problem in the next section.

\section{Time Integrals with Partial Mellin-Barnes}
\label{sec_timeint}

In this section we provide a systematic investigation of arbitrary nested time integrals in the PMB representation. It is clear from the previous section that the most general nested time integral has the following form:
\begin{align}
\label{eq_NTI}
  \mathbb{T}_{q_1\cdots q_V}(E_1,\cdots,E_V)=\int\prod_{\ell=1}^V\Big[\di\tau_{\ell}(-\tau_\ell)^{q_\ell-1}e^{\ii E_{\ell}\tau_\ell}\Big]\prod_{i,j}\theta(\tau_i-\tau_j).
\end{align}
Here we are again considering a $V$-fold time integral with arbitrary nesting. We require that all $\tau_i$ ($1\leq i\leq V$) appear in the $\theta$-factors so that the integral is fully nested. Also, we have used a factor $(-\tau_\ell)^{q_{\ell}-1}$ to account for a variety of external modes and couplings, as well as powers of time from the partial MB representation. In the notation of the previous section, we have:
\bge
\label{eq_qtos}
  q_\ell -1=p_\ell-2\sum\nolimits_\ell s.
\ede 

The difficulty with time ordering in (\ref{eq_NTI}) is easy to understand: A single time integral of exponential with power factors from $\tau=-\infty$ to $\tau=0$ gives rise to a $\Gamma$ function. However, if there is a time ordering, the integration limit for one time variable would be dependent on another integration variable. As a result, we get incomplete $\Gamma$ functions after finishing one layer of integral. Then we need to perform time integrals over incomplete $\Gamma$ functions with integration limits dependent on yet another time variable. This quickly becomes intractable with an increasing number of nested layers. 

Our strategy of solving this problem is again the Mellin-Barnes representation: whenever we perform a layer of nested time integration, we take the MB representation of the result so that the integrand for the next layer is still a simple exponential times a power. In this way, the nested time integrals can be done recursively layer by layer, until the last layer, which yields a simple $\Gamma$ factor. Along the way, we generate many layers of Mellin integrals, which can again be done by closing the contours properly. 

As we shall see below, this recursive integration is easiest if the time integral is nested with a partial order, which is not the case in the most general nested integrals. Thus, we should first use a simple relation $\theta(\tau_j-\tau_k)+\theta(\tau_k-\tau_j)=1$ to reorganize the original time integral such that the result is either partially ordered or factorized. This will be called a ``family-tree decomposition.'' Then, we apply the above procedure to the partially ordered integrals to get the explicit results for them. These steps will be carried out in detail below.

A side remark on notation and terminology: It will be helpful to use a diagrammatic representation for the nested time integral (\ref{eq_NTI}). We will use a directional line to denote a $\theta$-function where the direction of the arrow coincides with the direction of the time flow. Two factorized time variables (which are simply associated with a factor of 1) may be connected by a dashed line. So, for instance, we can write the relation $\theta(\tau_1-\tau_2)+\theta(\tau_2-\tau_1)=1$ as:
\bge
  \parbox{120mm}{\includegraphics[width=120mm]{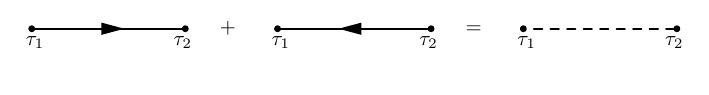}}
\ede
Also, to highlight the fact that these diagrams are not the original SK graphs for the inflation correlators, we will use ``site'' in place of ``vertex,'' and use ``line'' in place of ``propagators.'' Then, each site $\tau_i$ is associated with an energy variable $E_i$ and an exponent $q_i$, as is clear from (\ref{eq_NTI}). 

\subsection{Family-tree decomposition of nested integrals}
\label{sec_family}

Now, we describe our family-tree decomposition algorithm in detail.
We begin with the most general nested time integral (\ref{eq_NTI}). After finishing all the time integrals, the result $\mathbb{T}_{q_1\cdots q_V}(E_1,\cdots,E_V)$ is a function of $V$ energies $E_1,\cdots, E_V$ and $V$ exponents $q_1,\cdots,q_V$. In the following, we shall show that this integral can always be written as a sum over a finite number of terms. Each term is a product of several \emph{families}. Each family is a multi-variable hypergeometric function of several energy variables $E_i$. Of course, multi-variable hypergeometric functions are not well studied. It is most useful if we can find a fast converging series expansion of this hypergeometric function in terms of any given small energy ratios. Below, we will show that this can be done.

\paragraph{The reduction procedure.}
Our reduction procedure consists of the following simple steps:

\begin{description}
  \item[Step 1:] We start with a particular kinematic region of the integral $\mathbb{T}_{q_1\cdots q_V}(E_1,\cdots,E_V)$, where there is a \emph{largest} energy, say $E_i$, such that $E_i>E_j$ for all $j\neq i$. We want to find an analytical expression for $\mathbb{T}_{q_1\cdots q_V}(E_1,\cdots,E_V)$ as a series in $1/E_i$, which should be convergent in most of the region where $E_i$ remain the largest. We add a hat to the largest energy variable $\wh{E}_i$ to highlight the fact that we are considering a particular kinematic region. 
  
  So, if we choose $E_1$ to be the largest energy, we will write $\mathbb{T}_{q_1\cdots q_V}(\wh E_1,E_2,\cdots,E_V)$ to highlight this choice. If, instead, we want to consider the case where $E_2$ gets larger than $E_1$, then we should add the hat on $\wh{E}_2$. The degenerate case where there are multiple maximal energies will be considered in following subsections.  
\end{description}
   
\begin{description}  
  \item[Step 2:] We use the relation $\theta(\tau_j-\tau_k)+\theta(\tau_k-\tau_j)=1$ to flip the direction of time flows in some bulk lines, such that the original graph is broken into a sum of several terms. Each term can be represented as a graph, in which all sites are either partially ordered or factorized. As a result, each graph becomes a product of several integrals, each of which has a partial order structure, and is called a \emph{family}. As a part of the rule, we require that the maximal energy site has the earliest time in a family.
\end{description}
Let us define what is a (partially ordered) family. Clearly, a time-ordered line connects a site with an earlier time to another site with a later time. We call the earlier-time site the \emph{mother} of the later-time site, and call the later-time site the \emph{daughter} of the earlier-time site. Then, a partially ordered graph means that every site has a unique mother, except the maximal-energy site, which is the earliest-time site and motherless. On the other hand, a mother can have many daughters. In this way, all sites within a family integral genuinely belong to a family. Also, for a given site, we call all the sites flowing out of it the \emph{descendant sites}. Thus, the descendant sites of a given site consist of its daughters, granddaughters, great-granddaughters, etc.
  
Let us rephrase the above heuristic language into a more rigorous definition. We will use $\mathcal{C}_{q_1\cdots q_N}(\wh{E}_1,E_2,\cdots,E_N)$ to denote a family integral with $N$ sites, where we have highlighted the maximal energy $\wh{E}_1$ with a hat. Then, a family integral $\mathcal{C}_{q_1\cdots q_N}(\wh{E}_1,E_2,\cdots,E_N)$ has the following form:
\bge
\label{eq_Cint}
  \mathcal{C}_{q_1\cdots q_N}(\wh{E}_1,E_2,\cdots,E_N)=\int\prod_{\ell=1}^N\Big[\di\tau_{\ell}(-\tau_\ell)^{q_\ell-1}e^{\ii E_{\ell}\tau_\ell}\Big]\prod_{i,j}\theta(\tau_i-\tau_j),
\ede
with the following restrictions on the $\theta$-function factors:
\begin{enumerate}
  \item Every time variable $\tau_i$ $(1\leq i\leq N)$ appears in time-ordering $\theta$-function. (All sites belong to a family.)
  \item In a factor such as $\theta(\tau_j-\tau_k)$, let us call $\tau_j$ to be in the late position and $\tau_k$ in the early position. Then, it is required that every variable $\tau_i$ except the maximal energy site appears in the late position once and only once. (Every site has a unique mother except the maximal energy site.) On the other hand, early positions can be taken more than once by a given $\tau_i$. (A mother can give birth to more than one daughter.) 
  \item The maximal energy site $\tau_1$ appears in $\theta$ factors only in the early position. (The maximal energy site is motherless, but can have any (including zero) number of daughters.)
\end{enumerate}

\begin{description}
  \item[Step 3:] After taking Step 2, each resulting graph is a product of several fully factorized families. The maximal energy site sits in a particular family, which we call the maximal-energy family. As a consequence, families other than the maximal-energy family are independent of the maximal energy variable $\wh{E}_i$, and it becomes meaningless to ask for a series expansion in $\wh{E}_i$ for those families. We call them non-maximal energy families. Thus, for each of the non-maximal energy families, we should further assign a ``locally'' maximal energy, such that this energy is largest among all energies \emph{within} the family. Then, we further perform the reduction of Step 2 for all non-maximal energy families and we do this procedure recursively, until, within each family, the locally maximal energy site sits at the earliest time.
\end{description}

\begin{description} 
  \item[Step 4:] After taking the above steps, we fully reduce the original integral $\mathbb{T}_{q_1\cdots q_V}(\wh E_1,E_2,\cdots,E_V)$ into a sum of products of partially ordered families, and in each family, the locally maximal energy acquires the earliest time.
  
  It then remains to state the rule for directly writing down the answer for arbitrary families. The rule is the following: 1) Within each partially ordered family, we assign a summation variable $n_i$ for all sites except the (locally) maximal-energy site. Without loss of generality, we can always relabel the sites within a family such that the (locally) maximal energy is $E_1$. Then, for the $N$-site family defined in (\ref{eq_Cint}), the result is:
\end{description}
  \begin{keyeqn}
  \begin{align}
  \label{eq_family}
   &\mathcal{C}_{q_1\cdots q_N}(\wh{E}_1,E_2,\cdots,E_N)=\FR{1}{(\ii E_1)^{q_{1\cdots N}}}\wt{\mathcal{C}}_{q_1\cdots q_N}(\wh{E}_1,E_2,\cdots,E_N),\n\\
   &\wt{\mathcal{C}}_{q_1\cdots q_N}(\wh{E}_1,E_2\cdots,E_N)=\sum_{n_2,\cdots,n_N=0}^\infty \Gamma(q_{1\cdots N}+n_{2\cdots N})\prod_{j=2}^N\FR{(-\varrho_{j1})^{n_j}}{(\wt{q}_j+\wt{n}_j)n_j!}.
  \end{align}
  \end{keyeqn}
Here, the hatted energy $\wh{E}_1$ represents the maximal energy. In the first line, we stripped away a dimensionful factor $(\ii E_1)^{q_{1\cdots N}}$ so that the resulting integral $\wt{C}_{q_1\cdots q_N}(\wh{E}_1,E_2,\cdots,E_N)$ is dimensionless. In the second line, we have defined $\varrho_{jk}\equiv E_j/E_k$. Also, $\wt n_i$ is defined to be the sum of $n_i$-variables over the site $i$ and all its descendants. $\wt q$ is similarly defined.

This completes our reduction of the original time integral into a sum of products of hypergeometric series.

\begin{figure}
\centering 
\includegraphics[width=\textwidth]{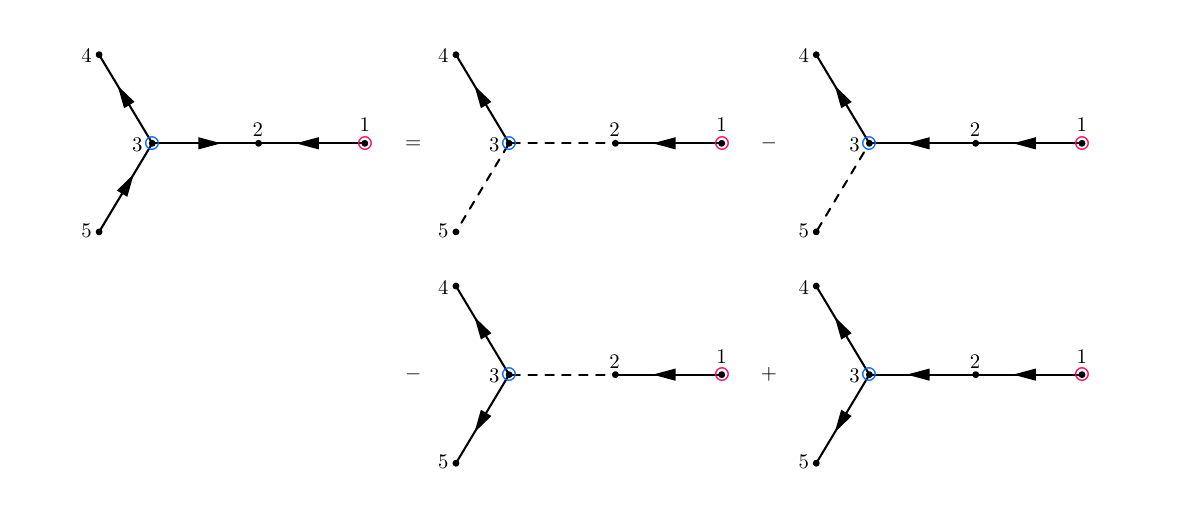}\caption{The diagrammatic representation of (\ref{eq_skeexpansion}), showing the reduction of a 5-layer time integral into partially ordered families. In this example, we choose $E_1>E_2,E_3,E_4,E_5$ and $E_3>E_4,E_5$. The maximal energy site (Site 1) is marked with a magenta circle and the locally maximal energy site (Site 3) is marked with a blue circle.} 
  \label{fig_fivesite}
\end{figure}

\paragraph{Example.} As often happens, it is better to demonstrate an algorithm with examples than mere abstract description. So, now, let us demonstrate the above reduction procedure with a concrete example. Suppose we want to compute a 5-layer time integral:
\begin{align}
\label{eq_5site}
  \mathbb{T}_{q_1\cdots q_5}(E_1,\cdots,E_5)\equiv\int_{-\infty}^0\prod_{\ell=1}^5\Big[\di\tau_{\ell}(-\tau_\ell)^{q_\ell-1}e^{\ii E_{\ell}\tau_\ell}\Big]
  \theta(\tau_2-\tau_1)\theta(\tau_2-\tau_3)\theta(\tau_4-\tau_3)\theta(\tau_3-\tau_5).
\end{align}
Furthermore, suppose that we want to consider the kinematic region where $E_1$ is the largest energy among all five energies. Thus, we want to express the final result as a series expansion of $1/E_1$. This is shown on the left hand side of Fig.\ \ref{fig_fivesite}, where the magenta-circled site represents the maximal-energy site. Then, according to the above procedure, we should use the relation $\theta(\tau_i-\tau_j)+\theta(\tau_j-\tau_i)=1$ to change the direction of several lines, such that 1) all sites become either partially ordered or factorized and 2) the maximal-energy site $E_1$ has the earliest time variable. This is done on the right hand side of Fig.\ \ref{fig_fivesite}. In each diagram on the right hand side, we get a product of one or several partially ordered families. 

In all but the last term on the right hand side of Fig.\ \ref{fig_fivesite}, we have families which do not contain the maximal energy site. Thus we should specify locally maximal site for each of them. The one-site family is trivial. The nontrivial non-maximal families appear in the first and third terms on the right hand side of Fig.\ \ref{fig_fivesite}, which can be expressed as $\mathcal{C}_{q_3q_4}(E_3,E_4)$ and $\mathcal{C}_{q_3q_4q_5}(E_3,E_4,E_5)$, respectively. Thus, we should further assign a maximal energy for these two sites. So, let us further work within the region where $E_3>E_4,E_5$, so that $E_3$ is the locally maximal energy, marked with a blue circle in Fig.\ \ref{fig_fivesite}. (On the other hand, the relation between $E_3$ and $E_{2}$ is irrelevant.) Then, we see that $E_3$ is already in the earliest time site in both families. So, we are done, and the result of our reduction procedure can be expressed as:
  \begin{align}
  \label{eq_skeexpansion}
    \mathbb{T}_{q_1\cdots q_5}(\wh E_1,\cdots,E_5)
    =&~\mathcal{C}_{q_1q_2}(\wh{E}_1,E_2)\mathcal{C}_{q_3q_4}(\wh{E}_3,E_4)\mathcal{C}_{q_5}(E_5)-\mathcal{C}_{q_1q_2q_3q_4}(\wh{E}_1,E_2,E_3,E_4)\mathcal{C}_{q_5}(E_5)\n\\
    &-\mathcal{C}_{q_1q_2}(\wh{E}_1,E_2)\mathcal{C}_{q_4q_3q_5}(E_4,\wh{E}_3,E_5)+\mathcal{C}_{q_1q_2q_3q_4q_5}^\text{(iso)}(\wh{E}_1,E_2,E_3,E_4,E_5).
  \end{align}
Here we have also added hats to the locally maximal energy $E_3$. In the last term, we added a superscript (iso) to show that this family has a cubic vertex. See the next subsection for details.

Next, we assign $n_2,\cdots,n_4$ for the sites with energy variables $E_2,\cdots,E_4$, respectively.
It is clear from (\ref{eq_skeexpansion}) that there are four independent nontrivial families (i.e., with more than one site) involved in this example. Applying the formula (\ref{eq_family}) for each of them, we get 
\begin{align}
\label{eq_C5site1}
  &\wt{\mathcal{C}}_{q_1q_2}(\wh{E}_1,E_2)
  =\sum_{n_2=0}^\infty\FR{(-1)^{n_2}\Gamma(q_{12}+n_2)}{(q_2+n_2)}\FR{\varrho_{21}^{n_2}}{n_2!},\\
  &\wt{\mathcal{C}}_{q_1q_2q_3q_4}(\wh{E}_1,E_2,E_3,E_4) 
   =\sum_{n_2,n_3,n_4=0}^\infty\FR{(-1)^{n_{234}}\Gamma(q_{1234}+n_{234})}{(q_{234}+n_{234})(q_{34}+n_{34})(q_{4}+n_{4})}\FR{\varrho_{21}^{n_2}}{n_2!}\FR{\varrho_{31}^{n_3}}{n_3!}\FR{\varrho_{41}^{n_4}}{n_4!},\\
  &\wt{\mathcal{C}}_{q_4q_3q_5}(E_4,\wh{E}_3,E_5)
  =\sum_{n_4,n_5=0}^\infty\FR{(-1)^{n_{45}}\Gamma(q_{345}+n_{45})}{(q_{4}+n_{4})(q_{5}+n_{5})}\FR{\varrho_{43}^{n_4}}{n_4!}\FR{\varrho_{53}^{n_5}}{n_5!},\\
\label{eq_C5site4}
  &\wt{\mathcal{C}}_{q_1q_2q_3q_4q_5}^\text{(iso)}(\wh{E}_1,E_2,E_3,E_4,E_5)\n\\
  &=\sum_{n_2,n_3,n_4,n_5=0}^\infty\FR{(-1)^{n_{2345}}\Gamma(q_{12345}+n_{2345})}{(q_{2345}+n_{2345})(q_{345}+n_{345})(q_{4}+n_{4})(q_{5}+n_{5})}\FR{\varrho_{21}^{n_2}}{n_2!}\FR{\varrho_{31}^{n_3}}{n_3!}\FR{\varrho_{41}^{n_4}}{n_4!}\FR{\varrho_{51}^{n_5}}{n_5!}.
\end{align}
On the other hand, the result for the one-site family is trivial: $\wt{\mathcal{C}}_q(E)=\Gamma(q)$. In fact, some of the above series can be summed to well-known hypergeometric functions, which we shall introduce below. In any case, we have found the series expression for the original 5-layer time integral $\mathbb{T}_{q_1\cdots q_5}(E_1,\cdots,E_5)$ without actually doing any integrals.

\begin{figure}
\centering 
\includegraphics[width=\textwidth]{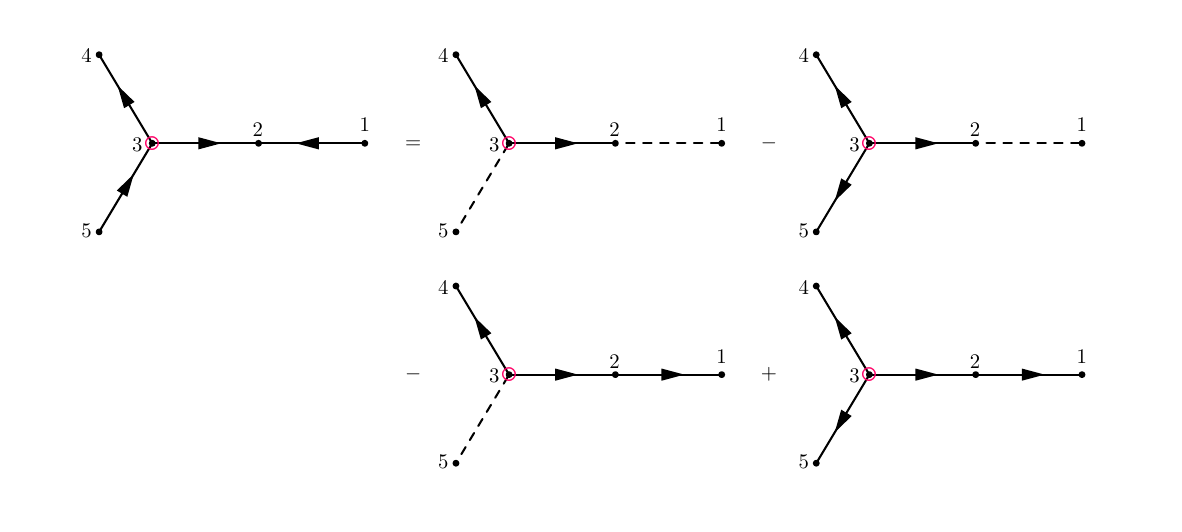}\caption{The diagrammatic representation of (\ref{eq_skeexpansion2}), showing the reduction of a 5-layer time integral into partially ordered families. In this example, we choose $E_3>E_i$ $(i=1,2,4,5)$. The maximal energy site (Site 3) is marked with a magenta circle.} 
  \label{fig_fivesite2}
\end{figure}

The above series solution has a validity range beyond which the summations no longer converge. This happens in particular when any energy $E_i$ ($i=2,3,4,5$) becomes larger than $E_1$. In principle, if we need the result when $E_1$ is no longer maximal, we need to take analytical continuation of the above series. This analytical continuation can be very conveniently implemented in our procedure. To see this, let us have a second look at the 5-site integral $\mathbb{T}_{q_1\cdots q_5}(E_1,\cdots,E_5)$ in (\ref{eq_5site}), but now choose $E_3$ as the maximal energy. Then, according to our procedure, we should do a new family-tree decomposition, as shown in Fig.\ \ref{fig_fivesite2}.
\begin{align}
  \label{eq_skeexpansion2}
    \mathbb{T}_{q_1\cdots q_5}(\wh E_1,\cdots,E_5)
    =&~\mathcal{C}_{q_1}(E_1)\mathcal{C}_{q_2q_3q_4}(E_2,\wh{E}_3,E_4)\mathcal{C}_{q_5}(E_5)-\mathcal{C}_{q_1}(E_1)\mathcal{C}_{q_2q_4q_5q_3}^\text{(iso)}(E_2,E_4,E_5,\wh{E}_3)\n\\
    &-\mathcal{C}_{q_1q_2q_3q_4}(E_1,E_2,\wh{E}_3,E_4)\mathcal{C}_{q_5}(E_5)+\mathcal{C}_{q_1q_2q_3q_4q_5}^\text{(iso)}(E_1,E_2,\wh{E}_3,E_4,E_5).
  \end{align}%
Clearly, we do not need to choose any locally maximal energy in this example. The explicit expressions for the above families can be written down directly according to the general formula (\ref{eq_family}):
\begin{align}
  &\wt{\mathcal{C}}_{q_2q_3q_4}(E_2,\wh{E}_3,E_4)=\sum_{n_2,n_4=0}^\infty\FR{(-1)^{n_{24}}\Gamma(q_{234}+n_{24})}{(q_2+n_2)(q_4+n_4)}\FR{\varrho_{23}^{n_2}}{n_2!}\FR{\varrho_{43}^{n_4}}{n_4!},\\
  &\wt{\mathcal{C}}_{q_2q_4q_5q_3}^\text{(iso)}(E_2,E_4,E_5,\wh{E}_3)=\sum_{n_2,n_4,n_5=0}^\infty\FR{(-1)^{n_{245}}\Gamma(q_{2345}+n_{245})}{(q_2+n_2)(q_4+n_4)(q_5+n_5)}\FR{\varrho_{23}^{n_2}}{n_2!}\FR{\varrho_{43}^{n_4}}{n_4!}\FR{\varrho_{53}^{n_5}}{n_5!},\\
    &\wt{\mathcal{C}}_{q_1q_2q_3q_4}(E_1,E_2,\wh{E}_3,E_4)=\sum_{n_1,n_2,n_4=0}^\infty\FR{(-1)^{n_{124}}\Gamma(q_{1234}+n_{124})}{(q_1+n_1)(q_{12}+n_{12})(q_4+n_4)}\FR{\varrho_{13}^{n_1}}{n_1!}\FR{\varrho_{23}^{n_2}}{n_2!}\FR{\varrho_{43}^{n_4}}{n_4!},\\
    &\wt{\mathcal{C}}_{q_1q_2q_3q_4q_4}^\text{(iso)}(E_1,E_2,\wh{E}_3,E_4,E_5)\n\\
    &=\sum_{n_1,n_2,n_4,n_5=0}^\infty\FR{(-1)^{n_{1245}}\Gamma(q_{12345}+n_{1245})}{(q_1+n_1)(q_{12}+n_{12})(q_4+n_4)(q_5+n_5)}\FR{\varrho_{13}^{n_1}}{n_1!}\FR{\varrho_{23}^{n_2}}{n_2!}\FR{\varrho_{43}^{n_4}}{n_4!}\FR{\varrho_{53}^{n_5}}{n_5!}.   
\end{align}
Thus we have found an expression for the original 5-site integral $\mathbb{T}_{q_1\cdots q_5}(E_1,\cdots,E_5)$ expanded as powers of $1/E_3$. Let us emphasize that (\ref{eq_skeexpansion}) and (\ref{eq_skeexpansion2}) are just different expansions for the same function $\mathbb{T}_{q_1\cdots q_5}(E_1,\cdots,E_5)$, with different validity regions.

\subsection{Partially ordered families: simple examples}
\label{sec_example}

Clearly, the only nontrivial step in our family-tree decomposition procedure is the last step, where we directly write down the answer for the family integral (\ref{eq_family}). The derivation of this result is best illustrated with examples. So in this subsection we will walk the readers through a few simple examples, before presenting a general proof in the next subsection.

\begin{figure}
\centering 
\subfigure[$\mathcal{C}_{q}(E)$]{\label{fig_onevertex}\includegraphics[height=25mm]{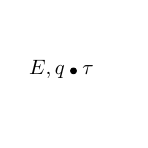}
\vspace{-5mm}} 
\hspace{25mm}
\subfigure[$\mathcal{C}_{q_1q_2}(\wh{E}_1,E_2)$]{\label{fig_twovertex}\includegraphics[height=25mm]{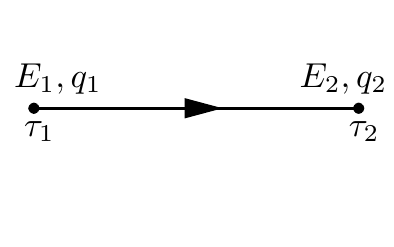}
\vspace{-5mm}} 
\caption{The one-site family and two-site family.} 
  \label{fig_1and2v}
\end{figure}

\paragraph{One-site family.} We begin with the simplest integral, the one-site family, shown in Fig.\ \ref{fig_onevertex}:
\bge
\label{eq_onesite}
  \wt{\mathcal{C}}_{q}(E)={(\ii E)^q}\int_{-\infty}^0 \di\tau\,(-\tau)^{q-1}e^{\ii E\tau}.
\ede
The application of the rule is trivial, and we have the following answer:
\bge
  \wt{\mathcal{C}}_{q}(E)=\Gamma(q).
\ede
The answer is obtained by a direct integration of (\ref{eq_onesite}). Since there is only one dimensionful variable $E$ involved in the problem, the final answer for the dimensionless family $\wt{\mathcal{C}}_q(E)$ must be independent of $E$.

\paragraph{Two-site family.} Next let us look at the simplest nontrivial example, namely the two-site family, shown in Fig.\ \ref{fig_twovertex}. The integral is: 
\begin{align}
\label{eq_twosite}
  \wt{\mathcal{C}}_{q_1q_2}(\wh{E}_1,E_2)=(\ii E_1)^{q_{12}}\int_{-\infty}^0 \di\tau_1\di\tau_2\,(-\tau_1)^{q_1-1}(-\tau_2)^{q_2-1}e^{\ii(E_1\tau_1+E_2\tau_2)}\theta(\tau_2-\tau_1). 
\end{align}
By design, we take $E_1>E_2$. Now let us try to find the answer for the above integral. It turns out useful to start from the integral of reversed time ordering:
\begin{align}
  \mathcal{R}\Big[\wt{\mathcal{C}}_{q_1q_2}(\wh{E}_1,E_2)\Big]\equiv(\ii E_1)^{q_{12}}\int_{-\infty}^0 \di\tau_1\di\tau_2\,(-\tau_1)^{q_1-1}(-\tau_2)^{q_2-1}e^{\ii(E_1\tau_1+E_2\tau_2)}\theta(\tau_1-\tau_2). 
\end{align}
Then, the integral over $\tau_2$ can be performed, with the result expressed in terms of an exponential integral $\mathrm{E}_p(z)$ whose definition is given in (\ref{eq_EInt}):
\begin{align}
\label{eq_S2intt1}
  \mathcal{R}\Big[\wt{\mathcal{C}}_{q_1q_2}(\wh{E}_1,E_2)\Big]=(\ii E_1)^{q_{12}}\int_{-\infty}^0 \di\tau_1\,(-\tau_1)^{q_{12}-1}e^{\ii E_1\tau_1}\mathrm{E}_{1-q_2}(-\ii E_2\tau_1) . 
\end{align}
At this point, we make use of the following MB representation of $\mathrm{E}_p(z)$:
\bge
\label{eq_MBofE}
  \mathrm{E}_p(z) =\int_{-\ii\infty}^{+\ii\infty}\FR{\di s}{2\pi\ii}\FR{\Gamma(s)z^{-s}}{s+p-1} .
\ede
The details of this MB representation are given in App.\ \ref{app_math}. As explained there, the pole in $s$ from the denominator $1/(s+p-1)$ should be interpreted as a left pole, in the sense that the integration contour should go around this pole from the right side. Now, using (\ref{eq_MBofE}) in (\ref{eq_S2intt1}), we get:
\begin{align}
  \mathcal{R}\Big[\wt{\mathcal{C}}_{q_1q_2}(\wh{E}_1,E_2)\Big]=(\ii E_1)^{q_{12}}\int_{s_2}\FR{\Gamma(s_2)(\ii E)^{-s_2}}{s_2-q_2}\int_{-\infty}^0 \di\tau_1\,(-\tau_1)^{q_{12}-1-s_2}e^{\ii E_1\tau_1}.
\end{align}
Then, the $\tau_1$ integral is trivial, which is simply given by $\mathcal{C}_{q_{12}-s_2}(E_1)=(\ii E_1)^{-q_{12}+s_2}\Gamma(q_{12}-s_2)$. So, finishing the $\tau_1$ integral, we get:
\begin{align}
  \mathcal{R}\Big[\wt{\mathcal{C}}_{q_1q_2}(\wh{E}_1,E_2)\Big]= \int_{s_2}\FR{\Gamma[s_2,q_{12}-s_2]\varrho_{21}^{-s_2}}{s_2-q_2} .
\end{align}
Now it remains to finish the Mellin integral over $s_2$. Given that $\varrho_{21}=E_2/E_1<1$, we should close the Mellin contour from the left side and collect the residues of all left poles. There are two sets of left poles, one at $s_2=-n_2$ with $n_2=0,1,2,\cdots$, which is from the $\Gamma$-factor $\Gamma(s_2)$, and the other at $s_2=q_2$ coming from the denominator. Collecting the residues at these poles, we get:
\begin{align}
  \mathcal{R}\Big[\wt{\mathcal{C}}_{q_1q_2}(\wh{E}_1,E_2)\Big]=\sum_{n_2=0}^\infty\FR{(-1)^{n_2+1}\Gamma[n_2+q_{12}]}{(n_2+q_2)}\FR{\varrho_{21}^{n_2}}{n_2!}+\FR{\Gamma[q_1,q_2]}{\varrho_{21}^{q_2}}.
\end{align}
Now, we recognize that the last term without any summation is the product of two one-site families:
\bge
  \FR{1}{(\ii E_1)^{q_{12}}}\FR{\Gamma[q_1,q_2]}{\varrho_{21}^{q_2}}= \FR{\Gamma(q_1)}{(\ii E_1)^{q_1}}\FR{\Gamma(q_2)}{(\ii E_2)^{q_2}}=\mathcal{C}_{q_1}(E_1)\mathcal{C}_{q_2}(E_2).
\ede
Then, given the relation:
\bge
  \mathcal{R}\Big[{\mathcal{C}}_{q_1q_2}(\wh{E}_1,E_2)\Big]+{\mathcal{C}}_{q_1q_2}(\wh{E}_1,E_2)=\mathcal{C}_{q_1}(E_1)\mathcal{C}_{q_2}(E_2),
\ede
we see that the original family integral (\ref{eq_twosite}) is:
\bge
\label{eq_Cq1q2series1}
  \wt{\mathcal{C}}_{q_1q_2}(\wh{E}_1,E_2)=\sum_{n_2=0}^\infty\FR{(-1)^{n_2}\Gamma[n_2+q_{12}]}{(n_2+q_2)}\FR{\varrho_{21}^{n_2}}{n_2!}.
\ede
This is exactly what we would get using the rule (\ref{eq_family}). Incidentally, the above summation can be directly done and the result is the well-known Gauss's hypergeometric function:
\bge
\label{eq_Cq1q2F}
  \wt{\mathcal{C}}_{q_1q_2}(\wh{E}_1,E_2)=\,{}_2\mathcal{F}_1\left[\bgm q_2,q_{12}\\ q_2+1\edm\middle|-\varrho_{21}\right]. 
\ede
Here we use the dressed version ${}_2\mathcal{F}_1$ instead of the original hypergeometric function ${}_2F_1$ for notational simplicity. The dressed hypergeometric functions are defined in App.\ \ref{app_math}. 

Now, had we chosen to expand the integral (\ref{eq_twosite}) in terms of $1/E_2$, we will get:
\begin{align}
\label{eq_Cq1q2series2}
  \mathcal{C}_2(E_1,\wh{E}_2)
  =&~\FR{\Gamma[q_1,q_2]}{(\ii E_1)^{q_1}(\ii E_2)^{q_2}}-\FR{1}{(\ii E_2)^{q_{12}}}\sum_{n_1}^\infty\FR{\Gamma[n_1+q_{12}]}{(n_1+q_1)}\FR{(-\varrho_{12})^{n_1}}{n_1!}\n\\
  =&~\FR{\Gamma[q_1,q_2]}{(\ii E_1)^{q_1}(\ii E_2)^{q_2}}-\FR{1}{(\ii E_2)^{q_{12}}}{}_2\mathcal{F}_1\left[\bgm q_1,q_{12}\\ q_1+1\edm\middle|-\varrho_{12}\right].
\end{align}
Clearly, the series expression in the first line in (\ref{eq_Cq1q2series2}) has a different region of convergence from the series expression in (\ref{eq_Cq1q2series1}). However, the two expressions are just two power-series expansions of the same function in two different limits, one at $E_2/E_1\to 0$ and the other at $E_1/E_2\to 0$. This becomes more transparent after finishing the summations of both series into hypergeometric functions. Indeed, equating (\ref{eq_Cq1q2F}) with the second line of (\ref{eq_Cq1q2series2}), we just get a transformation-of-variable formula for the hypergeometric function. Thus, our procedure provides a convenient way to derive many transformation-of-variable formulae for hypergeometric functions, which is particularly convenient for more complicated hypergeometric series, as we shall see below.

\begin{figure}
\centering 
\subfigure[$\mathcal{C}_{q_1q_2q_3}(\wh{E}_1,E_2,E_3)$]{\label{fig_threevertex1}\includegraphics[width=0.3\textwidth]{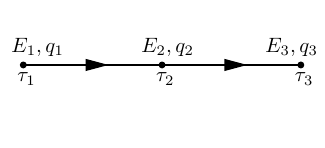}
\vspace{-5mm}} 
\hspace{20mm}
\subfigure[$\mathcal{C}_{q_1q_2q_3}(E_1,\wh{E}_2,E_3)$]{\label{fig_threevertex2}\includegraphics[width=0.3\textwidth]{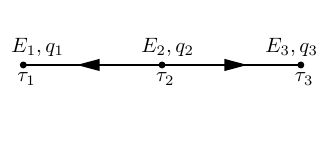}
\vspace{-5mm}} 
\caption{Two independent family integrals at 3-site level.} 
  \label{fig_threevertex}
\end{figure}

\paragraph{Three-site family.} Next we consider a slightly nontrivial case with three sites. There is only one topology at the tree level with 3 sites. However, after including the time ordering, there are two independent possibilities, depending on whether the latest site is on the side or in the middle. These two possibilities are shown in Fig.\ \ref{fig_threevertex}. Again, by construction, the latest site is chosen to be the maximal-energy site. So, for the case in Fig.\ \ref{fig_threevertex1}, we have:
\begin{align}
\label{eq_3site1}
  \wt{\mathcal{C}}_{q_1q_2q_3}(\wh{E}_1,E_2,E_3)=(\ii E_1)^{q_{123}}\int_{-\infty}^0 \prod_{i=1}^3\Big[\di\tau_i\,(-\tau_i)^{q_i-1}e^{\ii E_i\tau_i}\Big]\theta(\tau_3-\tau_2)\theta(\tau_2-\tau_1). 
\end{align}
We again start from the completely reversed integral: 
\begin{align}
  \mathcal{R}\Big[\wt{\mathcal{C}}_{q_1q_2q_3}(\wh{E}_1,E_2,E_3)\Big]=(\ii E_1)^{q_{123}}\int_{-\infty}^0 \prod_{i=1}^3\Big[\di\tau_i\,(-\tau_i)^{q_i-1}e^{\ii E_i\tau_i}\Big]\theta(\tau_1-\tau_2)\theta(\tau_2-\tau_3).
\end{align}
Now, we can repeat the above strategy and finish the three layers of time integrals in the order of $\tau_3,\tau_2,\tau_1$. The first two layers produce exponential integrals which can then be represented as Mellin integrals. The last layer is again a single-site integral which can be finished directly. Here we show the results after finishing each layer of time integral and taking the MB representation for exponential integrals:
\begin{align}
  &~\mathcal{R}\Big[\wt{\mathcal{C}}_{q_1q_2q_3}(\wh{E}_1,E_2,E_3)\Big]\n\\
  =&~(\ii E_1)^{q_{123}}\int_{s_3}\FR{\Gamma(s_3)(\ii E_{3})^{-s_3}}{s_3-q_3}\int_{-\infty}^0  \di\tau_1 \di\tau_2\,(-\tau_1)^{q_1-1}(-\tau_2)^{q_{23}-1-s_3}e^{\ii (E_1\tau_1+E_2\tau_2)}  \theta(\tau_1-\tau_2)\n\\
  =&~(\ii E_1)^{q_{123}}\int_{s_2,s_3}\FR{\Gamma[s_2,s_3](\ii E_{2})^{-s_2}(\ii E_{3})^{-s_3}}{(s_{23}-q_{23})(s_3-q_3)}\int_{-\infty}^0  \di\tau_1\,(-\tau_1)^{q_{123}-1-s_{23}}e^{\ii E_1\tau_1} \n\\
  =&~ \int_{s_2,s_3}\FR{\Gamma[s_2,s_3,q_{123}-s_{23}]\varrho_{21}^{-s_2}\varrho_{31}^{-s_3}}{(s_{23}-q_{23})(s_3-q_3)} .
\end{align}
We start to observe a pattern here: Let the maximal-energy site be $\tau_i$. When carrying out any but the last layer of time integrals, say $\tau_j$ with $j\neq i$, we are effectively generating a new layer of Mellin integral with Mellin variable $s_j$, a pole-generating factor $\Gamma(s_j)/(\wt s_j-\wt q_j)$, and a power of energy ratio $\varrho_{ji}^{-s_j}$. Here $\wt s_j$ is the sum of all Mellin variables assigned to the site $j$ and its descendant, and $\wt q_j$ is likewise defined.

Then it remains to carry out the Mellin integrals. Unlike the previous case, now we encounter pole-carrying factors involving more than one Mellin variable. In the current case, it is the denominator $1/(s_{23}-q_{23})$. To avoid any potential complication of such poles, our strategy is that we perform the Mellin integrals in the ``anti-chronological'' order. In the current case, we integrate out $s_3$ first, by collecting poles \emph{only} from $\Gamma(s_3)/(s_3-q_3)$. Only after this is done, we then perform the $s_2$-integral, by collecting poles from $\Gamma(s_2)/(s_{23}-q_{23})$. By this time, the $s_3$ variables in these factors have already been set to the poles. Thus, we never need to directly deal with poles involving a sum of several Mellin variables. Finishing the Mellin integral in this way, we get: 
\begin{align}
\label{eq_3siteRS}
  \FR{1}{(\ii E_1)^{q_{123}}}\mathcal{R}\Big[\wt{\mathcal{C}}_{q_1q_2q_3}(\wh{E}_1,E_2,E_3)\Big]
  =&~\FR{1}{(\ii E_1)^{q_{123}}}\sum_{n_2,n_3=0}^\infty\FR{(-1)^{n_{23}}\Gamma[n_{23}+q_{123}]}{(n_{23}+q_{23})(n_3+q_3)}\FR{\varrho_{21}^{n_2}}{n_2!}\FR{\varrho_{31}^{n_3}}{n_3!}\n\\
   &-\FR{\Gamma(q_1)}{(\ii E_1)^{q_{1}}}\FR{1}{(\ii E_2)^{q_{23}}}
   \sum_{n_3=0}^\infty\FR{(-1)^{n_3}\Gamma[n_3+q_{23}]}{(n_3+q_3)}\FR{\varrho_{32}^{n_3}}{n_3!}\n\\
   &-\FR{\Gamma(q_3)}{(\ii E_3)^{q_{3}}}\FR{1}{(\ii E_1)^{q_{12}}} 
   \sum_{n_2=0}^\infty\FR{(-1)^{n_2}\Gamma[n_2+q_{12}]}{(n_2+q_2)}\FR{\varrho_{21}^{n_2}}{n_2!}\n\\
   &+\FR{\Gamma(q_1)}{(\ii E_1)^{q_1}}\FR{\Gamma(q_2)}{(\ii E_2)^{q_2}}\FR{\Gamma(q_3)}{(\ii E_3)^{q_3}} .
\end{align}
Here we have restored all the dimemsionful energy factors to make clear the following point: The result of Mellin integral is effectively executing the identical transformation $\theta(\tau_i-\tau_j)=1-\theta(\tau_j-\tau_i)$ in a line-by-line fashion. Thus, with $N$ lines in a family, we will get $2^N$ terms. All but one of them are factorized. There is a unique unfactorized term with all lines reversed. In the current example it is the first term on the right hand side of (\ref{eq_3siteRS}). This is nothing but the original family integral. Thus:
\bge
\label{eq_C3E1}
  \wt{\mathcal{C}}_{q_1q_2q_3}(\wh{E}_1,E_2,E_3)=\sum_{n_2,n_3=0}^\infty\FR{(-1)^{n_{23}}\Gamma[n_{23}+q_{123}]}{(n_{23}+q_{23})(n_3+q_3)}\FR{\varrho_{21}^{n_2}}{n_2!}\FR{\varrho_{31}^{n_3}}{n_3!}.
\ede
Once again, this is exactly what we would get by applying the simple formula (\ref{eq_family}). It seems to us that this series does not sum to any widely known special function in general, but it can be represented as a (dressed) Kampé de Fériet function, whose definition is collected in App.\ \ref{app_math}:
\bge
  \wt{\mathcal{C}}_{q_1q_2q_3}(\wh{E}_1,E_2,E_3)={\;}^{2+1}\mathcal{F}_{1+1}\left[\bgm q_{123},q_{23}\\ q_{23}+1 \edm \middle| \bgm \text{\;-\;},q_3 \\ \text{\;-\;},q_3+1 \edm\middle|-\varrho_{21},-\varrho_{31}\right].
\ede

The lesson we learned from the above example is the following: To find the answer to a given family integral $\mathcal{C}$, all we need to do is to compute another integral $\mathcal{R}[\mathcal{C}]$ with all time orderings completely reversed. We compute $\mathcal{R}[\mathcal{C}]$ layer by layer. Each step generates an exponential integral of which we take the MB representation. The last layer of time integral is directly done, and we are left with an $(N-1)$-fold Mellin integral. We finish the Mellin integral by retaining poles only from $\Gamma$-factors. The result of this term is automatically a sign factor $(-1)^{N-1}$ times the original family $\mathcal{C}$.

With this lesson learned, we can bypass all steps detailed above, and write down the answers for arbitrary families. Now, let us go on to consider the three-site family in Fig.\ \ref{fig_threevertex2}, which corresponds to the following integral:
\begin{align}
  \wt{\mathcal{C}}_{q_1q_2q_3}(E_1,\wh{E}_2,E_3)=(\ii E_2)^{q_{123}}\int_{-\infty}^0 \prod_{i=1}^3\Big[\di\tau_i\,(-\tau_i)^{q_i-1}e^{\ii E_i\tau_i}\Big]\theta(\tau_1-\tau_2)\theta(\tau_3-\tau_2). 
\end{align}
The result after finishing all three layers of time integrals for the \emph{reversed diagram} $\mathcal{R}[\wt{\mathcal{C}}]$ is:
\begin{align}
  \mathcal{R}\Big[\wt{\mathcal{C}}_{q_1q_2q_3}(E_1,\wh{E}_2,E_3)\Big]=\int_{s_1,s_3}\FR{\Gamma[s_1,s_3,q_{123}-s_{13}]\varrho_{12}^{-s_1}\varrho_{32}^{-s_3}}{(s_1-q_1)(s_3-q_3)}.
\end{align}
Then, we finish the Mellin integral by picking up poles in $\Gamma[s_1,s_3]$ \emph{only}. Multiplying the result by a trivial sign factor of $(-1)^{3-1}=1$, we get the original family:
\begin{align}
  \wt{\mathcal{C}}_{q_1q_2q_3}(E_1,\wh{E}_2,E_3)=\sum_{n_1,n_3=0}^\infty\FR{(-1)^{n_{13}}\Gamma[n_{13}+q_{123}]}{(n_1+q_1)(n_3+q_3)}\FR{\varrho_{12}^{n_1}}{n_1!}\FR{\varrho_{32}^{n_3}}{n_3!}. 
\end{align}
Again, it agrees with what we would get by applying (\ref{eq_family}). Incidentally, the above two-fold hypergeometric series belongs to the well-known Appell series, which can be summed into the (dressed) Appell $F_2$-function:
\begin{align}
  \wt{\mathcal{C}}_{q_1q_2q_3}(E_1,\wh{E}_2,E_3)=\mathcal{F}_2\left[q_{123}\middle|\bgm q_1,q_3\\1+q_1,1+q_3\edm\middle|-\varrho_{12},-\varrho_{32}\right].
\end{align}
The definition of $\mathcal{F}_2$ is collected in App.\ \ref{app_math}.

\begin{figure}
\centering 
\subfigure[$\mathcal{C}_{q_1q_2q_3q_4}^\text{(cubic)}(E_1,E_2,E_3,\wh{E}_4)$]{\label{fig_cubic1}\includegraphics[width=0.3\textwidth]{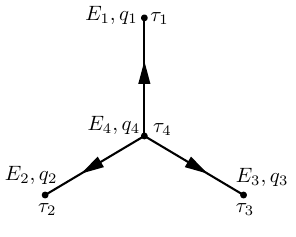}
\vspace{-5mm}} 
\hspace{20mm}
\subfigure[$\mathcal{C}_{q_1q_2q_3q_4}^\text{(cubic)}(\wh{E}_1,E_2,E_3,E_4)$]{\label{fig_cubic2}\includegraphics[width=0.3\textwidth]{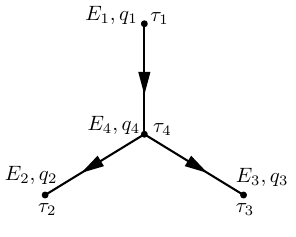}
\vspace{-5mm}} 
\caption{Two independent family integrals of 4-site graphs with a cubic vertex.} 
  \label{fig_cubic}
\end{figure}

\paragraph{Four-site family with a cubic vertex.} Finally let us look at four-site families. There are two possible tree topologies with 4 sites. One is the chain graph ${\mathcal{C}}_{q_1q_2q_3q_4}^\text{(n)}$, which is a direct generalization of the case considered above. We are not going to consider this chain graph any more. On the other hand, there is a new topology with a cubic vertex ${\mathcal{C}}_{q_1q_2q_3q_4}^\text{(iso)}$, as shown in Fig.\ \ref{fig_cubic}.\footnote{We are borrowing the nomenclature of organic chemistry, where an unbranched linear carbon chain is dubbed normal (n), while a chain with a ``cubic vertex'' is dubbed isometric (iso).} Again, there are two independent ways to assign the maximal-energy site, either at the middle site or at a boundary site, corresponding to Fig.\ \ref{fig_cubic1} and Fig.\ \ref{fig_cubic2}, respectively.

Consider Fig.\ \ref{fig_cubic1} first, where the maximal-energy site is chosen to be the middle site $\tau_4$. The corresponding (dimensionless) family integral is:
\begin{align}
  \wt{\mathcal{C}}_{q_1q_2q_3q_4}^\text{(iso)}(E_1,E_2,E_3,\wh{E}_4)
  =(\ii E_4)^{q_{1234}}\int_{-\infty}^0 \prod_{i=1}^4\Big[\di\tau_i\,(-\tau_i)^{q_i-1}e^{\ii E_i\tau_i}\Big]\theta(\tau_1-\tau_4)\theta(\tau_2-\tau_4)\theta(\tau_3-\tau_4). 
\end{align}
As always, we compute the corresponding reversed time-ordering integral:
\begin{align}
  \mathcal{R}\Big[\wt{\mathcal{C}}_{q_1q_2q_3q_4}^\text{(iso)}(E_1,E_2,E_3,\wh{E}_4)\Big]
=\int_{s_1,s_2,s_3}\FR{\Gamma[s_1,s_2,s_3,q_{1234}-s_{123}]\varrho_{14}^{-s_1}\varrho_{24}^{-s_2}\varrho_{34}^{-s_3}}{(s_1-q_1)(s_2-q_2)(s_3-q_3)}. 
\end{align}
Then, the original integral is obtained by finishing the three-fold Mellin integral in which we only collect poles from $\Gamma[s_1,s_2,s_3]$, and multiplying the result by $(-1)^{4-1}=-1$. The result is:
\begin{align}
  \wt{\mathcal{C}}_{q_1q_2q_3q_4}^\text{(iso)}(E_1,E_2,E_3,\wh{E}_4)=\sum_{n_1,n_2,n_3=0}^\infty\FR{(-1)^{n_{123}}\Gamma[n_{123}+q_{1234}]}{(n_1+q_1)(n_2+q_2)(n_3+q_3)}\FR{\varrho_{14}^{n_1}}{n_1!}\FR{\varrho_{24}^{n_2}}{n_2!}\FR{\varrho_{34}^{n_3}}{n_3!}. 
\end{align}
This three-variable series is not covered in commonly known special functions. It is however covered by the so-called (dressed) Lauricella's $F_A$ function:
\begin{align}
  \wt{\mathcal{C}}_{q_1q_2q_3q_4}^\text{(iso)}(E_1,E_2,E_3,\wh{E}_4)
  =\mathcal{F}_A\left[q_{1234}\middle|\bgm q_1,q_2,q_3\\ q_1+1,q_2+1,q_3+1\edm \middle|-\varrho_{14},-\varrho_{24},-\varrho_{34}\right].
\end{align}
The definition of this function is collected in App.\ \ref{app_math}.

Next let us look at Fig.\ \ref{fig_cubic2} where the maximal-energy site is on the side. We take it to be $E_1$. Then, the corresponding family integral is given by:
\begin{align}
  \wt{\mathcal{C}}_{q_1q_2q_3q_4}^\text{(iso)}(\wh{E}_1,E_2,E_3,E_4)
  =(\ii E_1)^{q_{1234}}\int_{-\infty}^0 \prod_{i=1}^4\Big[\di\tau_i\,(-\tau_i)^{q_i-1}e^{\ii E_i\tau_i}\Big]\theta(\tau_2-\tau_4)\theta(\tau_3-\tau_4)\theta(\tau_4-\tau_1). 
\end{align}
Without mentioning any details of intermediate steps, we directly provide the final answer to this integral:
\begin{align}
\label{eq_S4iso}
  \wt{\mathcal{C}}_{q_1q_2q_3q_4}^\text{(iso)}(\wh{E}_1,E_2,E_3,E_4)=\sum_{n_2,n_3,n_4=0}^\infty\FR{(-1)^{n_{234}}\Gamma[n_{234}+q_{1234}]}{(n_2+q_2)(n_3+q_3)(n_{234}+q_{234})}\FR{\varrho_{21}^{n_2}}{n_2!}\FR{\varrho_{31}^{n_3}}{n_3!}\FR{\varrho_{41}^{n_4}}{n_4!}. 
\end{align}

\subsection{General family integrals}
\label{sec_general}

Above we have examined enough number of examples. By now, it is clear why we need to do family-tree decomposition: Our performance of nested integrals requires a partial order of the nested time variables, and this partial order can always be achieved by the family-tree decomposition. Once we have a partially ordered integral, we can always carry out the completely reversed integral, from the originally latest sites (earliest sites in the reversed integral), and to their mothers, and to grandmothers, etc., until the last layer which is the maximal energy site. In this way, a full derivation of the general equation (\ref{eq_family}) becomes a matter of mathematical induction. Below we complete this proof.

We begin with a general partially ordered family with $N$ sites, $\mathcal{C}_{q_1\cdots q_N}(\wh{E_1},E_2,\cdots,E_N)$, where we assume that the maximal energy site is $\tau_1$. Its integral representation is given in (\ref{eq_Cint}). As in the previous section, we work with the completely reversed integral $\mathcal{R}[\mathcal{C}_{q_1\cdots q_N}(\wh{E_1},E_2,\cdots,E_N)]$, and we integrate every time variable from $-\infty$ to the time variable of her mother. 

Suppose that we have finished all the time integrals for the descendants of the site $\tau_j$, and now we want to finish the time integral at $\tau_j$. Our induction assumption is that, after all the descendants of $\tau_j$ integrated out, the integral over the variable $\tau_j$ has the following form:
\begin{align}
\label{eq_ItauM}
  \mathcal{I}_j(\tau_M)=\int_{-\infty}^{\tau_M}\di\tau_j\,(-\tau_j)^{\wt{q}_j-1}e^{\ii E_j\tau_j}\prod_{i\in \mathfrak{D}(j)}\int_{s_i}\FR{\Gamma[s_i](\ii E_i)^{-s_i}}{\wt{s}_i-\wt{q}_i}(-\tau_j)^{-{s}_i},
\end{align}
where $\tau_M$ is the time variable of $\tau_j$'s mother, and $\mathfrak{D}(j)$ denotes the set of labels for all $\tau_j$'s descendants. Now, finishing the $\tau_j$ integral, we get:
\begin{align}
  \mathcal{I}_j(\tau_M)
  =&~ \prod_{i\in \mathfrak{D}(j)}\bigg[\int_{s_i}\FR{\Gamma[s_i](\ii E_i)^{-s_i}}{\wt{s}_i-\wt{q}_i}\bigg](-\tau_M)^{\wt q_j-\sum s_i}\mathrm{E}_{1-\wt q_j+\sum  s_i}(-\ii E_j\tau_M) \n\\
  =&~\prod_{i\in \mathfrak{D}(j)}\bigg[\int_{s_i}\FR{\Gamma[s_i](\ii E_i)^{-s_i}}{\wt{s}_i-\wt{q}_i}\bigg]\int_{s_j}\FR{\Gamma(s_j)(\ii E_j)^{-s_j}}{s_j-\wt q_j+\sum\limits_{i\in\mathfrak{D}(j)}s_i} (-\tau_M)^{\wt q_j-s_j-\sum s_i}\n\\
  =&~ (-\tau_M)^{\wt q_j}\prod_{i\in\{j\}\cap \mathfrak{D}(j)}\bigg[\int_{s_i}\FR{\Gamma[s_i](\ii E_i)^{-s_i}}{\wt{s}_i-\wt{q}_i}(-\tau_M)^{- s_i}\bigg] .
\end{align}
Here we have used the fact that $s_j+\sum\limits_{i\in\mathfrak{D}(j)}s_i=\wt s_j$. Also, we have abbreviated $\sum\limits_{i\in\mathfrak{D}(j)}s_i$ as $\sum s_i$ when it appears as an upper or lower index.

Now, to go one step further, we should finish the time integral over $\tau_M$, which we denote as $\mathcal{I}_M(\tau_G)$ and $\tau_G$ is the time variable of $\tau_j$'s grandmother. To this end, we take products of above integral $\mathcal{I}_j(\tau_M)$ over all daughters $\tau_j$ of $\tau_M$. Then, together with $\tau_M$'s own factor $(-\tau_M)^{q_{M}-1}e^{\ii E_M\tau_M}$, we get:
\begin{align}
  \mathcal{I}_M(\tau_G)
  =&\int_{-\infty}^{\tau_G}\di\tau_M \,(-\tau_M)^{q_{M}-1}e^{\ii E_M\tau_M}\prod_{j\in\text{daughters of $\tau_M$}}(-\tau_M)^{\wt q_j}\prod_{i\in\{j\}\cap \mathfrak{D}(j)}\bigg[\int_{s_i}\FR{\Gamma[s_i](\ii E_i)^{-s_i}}{\wt{s}_i-\wt{q}_i}(-\tau_M)^{- s_i}\bigg]\n\\
  =&\int_{-\infty}^{\tau_G}\di\tau_M \,(-\tau_M)^{\wt q_{M}-1}e^{\ii E_M\tau_M}\prod_{j\in\mathfrak{D}(M)} \bigg[\int_{s_j}\FR{\Gamma[s_j](\ii E_j)^{-s_j}}{\wt{s}_j-\wt{q}_j}(-\tau_M)^{- s_j}\bigg].
\end{align}
This is identical to (\ref{eq_ItauM}) upon a ``generation shift'' $j\to M$ and $M\to G$. So we have shown that the original induction assumption (\ref{eq_ItauM}) persists to all generations as long as it holds at one generation. On the other hand, it is trivial to check that the induction assumption holds for the initial step, i.e., at any site that has no descendant. Thus, we have proved that the induction assumption (\ref{eq_ItauM}) holds for all sites. In particular, (\ref{eq_ItauM}) holds for the maximal energy site $\tau_1$ if we take $\tau_j=\tau_1$ and $\tau_M=0$. Then, completing this final layer of time integral over $\tau_1$, we get, for the whole reversed family,
\begin{align}
  \mathcal{R}\Big[\mathcal{C}_{q_1\cdots q_N}(\wh{E_1},E_2,\cdots,E_N)\Big]
  =&\int_{-\infty}^{0}\di\tau_1\,(-\tau_1)^{\wt{q}_1-1}e^{\ii E_1\tau_1}\prod_{i=2}^N\int_{s_i}\FR{\Gamma[s_i](\ii E_i)^{-s_i}}{\wt{s}_i-\wt{q}_i}(-\tau_1)^{-{s}_i}
  \n\\
  =&~ \prod_{i=2}^N\int_{s_i}\FR{\Gamma[s_i](\ii E_i)^{-s_i}}{\wt{s}_i-\wt{q}_i} \FR{\Gamma[q_{1\cdots N}-s_{2\cdots N}]}{(\ii E_1)^{q_{1\cdots N}-s_{2\cdots N}}}\n\\
  =&~\FR{1}{(\ii E_1)^{q_{1\cdots N}}}\prod_{i=2}^N\bigg[\int_{s_i}\FR{\Gamma[s_i]\varrho_{i1}^{-s_i}}{\wt{s}_i-\wt{q}_i}\bigg] \Gamma[q_{1\cdots N}-s_{2\cdots N}].
\end{align}
As shown many times in the previous subsection, the original family $\mathcal{C}_{q_1\cdots q_N}(\wh{E_1},E_2,\cdots,E_N)$ is recovered by picking up all poles from $\Gamma[s_i]$, and including an overall factor $(-1)^{N-1}$ which comes from reversing the directions of $N-1$ bulk lines. Thus, we get:
\begin{align}
  \mathcal{C}_{q_1\cdots q_N}(\wh{E_1},E_2,\cdots,E_N)  
  =&~\FR{(-1)^{N-1}}{(\ii E_1)^{q_{1\cdots N}}}\sum_{n_2,\cdots,n_N=0}^\infty\Gamma(q_{1\cdots N}+n_{2\cdots N}) \prod_{i=2}^N  \FR{(-\varrho_{i1})^{n_i}}{(-\wt n_i-\wt{q}_i)n_i!}.
\end{align}
This is exactly the original family formula (\ref{eq_family}). Thus we have completed the proof.

\subsection{Alternative representation}
\label{sec_alter}

The MB representation of a function is not unique. In the previous computations, we have chosen a relatively simple representation (\ref{eq_MBofE}) for the exponential function $\mathrm{E}_p(z)$. This representation allows us to find simple expression for the dimensionless family integrals as Taylor series of energy ratios. On the other hand, there exist other MB representations which may be useful in certain cases. One example is the following partially resolved MB representation, which is particularly useful to improve the convergence of the hypergeometric series when there are several energies comparable to the maximal energy:
\begin{align}
\label{eq_EpPRMB}
  \mathrm{E}_p(z)=e^{-z}\int_{-\ii\infty}^{+\ii\infty}\FR{\di s}{2\pi\ii}\Gamma\bgb p+s,1+s,-s \\ p \edb z^{-s-1}.
\end{align}
This result can be derived from the MB representation for a confluent hypergeometric function, as discussed in App.\ \ref{app_math}. We note that (\ref{eq_EpPRMB}) is not a complete MB representation of $\mathrm{E}_p(z)$, as there is an exponential factor $e^{-z}$ left.\footnote{Incidentally, this is very similar to the partially resolved MB representation for the Whittaker function used in \cite{Qin:2022fbv}.} As we shall see, this remaining exponential factor will help us to circumvent the problem of convergence of hypergeometric series with several maximal energies. Let us take monotonic three-site family integral (\ref{eq_3site1}) as an example, namely Fig.\ \ref{fig_threevertex1}, but we do not divide out the dimensionful factor $(\ii E_1)^{q_{123}}$, nor do we assign a maximal energy variable:
\begin{align}
   {\mathcal{C}}_{q_1q_2q_3}({E}_1,E_2,E_3)= \int_{-\infty}^0 \prod_{i=1}^3\Big[\di\tau_i\,(-\tau_i)^{q_i-1}e^{\ii E_i\tau_i}\Big]\theta(\tau_3-\tau_2)\theta(\tau_2-\tau_1).
\end{align}
As before, we compute the integral with all time orderings reversed, but with the new representation (\ref{eq_EpPRMB}). The result is:
\begin{align}
   \mathcal{R}\Big[ {\mathcal{C}}_{q_1q_2q_3}(E_1,E_2,E_3)\Big] 
  =&\int_{s_2,s_3}\,\Gamma\bgb 1+s_3,1-q_3+s_3,-s_3 \\ 1-q_3 \edb\Gamma\bgb 1+s_2,2-q_{23}+s_{23},-s_2 \\ 2-q_{23}+s_3\edb \n\\
   &\times (\ii E_3)^{-1-s_3}(\ii E_{23})^{-1-s_2} \FR{\Gamma[q_{123}-2-s_{23}]}{(\ii E_{123})^{q_{123}-2-s_{23}}}.
\end{align}
Similar to the Mellin integrals in the previous representation, each Mellin variable $s_i$ gets two sets of left poles from the $\Gamma$ factors, one at $s_i=-1-n_i$ ($n_i=0,1,\cdots$) from $\Gamma(1+s_i)$, and the other more complicated, involving both $\wt q_i$ and other Mellin variables from the descendants of $s_i$.
We are not going to present a detailed analysis here, but only mention that, similar to the previous case, the original family integral $\wt{\mathcal{C}}_{q_1q_2q_3}(E_1,E_2,E_3)$ is recovered by picking up poles from all $\Gamma(1+s_i)$ factors only and multiplying the result with an appropriate sign factor. Thus:
\begin{align}
\label{eq_C3Alter}
   {\mathcal{C}}_{q_1q_2q_3}(E_1,E_2,E_3) 
  =\FR{1}{(\ii E_{123})^{q_{123}}}\sum_{n_2,n_3=0}^{\infty}(-1)^{n_{23}}\Gamma\bgb n_{23}+q_{123},-n_{23}-q_{23},-n_3-q_3 \\ 1-n_3-q_{23},1-q_3\edb \varrho_{23T}^{n_2} \varrho_{3T}^{n_3},
\end{align}
where we have defined $\varrho_{23T}\equiv E_{23}/E_{123}$ and $\varrho_{3T}\equiv E_3/E_{123}$. Thus we see that, instead of using the inverse of the maximal energy as the expansion parameter, in this representation, we are using the inverse of the \emph{total energy} $E_{123}$ as the expansion parameter. Although it has a somewhat more complicated looking than the previous representation, this representation is a safer choice in certain cases, in particular in the kinematic region of several equal or comparable maximal energies. In any case, it is easy to check numerically that (\ref{eq_C3Alter}) and (\ref{eq_C3E1}) agree with each other perfectly whenever both series converge. 

The lesson here is that we can make use of the flexibility of MB representations to get different series solutions for the nested time integrals, expanded either in the inverse power of some energy variable of a given site, in the inverse power of the sum of several energy variables, or even in the inverse power of the total energy. Although the final results may look quite different, these results are just different expansions of the same function. We can thus obtain a large number of transformation-of-variable relations for these multi-variable hypergeometric functions. We leave a more systematic investigation of this topic to future works.

\subsection{Discussions}

We end this section with a discussion of the nested time integral.

\paragraph{Pole structure.} In Sec.\ \ref{sec_tree}, we mentioned that the time integral in the PMB representation only contains right poles for all Mellin variables, which was proved in \cite{Qin:2023bjk}. Now that we have explicit results for arbitrary nested time integrals, it is straightforward to check this statement. Indeed, we can rewrite our general result for the family integral (\ref{eq_family}) in the following way:
\bge
\label{eq_Cgamma}
  \wt{\mathcal{C}}_{q_1\cdots q_N}(\wh{E}_1,E_2\cdots,E_N)=\sum_{n_2,\cdots,n_N=0}^\infty \Gamma(q_{1\cdots N}+n_{2\cdots N})\prod_{j=2}^N\Gamma\bgb \wt q_j+\wt n_j \\ \wt q_j+\wt n_j+1\edb\FR{(-\varrho_{j1})^{n_j}}{n_j!}.
\ede
Then it is clear that all the exponents $q_i$ $(i=1,\cdots,N)$ have the positive coefficients when appearing in the arguments of $\Gamma$ factors. Now, if we use (\ref{eq_qtos}) to rewrite all $q$'s in terms of Mellin variables $s$, we will see that all Mellin variables $s$ have \emph{negative} coefficients when appearing in the arguments of all $\Gamma$ factors in (\ref{eq_Cgamma}). So, we have confirmed with our explicit results that nested time integrals only have right poles in all Mellin variables.

With the explicit results for the nested time integrals, it is also easy to confirm that the Mellin integrand for any tree graph in the PMB representation is well balanced for all Mellin variables, an important fact for the computation of Mellin integrals, as mentioned in Sec.\ \ref{sec_tree}. To see this, we only need to derive (\ref{eq_RightS}) from our result.

From our result for the time integral in (\ref{eq_family}), it is trivial to see that the exponents $q_i$ ($i=1,\cdots, N$) appear in the $\Gamma$ factor $\Gamma(q_{1\cdots N}+n_{2\cdots N})$ as a total sum. Then, let us look at (\ref{eq_qtos}), which says that the value of $q_\ell$ at Vertex $\ell$ receives contributions from all Mellin variables ending at this vertex. Note that, by construction, every Mellin variable is associated to one and only one vertex. Thus, (\ref{eq_qtos}) tells us that summing over all $q_\ell$ is equivalent to summing over all Mellin variables. As a result, the argument of the $\Gamma$ factor $\Gamma(q_{1\cdots N}+n_{2\cdots N})$ becomes:
\bge
  q_{1\cdots N}+n_{2\cdots N}=-2\sum_{i=1}^I(s_i+\bar s_i)+p_{1\cdots N}+n_{2\cdots N}+N,
\ede
which agrees nicely with the general structure in (\ref{eq_RightS}). So, we see that the Mellin variables are indeed balanced. 

\paragraph{Hard limits.} It is interesting to look at different kinematic limits of our result (\ref{eq_family}). First, it is simple to take a hard limit where one energy $E_1$ is much greater than any other energies. Obviously, in this limit, we should work with the expression where $E_1$ is chosen as the maximal energy site. Then, in the series expansion (\ref{eq_family}), only the leading term with $n_2=\cdots=n_N=0$ survives the limit. So we get:
  \begin{align}  
  \label{eq_hardlim}
   \lim_{E_1\to \infty}{\mathcal{C}}_{q_1\cdots q_N}(E_1,E_2,\cdots,E_N)=\FR{ \Gamma(q_{1\cdots N})}{(\ii E_1)^{q_{1\cdots N}}}\prod_{j=2}^N\FR{1}{\wt{q}_j}.
  \end{align}
Apart from the simple numerical factor $\prod\limits_j 1/\wt q_j$, this is very similar to the result of one-site family $\mathcal{C}_q(E)=\Gamma(q)/(\ii E)^{q}$ with $E=E_{1\cdots N}$ and $q=q_{1\cdots N}$. Here we have used the fact that $E_1\simeq E_{1\cdots N}$ in the $E_1\to \infty$ limit. So, in this hard limit, the time integral behaves as if we have pinched all $N$ nested vertices together, with all exponents $q_i$ ($i=1,\cdots,N$) summed. Thus, this hard limit can in a sense be thought of as an EFT limit, where all internal lines in the family integral shrink into local vertices. From the viewpoint of the cosmological bootstrap \cite{Arkani-Hamed:2018kmz,Qin:2022fbv}, we know that the EFT part is related to the particular solution of the bootstrap equation with a local source term. This local source term originates exactly from the time-ordering part of the internal propagators. So, there is a close relation between the local EFT limit and the nested integrals, and it is not surprising to get (\ref{eq_hardlim}). 

However, we can make a new interesting observation from (\ref{eq_hardlim}): In the original SK integral (\ref{eq_SKint}) for a correlator, we need to sum over all SK indices, which involve all kinds of propagators with arbitrary nesting. This means that the site of $E_1$ can be nested arbitrarily with other sites. Then, coupled with (\ref{eq_hardlim}), we see that the $E_1\to \infty$ limit can generate a power $1/E_1^{q_1\cdots q_N}$ involving exponents $q_i$ at any other site. Note that the variable $q_i$ contains the Mellin variables of internal lines ending at Site $i$, and we see that the power $1/E_1^{q_1\cdots q_N}$ could be dependent on Mellin variables of any internal lines \emph{not} ending at Site 1. Thus, if we finish the Mellin integrals by picking up left poles for those Mellin variables, they can introduce noninteger powers of $1/E_1$. This is exactly the source of \emph{local signals}. The local signals have been considered mainly for single exchange graphs in previous works \cite{Tong:2021wai,Qin:2022fbv}. Here we see that, in the hard limit $E_1\to \infty$, the local signal from $E_1$ can in principle be generated by \emph{any} internal massive propagators \emph{not} ending on Site 1. So, the local signal is more subtle and more complicated than the nonlocal signal. This topic will be further explored in a separate work \cite{localsignal}. 

\paragraph{Soft limits: internal vertices.} Now let us look at an opposite limit where one or several energies approach zero, $E_i\to 0$. This is a soft limit. Note that, in our general expression for the nested time integral (\ref{eq_NTI}), we have assigned an exponential factor $e^{\ii E_i\tau_i}$ for each site at time $\tau_i$. This factor is generally from the bulk-to-boundary propagator of a massless or conformal scalar (or from a massless graviton if nonzero spins are considered). In realistic tree graphs, there are certainly vertices on which only bulk massive propagators end, and no bulk-to-boundary lines are attached. We call such a vertex an \emph{internal vertex} following \cite{Qin:2023nhv}. Clearly, we have $E_i=0$ for such a vertex. So, it is necessary to know how to take soft limits if we want to consider graphs with internal vertices.
 
Fortunately, our series expression for the time integral makes it very convenient to take a soft limit. For instance, suppose that we want to set $E_4=0$ in the 4-site family (\ref{eq_S4iso}) which corresponds to Fig.\ \ref{fig_cubic2}. Then, the form of the hypergeometric series in (\ref{eq_S4iso}) allows us to set $\varrho_{41}=E_4/E_1=0$ directly without encountering any singularities. Then, in the summation over $n_4$, only the term with $n_4=0$ survives the $\varrho_{41}\to 0$ limit, and we get:
\begin{align} 
\label{eq_4site0}
  \mathcal{C}_{q_1q_2q_3q_4}^\text{(iso)}(\wh{E}_1,E_2,E_3,E_4=0)=\sum_{n_2,n_3=0}^\infty\FR{(-1)^{n_{23}}\Gamma[n_{23}+q_{1234}]}{(n_2+q_2)(n_3+q_3)(n_{23}+q_{234})}\FR{\varrho_{21}^{n_2}}{n_2!}\FR{\varrho_{31}^{n_3}}{n_3!}. 
\end{align}

\begin{figure}
\centering  
\includegraphics[width=\textwidth]{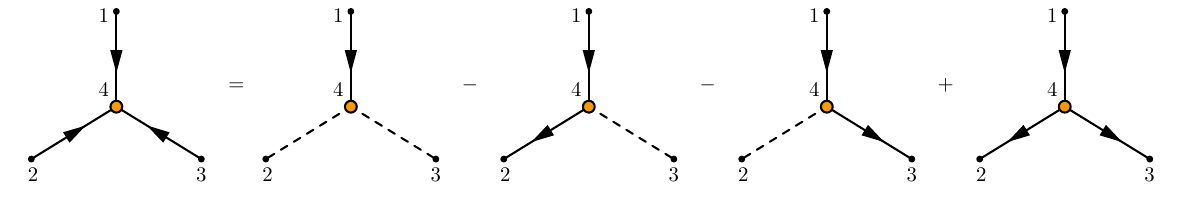}
\caption{The family-tree decomposition of a 4-site nested time integral with an internal vertex. The internal vertex (Vertex 4) with $E_4=0$ is marked with an orange dot.} 
  \label{fig_isored}
\end{figure}

We take this opportunity to make a general comment on the computation of graphs with internal vertices, as briefly mentioned in Footnote \ref{fn_internal}. We illustrate the point with a concrete example. Suppose we want to compute the following integral with $E_4=0$:
\begin{align}
\label{eq_T4internal}
  \mathbb{T}_{q_1q_2q_3q_4}(E_1,E_2,E_3)= \int_{-\infty}^0 \prod_{i=1}^4\Big[\di\tau_i\,(-\tau_i)^{q_i-1}e^{\ii E_i\tau_i}\Big]\theta(\tau_4-\tau_1)\theta(\tau_4-\tau_2)\theta(\tau_4-\tau_3).
\end{align}
Note that we set $E_4=0$ on the right hand side. Suppose we want the result for this integral with $E_1$ chosen as the maximal energy. Then, according to our reduction procedure, we should proceed with the following family-tree decomposition:
\begin{align}
  \mathbb{T}_{q_1q_2q_3q_4}(E_1,E_2,E_3)
  =&~\mathcal{C}_{q_1q_4}(\wh{E}_1,0)\mathcal{C}_{q_2}(E_2)\mathcal{C}_{q_3}(E_3)-\mathcal{C}_{q_1q_4q_2}(\wh{E}_1,0,E_2)\mathcal{C}_{q_3}(E_3)\n\\
  &-\mathcal{C}_{q_1q_4q_3}(\wh{E}_1,0,E_3)\mathcal{C}_{q_2}(E_2)
  +\mathcal{C}_{q_1q_4q_2q_3}^\text{(iso)}(\wh{E}_1,0,E_2,E_3).
\end{align}
This is shown diagrammatically in Fig.\ \ref{fig_isored}. (Note that the last graph in Fig.\ \ref{fig_isored} is exactly the previous example in (\ref{eq_4site0}).) Then, applying the general formula (\ref{eq_family}) to all families here, and setting the summation variable $n_4=0$, we get:
\begin{align}
\label{eq_T4internalRes}
  &\mathbb{T}_{q_1q_2q_3q_4}(E_1,E_2,E_3)\n\\
  =&~\FR{\Gamma[q_{14},q_{2},q_3]/q_4}{(\ii E_1)^{q_{14}}(\ii E_2)^{q_{2}}(\ii E_3)^{q_{3}}}-\bigg\{\FR{\Gamma[q_3]}{(\ii E_1)^{q_{142}}(\ii E_3)^{q_{3}}}\sum_{n_2=0}^\infty\FR{(-1)^{n_2}\Gamma[q_{142}+n_2]}{(n_2+q_2)(n_2+q_{24})}\FR{\varrho_{21}^{n_2}}{n_2!}\n\\
  &+(2\leftrightarrow 3)\bigg\}+\FR{1}{(\ii E_1)^{q_{1234}}}\sum_{n_2,n_3=0}^\infty\FR{(-1)^{n_{23}}\Gamma[n_{23}+q_{1234}]}{(n_2+q_2)(n_3+q_3)(n_{23}+q_{234})}\FR{\varrho_{21}^{n_2}}{n_2!}\FR{\varrho_{31}^{n_3}}{n_3!}.
\end{align}
On the other hand, we can also compute the integral (\ref{eq_T4internal}) directly. First, we integrate out $\tau_1$, $\tau_2$, and $\tau_3$, and the result is:
\begin{align}
\label{eq_T4directInt} 
  \mathbb{T}_{q_1q_2q_3q_4}(E_1,E_2,E_3)= \prod_{i=1}^3\bigg[\int_{s_i}\FR{\Gamma[s_i](\ii E_i)^{-s_i}}{(s_i-q_i)}\bigg]\int_{-\infty}^0 \di\tau_4(-\tau_4)^{q_{1234}-1-s_{123}} .
\end{align}
Now, the final layer integral contains no energy variable since it is from an internal vertex. Finishing this integral, we get a $\de$ function:
\bge
\label{eq_inttau4}
  \int_{-\infty}^0 \di\tau_4(-\tau_4)^{q_{1234}-1-s_{123}}=(2\pi)\de\big[\ii(q_{1234}-s_{123})\big].
\ede 
Now we need to choose a maximal energy. Suppose we choose $E_1$ without loss of generality. Then, we use the above $\de$ function to integrate out $s_1$. Then, (\ref{eq_T4directInt}) becomes:
\begin{align} 
  \mathbb{T}_{q_1q_2q_3q_4}(E_1,E_2,E_3)=  \int_{s_2,s_3}\FR{\Gamma[q_{1234}-s_{23},s_2,s_3](\ii E_1)^{s_{23}-q_{1234}}(\ii E_2)^{-s_2}(\ii E_3)^{-s_3}}{(q_{234}-s_{23})(s_2-q_2)(s_3-q_3)}   .
\end{align}
We can finish this integral by collecting the residues of all left poles of the integrand, as before. The result exactly agrees with (\ref{eq_T4internalRes}). 

There are two lessons to be learnt here. First, when computing a specific nested integral, if we decide to do the time integral directly, we do not have to be as rigid as when we derive the family-tree decomposition. Instead, we can always do the nested integral so long as the integral has a partial order, and the latest (or earliest) site does not have to be the maximal-energy site. The choice of maximal energy can be delayed until we perform the Mellin integral, where we do need a maximal energy to decide how to make the series expansion. On the other hand, the advantage of family-tree decomposition is that we do not have to compute the integral at all; So long as we follow this reduction procedure, we can write down the answer directly. 

The second lesson is about the $\de$ function generated from an energy-less time integral, such as the one in (\ref{eq_inttau4}). When we use (\ref{eq_inttau4}) to integrate out a Mellin variable, say $s_1$, we are effectively setting $s_1=q_{1234}-s_{23}$ everywhere in the integrand. Then, all previous left (right) poles of $s_1$ now become right (left) poles of $s_{23}$. However, this left-right flip is harmless at least for tree graphs. The general rule is that, whenever we have a $\de$ function from an internal vertex, we choose a maximal energy among all energies connected at this vertex, and we use the $\de$ function to integrate out the Mellin variable associated with the maximal energy. Then, we still pick up left poles of other Mellin variables to finish the Mellin integral. In this way, we will end up with a series expansion in terms of small energy ratios, as shown in the above example.

\paragraph{Multiple maximal energies.} Finally, there is a more difficult parameter region where the energies at more than one sites become equal or comparable. This case is tractable if the equal energies are not maximal. The only tricky situation is when the equal energies are maximal, so that, in the series solution (\ref{eq_family}), there is at least one energy ratio $\varrho_{j1}$ approaching 1. At this point, the series representation is likely divergent. There are several things one can try in this case. First, it is always possible to finish any one layer of summation in the general formula (\ref{eq_family}) in terms of a (generalized) hypergeometric function ${}_p\mathcal{F}_q$. Then, one can study the behavior of this hypergeometric function with argument equal to 1. Such a pure analytical strategy can sometimes be extended to two-variable summation as well. Second, one can switch to the partially resolved representation (\ref{eq_EpPRMB}) discussed in Sec.\ \ref{sec_alter}, so that the result is expanded in powers of $1/E_{1\cdots N}$ instead of the inverse of any single energy variable. This helps to improve the convergence of the series in many cases. As mentioned above, this is a practical way to discover many transformation-of-variable formula for multi-variable hypergeometric functions, and thus could be particularly useful. Third, when all previous methods fail (such as when all energies become nearly equal), we can do numerical interpolation to sew together disconnected parameter region with convergent series expressions. We leave this somewhat mathematically oriented problem to a future work.

\section{General Two Massive Exchanges}
\label{sec_2mass}

\begin{figure}
\centering  
\includegraphics[width=0.42\textwidth]{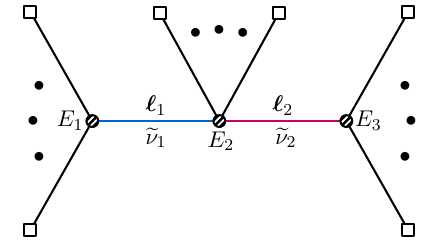}
\caption{A general tree graph with two massive exchanges.} 
  \label{fig_6pt}
\end{figure}

With the nested time integral done in the last section, in principle, we are able to compute arbitrary tree-level inflation correlators with any number of massive exchanges. In this section, we illustrate this procedure with a concrete example, namely a general tree graph with two massive exchanges, as shown in Fig.\ \ref{fig_6pt}. We follow the diagrammatic representation of \cite{Chen:2017ryl}. In particular, the external (bulk-to-boundary) propagators can be either conformal scalars with $m^2=2$, massless scalar fields such as the inflaton, or the massless spin-2 graviton. The conformal scalar is technically easiest and is often used as a starting point in a theoretical analysis of inflation correlators. The cases of massless scalar and tensor modes are more relevant to CC phenomenology. On the other hand, the two internal (bulk) propagators represent two massive scalar fields which can be either identical or distinct. There is no difficulty in generalizing the bulk lines to massive fields with spins or with helical chemical potentials, but we choose to work with scalars of principle series ($m_1,m_2>3/2$) for definiteness. Thus, we assign two mass parameters $\wt\nu_{1,2}\equiv\sqrt{m_{1,2}^2-9/4}$ for the two lines, respectively. 

\subsection{Three-vertex seed integral}

Following the diagrammatic rule of SK formalism \cite{Chen:2017ryl}, one can show that the correlators in the form of Fig.\ \ref{fig_6pt} can in general be reduced to the following \emph{three-vertex seed integral}: 
\begin{keyeqn}
\begin{align}
\label{eq_6ptseed}
  \mathcal{I}_{\aa_1\aa_2\aa_3}^{p_1p_2p_3}
  \equiv & -\ii\aa_1\aa_2\aa_3 E_1^{p_1+1}E_2^{p_2+1}E_3^{p_3+1}\ell_1^3\ell_2^3\int_{-\infty}^0\di\tau_1\di\tau_2\di\tau_3(-\tau_1)^{p_1}(-\tau_2)^{p_2}(-\tau_3)^{p_3}\n\\
  &\times e^{\ii\aa_1 E_1\tau_1+\ii\aa_2 E_2\tau_2+\ii\aa_3 E_3\tau_3}D_{\aa_1\aa_2}^{(\wt\nu_1)}(\ell_1;\tau_1,\tau_2)D_{\aa_2\aa_3}^{(\wt\nu_2)} (\ell_2;\tau_2,\tau_3).
\end{align}
\end{keyeqn}
Here, as before, $E_i$ represents the total energies of \emph{bulk-to-boundary} lines at the Vertex $i$, while $\bm \ell_1$ and $\bm\ell_2$ represent the 3-momenta of the two internal lines, respectively. As before, we have included power factors of the form $(-\tau_i)^{p_i}$ to allow for different choices of external states and coupling types. Finally, the two bulk massive propagators $D_{\aa\bb}^{(\wt\nu_1)}$ and $D_{\bb\cc}^{(\wt\nu_2)}$ are given in (\ref{eq_Dmp}), (\ref{eq_Dpm}), and (\ref{eq_Dpmpm}). To minimize unnecessary complications, we will take $p_1,p_2,p_3\in\mathbb{R}$. Generalization to complex values of $p_{i}$ is straightforward, although the expressions will be lengthier. 

The $E$ and $\ell$ factors in front of the integral in (\ref{eq_6ptseed}) are included to make the integral dimensionless. The reason we introduce this special combination of energy variables is the following: We can define the dimensionless integration variables $z_i=E_i\tau_i$ $(i=1,2,3)$, and use the momentum ratios $r_1=\ell_1/E_1$, $r_2=\ell_1/E_2$, $r_3=\ell_2/E_2$, $r_4=q_\ell/E_3$. Then, one can easily verify:
\bge
\ell_1^3D_{\aa\bb}^{(\wt\nu_1)}(\ell_1;\tau_1,\tau_2)= \wh D_{\aa\bb}^{(\wt\nu_1)}(r_1z_1,r_2z_2),
\ede
and a similar relation for the $\ell_2$-propagator. Then, the seed integral is manifestly dimensionless and depends only on dimensionless energy ratios:
\begin{align}
  \mathcal{I}_{\aa_1\aa_2\aa_3}^{p_1p_2p_3}(r_1,r_2,r_3,r_4)
  \equiv & \int_{-\infty}^0\prod_{i=\ell}^3\Big[\ii\aa_\ell\di z_\ell(-z_\ell)^{p_\ell} e^{\ii\aa_\ell z_\ell}\Big]\wh D_{\aa_1\aa_2}^{(\wt\nu_1)}(r_1z_1,r_2z_2)\wh D_{\aa_2\aa_3}^{(\wt\nu_2)} (r_3z_2,r_4z_3).
\end{align}
There are simple kinematic constraints for the range of $r_i$ variables from the momentum conservation at each vertex. For instance, let there be $N$ external lines ending at Site 1 with 3-momenta $\bm k_1,\cdots,\bm k_N$. Then, by definition, $E_1=k_1+\cdots+k_N\geq |\bm k_1+\cdots+\bm k_N|=\ell_1$. Thus we have $0<r_1<1$. Similarly, we have $0<r_4<1$. On the other hand, the constraints on $r_2$ and $r_3$ are much weaker. In general, these two values can take any nonnegative real values.

Many correlators with two massive exchanges can be expressed in terms of $\mathcal{I}_{\aa_1\aa_2\aa_3}^{p_1p_2p_3}$. For example, when the external leges are conformal scalars $\phi_c$ with cubic and quartic direct couplings with two massive scalars $\si_{1,2}$, we can form a 6-point correlator. The Lagrangian is:
\begin{align}
\label{eq_ld6pt}
  \ld\supset -\FR12a^2(\pd_\mu\phi_c)^2-\FR12a^4m_c^2\phi_c^2-\FR12\sum_{i=1}^2\Big[a^2(\pd_\mu\si_i)^2+a^4m_i^2\si_i+a^4\mu_i\phi_c^2\si_i\Big]-\FR12 a^4\lam\phi_c^2\si_1\si_2.
\end{align}
Here $m_c=\sqrt 2$ is the mass of the conformal scalar $\phi_c$, and $m_{1,2}>3/2$ are the mass of two scalars $\si_{1,2}$, respectively. We also include two cubic couplings with dimension-1 coupling constants $\mu_i$ $(i=1,2)$ and a quartic coupling with dimensionless coupling $\lam$. The powers of scale factors $a=-1/\tau$ are introduced to make the Lagrangian scale-invariant, and the spacetime indices in (\ref{eq_ld6pt}) are contracted by Minkowski metric $\eta_{\mu\nu}$. Then, the 6-point correlator is shown in Fig.\ \ref{fig_6pt} with all black dots removed. With the diagrammatic rule, it is easy to see that the corresponding SK integral reduces to the seed integral in the following way: 
\begin{align}
  \mathcal{G}(\bm k_1,\cdots,\bm k_6)=-\mu_1\mu_2\lam\FR{(-\tau_f)^6k_{12}k_{34}k_{56}}{64k_1\cdots k_6\ell_1^3\ell_2^3}\sum_{\aa_1,\aa_2,\aa_3=\pm}\mathcal{I}_{\aa_1\aa_2\aa_3}^{-2,-2,-2}\bigg(\FR{\ell_1}{k_{12}},\FR{\ell_1}{k_{56}},\FR{\ell_2}{k_{56}},\FR{\ell_2}{k_{34}}\bigg),
\end{align}
where we have introduced a final-time cutoff $\tau_f$.
We note that this expression is for a single graph $\mathcal{G}(\bm k_1,\cdots,\bm k_6)$ rather than the whole correlator $\mathcal{T}(\bm k_1,\cdots,\bm k_6)$ at the same perturbative order. The correlator $\mathcal{T}$ can be obtained from the graph $\mathcal{G}$ by including suitable permutations, which we do not spell out here.

As another example, we can consider the 4-point correlators of massless inflaton $\varphi$ with two massive exchanges, as shown in Fig.\ \ref{fig_4pt2massive}. We assume that the inflaton $\varphi$ is coupled derivatively, to respect the approximate shift symmetry of the inflaton field, and also to produce a nontrivial result.\footnote{Directly coupled 2-point vertex (the mass mixing) is trivial in the sense that it can be rotated away by diagonalizing the mass matrix.} The relevant Lagrangian is:
\bge
\label{eq_Ld4pt}
  \ld\supset -\FR12a^2(\pd_\mu\varphi)^2-\sum_{i=1}^2\bigg[\FR12a^2(\pd_\mu\si_i)^2+\FR12a^4m_i^2\si_i+a^3\mu_i\varphi'\si_i\bigg]-\FR{\lam}2 a^2 (\varphi')^2\si_1\si_2.
\ede
Then, the 4-point graph in Fig.\ \ref{fig_4pt2massive} can be expressed as: 
\begin{align}
\label{eq_4ptT}
  \mathcal{G}(\bm k_1,\cdots,\bm k_4)=-\FR{\mu_1\mu_2\lam}{16k_1^3k_2^3k_3k_4k_{34}}\sum_{\aa_1,\aa_2,\aa_3=\pm}\mathcal{I}_{\aa_1\aa_2\aa_3}^{-2,0,-2}\Big(1,\FR{k_1}{k_{34}},\FR{k_2}{k_{34}},1\Big).
\end{align}
Thus the computation of the 4-point correlator (\ref{eq_4ptT}) requires us to take simultaneous folded limit $r_1\to 1$ and $r_4\to 1$ which is a bit nontrivial. We shall take this limit in the next section.

\subsection{Computing the seed integral}

Now we are going to compute the seed integral (\ref{eq_6ptseed}). The computation is rather lengthy and tedious. Here we only outline the main steps, and collect more details in App.\ \ref{app_detail}

The seed integral $\mathcal{I}_{\aa_1\aa_2\aa_3}^{p_1p_2p_3}(r_1,r_2,r_3,r_4)$ in  (\ref{eq_6ptseed}) has 8 SK branches, depending on the values of the 3 SK indices $\aa_1,\aa_2,\aa_3=\pm$. We only need to compute 4 integrals: $\mathcal{I}_{+++}^{p_1p_2p_3}$, $\mathcal{I}_{++-}^{p_1p_2p_3}$, $\mathcal{I}_{+-+}^{p_1p_2p_3}$, and $\mathcal{I}_{-++}^{p_1p_2p_3}$. Since all exponents $p_i$ ($i=1,2,3$) are real, the other four can be obtained by taking complex conjugation. As a general rule, for graphs with real couplings, flipping the sign of \emph{all} SK indices simultaneously brings an integral to its complex conjugate. 

As usual, the main difficulty of the computation comes from the time orderings. Thus, our first step is to rewrite the time-ordered propagator $D_{++}$ in a more suitable form. For definiteness, we work in the region where $E_2>E_1$ and $E_2>E_3$. Then, according to the discussion of the previous section, whenever we have a time ordering between $\tau_1$ and $\tau_2$, or between $\tau_3$ and $\tau_2$, we should let $\tau_2$ take the earlier position. Thus, we use the following expression for the two $D_{++}$ propagators:
\begin{align}
  D_{++}^{(\wt\nu_1)}(\ell_1;\tau_1,\tau_2)
  =&~D_{+-}^{(\wt\nu_1)}(\ell_1;\tau_1,\tau_2)+\Big(D_{-+}^{(\wt\nu_1)}(\ell_1;\tau_1,\tau_2)- D_{+-}^{(\wt\nu_1)}(\ell_1;\tau_1,\tau_2)\Big)\theta(\tau_1-\tau_2),\\
  D_{++}^{(\wt\nu_2)}(\ell_2;\tau_2,\tau_3)
  =&~D_{-+}^{(\wt\nu_2)}(\ell_2;\tau_2,\tau_3)+\Big(D_{+-}^{(\wt\nu_2)}(\ell_2;\tau_2,\tau_3)- D_{-+}^{(\wt\nu_2)}(\ell_2;\tau_2,\tau_3)\Big)\theta(\tau_3-\tau_2).
\end{align}
After taking this representation, the expressions for the four integrals $\mathcal{I}_{+++}^{p_1p_2p_3}$, $\mathcal{I}_{++-}^{p_1p_2p_3}$, $\mathcal{I}_{+-+}^{p_1p_2p_3}$, and $\mathcal{I}_{-++}^{p_1p_2p_3}$ can be obtained directly. Here we show one example of $\mathcal{I}_{+++}^{p_1p_2p_3}$. The complete list is given in (\ref{eq_Ippp}) to (\ref{eq_Ipmp}).
\begin{align} 
  \mathcal{I}_{+++}^{p_1p_2p_3}
  =& -\ii E_1^{p_1+1}E_2^{p_2+1}E_3^{p_3+1}\ell_1^3\ell_2^3\int_{-\infty}^0\di\tau_1\di\tau_2\di\tau_3(-\tau_1)^{p_1}(-\tau_2)^{p_2}(-\tau_3)^{p_3}e^{\ii (E_1\tau_1+ E_2\tau_2+ E_3\tau_3)}\n\\
   &\times \Big[D_{+-}^{(\wt\nu_1)}(\ell_1;\tau_1,\tau_2)+\Big(D_{-+}^{(\wt\nu_1)}(\ell_1;\tau_1,\tau_2)- D_{+-}^{(\wt\nu_1)}(\ell_1;\tau_1,\tau_2)\Big)\theta(\tau_1-\tau_2)\Big]\n\\
   &\times \Big[D_{-+}^{(\wt\nu_2)}(\ell_2;\tau_2,\tau_3)+\Big(D_{+-}^{(\wt\nu_2)}(\ell_2;\tau_2,\tau_3)- D_{-+}^{(\wt\nu_2)}(\ell_2;\tau_2,\tau_3)\Big)\theta(\tau_3-\tau_2)\Big].
\end{align}

\begin{figure}
\centering  
\includegraphics[width=0.32\textwidth]{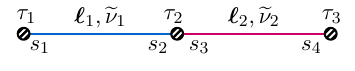}
\caption{The Mellin variables for the two massive propagators in the computation of the three-vertex seed integral.} 
  \label{fig_6ptMellin}
\end{figure}

Next, we expand all the integrals and classify the terms according to whether the adjacent two time variables are time-ordered (T) or factorized (F).
\begin{align}
  \mathcal{I}_{+++}^{p_1p_2p_3}
  =&~\mathcal{I}_{+++}^\text{(FF)}+\mathcal{I}_{+++}^\text{(FT)}+\mathcal{I}_{+++}^\text{(TF)}+\mathcal{I}_{+++}^\text{(TT)},\\
  \mathcal{I}_{++-}^{p_1p_2p_3}
  =&~\mathcal{I}_{++-}^\text{(F)}+\mathcal{I}_{++-}^\text{(T)},\\
  \mathcal{I}_{++-}^{p_1p_2p_3}
  =&~\mathcal{I}_{-++}^\text{(F)}+\mathcal{I}_{-++}^\text{(T)}.
\end{align} 
The explicit expressions of these integrals are given in (\ref{eq_IFF})-(\ref{eq_ImppN}).
Then, we define the following 4 integrals:
\begin{align}
\label{eq_IFFdef}
  \mathcal{I}^\text{(FF)}=&~\Big[\mathcal{I}_{+++}^\text{(FF)}+\mathcal{I}_{++-}^\text{(F)}+\mathcal{I}_{-++}^\text{(F)}+\mathcal{I}_{+-+}\Big]+\text{c.c.};\\
  \mathcal{I}^\text{(FT)}=&~\Big[\mathcal{I}_{+++}^\text{(FT)}+\mathcal{I}_{-++}^\text{(T)} \Big]+\text{c.c.};\\
  \mathcal{I}^\text{(TF)}=&~\Big[\mathcal{I}_{+++}^\text{(TF)}+\mathcal{I}_{++-}^\text{(T)} \Big]+\text{c.c.};\\
\label{eq_ITTdef}
  \mathcal{I}^\text{(TT)}=&~\mathcal{I}_{+++}^\text{(TT)}+\text{c.c.}.
\end{align}
After this regrouping of terms, each of the four integrals in $\{\mathcal{I}^\text{(FF)},\mathcal{I}^\text{(FT)},\mathcal{I}^\text{(TF)},\mathcal{I}^\text{(TT)}\}$ has a definite nesting structure in its time integral, which can then be readily computed using the  PMB representation. The procedure is by now standard: We first use the MB representations for the two massive propagators:
\begin{align} 
\label{eq_Dnu1}
    D_{\pm\mp}^{(\wt\nu_1)}(\ell_1;\tau_1,\tau_2) 
    =&~ \FR{1}{4\pi}
    \int_{s_1,s_2}
    e^{\mp\ii\pi(s_1-s_2)}\Big(\FR{\ell_1}2\Big)^{-2s_{12}}
    (-\tau_1)^{-2s_1+3/2}(-\tau_2)^{-2s_2+3/2}\n\\
    &\times \Gamma\Big[s_1-\FR{\ii\wt\nu_1}2,s_1+\FR{\ii\wt\nu_1}2,s_2-\FR{\ii\wt\nu_1}2,s_2+\FR{\ii\wt\nu_1}2\Big],\\ 
\label{eq_Dnu2}
    D_{\pm\mp}^{(\wt\nu_2)}(\ell_2;\tau_2,\tau_3) 
    =&~ \FR{1}{4\pi}
    \int_{s_3,s_4}
    e^{\mp\ii\pi(s_3-s_4)}\Big(\FR{\ell_2}2\Big)^{-2s_{34}}
    (-\tau_2)^{-2s_3+3/2}(-\tau_3)^{-2s_4+3/2}\n\\
    &\times \Gamma\Big[s_3-\FR{\ii\wt\nu_2}2,s_3+\FR{\ii\wt\nu_2}2,s_4-\FR{\ii\wt\nu_2}2,s_4+\FR{\ii\wt\nu_2}2\Big].
\end{align} 
The assignment of the four Mellin variables $s_1,\cdots,s_4$ are shown in Fig.\ \ref{fig_6ptMellin}.
Then, the time integrals can be directly done for all branches using results of the previous section. It then remains to finish the integrals over the four Mellin variables $s_1,\cdots,s_4$. We work in the region where the two bulk momenta $\bm\ell_1$ and $\bm\ell_2$ are both softer than all energies $E_1$, $E_2$, and $E_3$. In this region, we have $0<r_i<1$ ($i=1,2,3,4$), which means that we should pick up all left poles to finish the Mellin integrals.\footnote{Slightly stronger condition than $0<r_i<1$ may be required for the convergence of the final result, but we expect that the radius of convergence for the final result should be $\order{1}$. We will not be concerned with the precise value of the radius of convergence.} 

As already explained in Sec.\ \ref{sec_tree}, the left poles of the Mellin integrand are all from the $\Gamma$ factors in (\ref{eq_Dnu1}) and (\ref{eq_Dnu2}), and are given by:
\begin{align}
\label{eq_leftPoles}
  &s_1=-n_1-\FR{\ii\aa_1\wt\nu_1}2,
  &&s_2=-n_2-\FR{\ii\aa_2\wt\nu_1}2,
  &&s_3=-n_3-\FR{\ii\aa_3\wt\nu_2}2,
  &&s_4=-n_4-\FR{\ii\aa_4\wt\nu_2}2.
\end{align}
Here $n_i=0,1,2,\cdots$ $(i=1,\cdots,4)$, and $\aa_i=\pm$ are not SK indices. Thus, by collecting the residues of the integrand at all these poles, we get the final answer for the three-vertex seed integral in (\ref{eq_6ptseed}). Similar to the case of single-exchange graph studied in previous works, it turns out convenient to express the final answer of  $\mathcal{I}_{\aa\bb\cc}^{p_1p_2p_3}$ with all indices $\aa,\bb,\cc$ summed. Then, the result can be written as a sum of four distinct terms, plus trivial permutations:
\begin{keyeqn}
\begin{align}
\label{eq_seedIntResult}
  \sum_{\aa,\bb,\cc=\pm}\mathcal{I}^{p_1p_2p_3}_{\aa\bb\cc}\big(\{r_i\}\big) =&~\bigg\{\Big[\mathcal{I}_\text{SS}\big(\{r_i\}\big) +\mathcal{I}_\text{SB}\big(\{r_i\}\big) +\mathcal{I}_\text{BS}\big(\{r_i\}\big) +\mathcal{I}_\text{BB}\big(\{r_i\}\big) \Big]\n\\
  &+ (\wt\nu_1\rightarrow -\wt\nu_1) + (\wt\nu_2\rightarrow -\wt\nu_2)+\bgp\wt\nu_1\rightarrow -\wt\nu_1\\ \wt\nu_2\rightarrow -\wt\nu_2\edp\bigg\} + \text{c.c.}.
\end{align}  
\end{keyeqn}
Here we use the shorthand notation $\mathcal{I}^{p_1p_2p_3}_{\aa\bb\cc} (\{r_i\} )\equiv\mathcal{I}^{p_1p_2p_3}_{\aa\bb\cc} (r_1,r_2,r_3,r_4)$. Also, we are using the terminology of CC physics: The term $\mathcal{I}_\text{SS}$, with the subscript SS denoting ``signal-signal,'' is nonanalytic in all four momentum ratios $r_1,\cdots,r_4$ when these ratios go to 0, and thus corresponds to the CC signals generated by both bulk lines:
\begin{align}
  \mathcal{I}_\text{SS}  =&\sum_{\aa_1,\aa_2=\pm}\mb{A}^{p_1p_2p_3}_{\aa_1\aa_2}(r_1,r_2,r_3,r_4)\Big(\FR{r_1}2\Big)^{\ii\wt\nu_1}\Big(\FR{r_2}{2}\Big)^{\ii\aa_1\wt\nu_1}\Big(\FR{r_3}{2}\Big)^{\ii\aa_2\wt\nu_2}\Big(\FR{r_4}2\Big)^{\ii\wt\nu_2},
\end{align}
where the function $\mb{A}_{\aa_1\aa_2}^{p_1p_2p_3}(r_1,r_2,r_3,r_4)$ is fully analytic in the limit $r_i\to 0$ ($i=1,\cdots,4$), whose explicit expression will be given below in (\ref{eq_boldA}).

Next, the term $\mathcal{I}_\text{SB}$ denotes the ``signal-background,'' which contains CC signals from the $\wt\nu_1$-leg, and thus is nonanalytic in both $r_1$ and $r_2$ as $r_{1,2}\to 0$, but is fully analytic\footnote{There could be factors such as $(r_3/r_4)^{p_3}$ which we do not count as nonanalytic.} in both $r_3$ and $r_4$ as $r_{3,4}\to 0$. Likewise, the term  $\mathcal{I}_\text{SB}$, denoting the ``background-signal,'' is a trivial permutation of $\mathcal{I}_\text{SB}$:
\begin{align}
  \mathcal{I}_\text{SB} =&\sum_{\aa=\pm}\mb{B}_{\wt\nu_1\wt\nu_2|\aa}^{p_1p_2p_3}(r_1,r_2,r_3,r_4)\Big(\FR{r_1}2\Big)^{\ii\wt\nu_1}\Big(\FR{r_2}2\Big)^{\ii\aa\wt\nu_1},\\
  \mathcal{I}_\text{BS} =&\sum_{\aa=\pm}\mb{B}_{\wt\nu_2\wt\nu_1|\aa}^{p_1p_2p_3}(r_4,r_3,r_2,r_1)\Big(\FR{r_4}2\Big)^{\ii\wt\nu_2}\Big(\FR{r_3}2\Big)^{\ii\aa\wt\nu_2}.
\end{align}
Again, the function $\mb{B}_{\wt\nu_1\wt\nu_2|\aa}^{p_1p_2p_3}(r_1,r_2,r_3,r_4)$ is fully analytic in all $r_i$ as $r_i\to 0$ ($i=1,\cdots,4$), whose explicit expression will be given below in (\ref{eq_boldB}).

Finally, the term $\mathcal{I}_\text{BB}$ denotes the ``background-background,'' and is fully analytic in all momentum ratios $r_i$ as $r_i\to 0$. 
\begin{align}
  \mathcal{I}_\text{BB}
=& ~ \FR{\sin(\ii\pi\wt\nu_1)\sin(\ii\pi\wt\nu_2)e^{-\ii p_{123}\pi/2}}{4\pi^2} 
 \sum_{n_1,\cdots,n_4=0}^\infty\FR{r_2^3 r_3^3}{n_1!n_2!n_3!n_4!}  \Big(\FR{r_2}{r_1}\Big)^{p_1+1}\Big(\FR{r_3}{r_4}\Big)^{p_3+1} \Big(\FR{r_2}{2}\Big)^{2n_{12}}\Big(\FR{r_3}{2}\Big)^{2n_{34}}\n\\
 &\times \Gamma\Big[-n_1-\ii \wt\nu_1,-n_2+\ii\wt\nu_1,-n_3+\ii\wt\nu_2,-n_4-\ii\wt\nu_2\Big] \n\\
 &\times  \mathcal{F}_2 \left[ p_{123}+2n_{1234}+9 \middle|\bgm p_1+2n_1+\ii\wt\nu_1+\fr52,p_3+2n_4+\ii\wt\nu_2+\fr52 \\ p_1+2n_1+\ii\wt\nu_1+\fr72 , p_3+2n_4+\ii\wt\nu_2+\fr72 \edm \middle|-\FR{r_2}{r_1},-\FR{r_3}{r_4}\right],
\end{align}
where the summation is taken over $n_1,\cdots,n_4=0,1,\cdots$, and $\mathcal{F}_2$ denotes the dressed Appell $F_2$ function, which is defined in (\ref{eq_dressedF2}).

Now we give explicit expressions for the functions $\mb{A}_{\aa_1\aa_2}^{p_1p_2p_3}$ and $\mb{B}_{\wt\nu_1\wt\nu_2|\aa}^{p_1p_2p_3}$. The function $\mb{A}_{\aa_1\aa_2}^{p_1p_2p_3}$ is given by:
\begin{align}
\label{eq_boldA}
    \mb{A}_{\aa_1\aa_2}^{p_1p_2p_3}(r_1,r_2,r_3,r_4) \equiv & ~ \mathcal{C}^{p_1p_2p_3}_{\wt\nu_1\wt\nu_2|\aa_1\aa_2} (r_1r_2r_3r_4)^{\frac32} \mb{F}(p_1+\ii\wt\nu_1+\fr52,-\ii\wt\nu_1;r_1^2)\mb{F}(p_3+\ii\wt\nu_2+\fr52,-\ii\wt\nu_2;r_4^2)\n\\
    &\times \mathcal{F}_4 \left[ \bgm \fr{\ii\aa_1\wt\nu_1+\ii\aa_2\wt\nu_2+p_2+4}2,\fr{\ii\aa_1\wt\nu_1+\ii\aa_2\wt\nu_2+p_2+5}2\\1+\ii\aa_1\wt\nu_1,1+\ii\aa_2\wt\nu_2\edm \middle| r_2^2,r_3^2\right].
\end{align}
Here the coefficient $\mathcal{C}^{p_1p_2p_3}_{\wt\nu_1\wt\nu_2|\aa_1\aa_2}$ is defined by:
\begin{align}
    \mathcal{C}^{p_1p_2p_3}_{\wt\nu_1\wt\nu_2|\aa_1\aa_2} \equiv \FR{2^{p_2-1+\ii\aa_1\wt\nu_1+\ii\aa_2\wt\nu_2}}{\pi^{1/2}\operatorname{sin}(\ii\aa_1\pi\wt\nu_1)\operatorname{sin}(\ii\aa_2\pi\wt\nu_2)}\Big[&-e^{-\ii\pi(\ii\aa_1\wt\nu_1+\ii\aa_2\wt\nu_2+p_{123}/2)}+\ii e^{-\ii\pi(\ii\aa_1\wt\nu_1+p_{12}/2-p_3/2)} \n\\
   &+\ii e^{-\ii\pi(\ii \aa_2\wt\nu_2-p_1/2+p_{23}/2)}+ e^{\ii\pi(-p_2/2+p_{13}/2)}\Big].
\end{align}
The function $\mb{F}(a,b;z)$ in (\ref{eq_boldA}) is defined in terms of the Gaussian hypergeometric function, given in (\ref{eq_boldF}), and the function $\mathcal{F}_4$ denotes the dressed Appell $F_4$ function, which is defined in (\ref{eq_dressedF4}). Finally, the function $\mb{B}_{\wt\nu_1\wt\nu_2|\aa}^{p_1p_2p_3}$ is given by:
\begin{align}
\label{eq_boldB}
    \mb{B}_{\wt\nu_1\wt\nu_2|\aa}^{p_1p_2p_3}&\,(r_1,r_2,r_3,r_4) 
    \equiv  \FR{e^{-\ii\pi(p_{23}+\ii\aa\wt\nu_1-1/2)/2}}{4\pi^2} \sin\big[\fr{\pi}2(\ii\aa\wt\nu_1+p_1-\fr32)\big]\sin(\ii\pi\wt\nu_2)(r_1 r_2r_3^2)^{\frac32} \n\\
  &\times \sum_{n_1,n_2,n_3=0}^
    {\infty} \FR{(-1)^{n_{123}}}{n_1!n_2!n_3!}\FR{\Gamma[-n_1-\ii\wt\nu_2,-n_2+\ii\wt\nu_2]}{p_3+n_3+2n_2-\ii\wt\nu_2+\fr52} \Big(\FR{r_3}2\Big)^{2n_{12}} \Big(\FR{r_3}{r_4}\Big)^{n_3+p_3+1}\n\\
  &\times \mb{F}(p_1+\ii\wt\nu_1+\fr52,-\ii\wt\nu_1;r_1^2) 
    \mb{F}(n_3+2n_{12}+p_{23}+\ii\aa\wt\nu_1+\fr{13}2,-\ii\aa\wt\nu_1;r_2^2). 
\end{align}

We can make a finer classification of terms with signals in (\ref{eq_seedIntResult}), according to whether the signal is local or nonlocal. Once again, the nonlocal signal means a piece in the correlator which is nonanalytic in the bulk momentum $\ell_1$ or $\ell_2$ as $\ell_{1,2}\to 0$. Using our momentum ratios, this means that a nonlocal signal contains noninteger powers of either $r_1r_2$ or $r_3r_4$. On the other hand, a local signal means a piece analytic as $\ell_{1,2}\to 0$ but nonanalytic in other momentum ratios. In our case, local signals come from noninteger powers of $r_1/r_2$ or $r_3/r_4$. Thus, we can further write:
\begin{align}
  \mathcal{I}_\text{SS}=&~\mathcal{I}_\text{LL}+\mathcal{I}_\text{LN}+\mathcal{I}_\text{NL}+\mathcal{I}_\text{NN};\\
  \mathcal{I}_\text{SB}=&~\mathcal{I}_\text{LB}+\mathcal{I}_\text{NB};\\
\mathcal{I}_\text{BS}=&~\mathcal{I}_\text{BL}+\mathcal{I}_\text{BN}.
\end{align}
Then, explicitly, we have: 
{\allowdisplaybreaks
\begin{align}
    \mathcal{I}_\text{LL}=&~ \mb{A}^{p_1p_2p_3}_{--}(r_1,r_2,r_3,r_4)\Big(\FR{r_1}{r_2}\Big)^{\ii\wt\nu_1}\Big(\FR{r_4}{r_3}\Big)^{\ii\wt\nu_2},\\
    \mathcal{I}_\text{LN}=&~ \mb{A}^{p_1p_2p_3}_{-+}(r_1,r_2,r_3,r_4)\Big(\FR{r_1}{r_2}\Big)^{\ii\wt\nu_1}\Big(\FR{r_3 r_4}{4}\Big)^{\ii \wt\nu_2},\\
    \mathcal{I}_\text{NL}=&~ \mb{A}^{p_1p_2p_3}_{+-}(r_1,r_2,r_3,r_4)\Big(\FR{r_1 r_2}{4}\Big)^{\ii\wt\nu_1}\Big(\FR{r_4}{r_3}\Big)^{\ii\wt\nu_2},\\
    \mathcal{I}_\text{NN}=&~ \mb{A}^{p_1p_2p_3}_{++}(r_1,r_2,r_3,r_4)\Big(\FR{r_1r_2}{4}\Big)^{\ii\wt\nu_1}\Big(\FR{r_3r_4}{4}\Big)^{\ii\wt\nu_2},\\
  \mathcal{I}_\text{LB}=&~\mb{B}_{\wt\nu_1\wt\nu_2|-}^{p_1p_2p_3}(r_1,r_2,r_3,r_4)\Big(\FR{r_1}{r_2}\Big)^{\ii\wt\nu_1},\\
  \mathcal{I}_\text{NB}=&~\mb{B}_{\wt\nu_1\wt\nu_2|+}^{p_1p_2p_3}(r_1,r_2,r_3,r_4)\Big(\FR{r_1r_2}{4}\Big)^{\ii\wt\nu_1},\\
  \mathcal{I}_\text{BL}=&~\mb{B}_{\wt\nu_2\wt\nu_1|-}^{p_1p_2p_3}(r_4,r_3,r_2,r_1)\Big(\FR{r_4}{r_3}\Big)^{\ii\wt\nu_2},\\
  \mathcal{I}_\text{BN}=&~\mb{B}_{\wt\nu_2\wt\nu_1|+}^{p_1p_2p_3}(r_4,r_3,r_2,r_1)\Big(\FR{r_3r_4}{4}\Big)^{\ii\wt\nu_2}.
\end{align}
}
We have checked numerically that our analytical result (\ref{eq_seedIntResult}) agrees well with a direct numerical integration of (\ref{eq_6ptseed}). 

\section{Four-Point Correlator with Two Massive Exchanges}
\label{sec_4point}

\begin{figure}
\centering  
\includegraphics[width=0.4\textwidth]{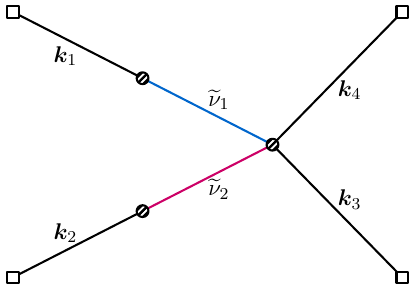}
\caption{The tree-level 4-point inflaton correlator with two massive exchanges.} 
  \label{fig_4pt2massive}
\end{figure}

Now, we consider a special case of the general two massive exchanges from the last section. We will compute a four-point graph in Fig.\ \ref{fig_4pt2massive}, in which we have four external massless scalar $\varphi$ derivatively coupled to two principal massive scalars in the bulk. The Lagrangian is given in (\ref{eq_Ld4pt}). We use this example to show how to take folded limits in the three-vertex seed integral. The correlator corresponding to Fig.\ \ref{fig_4pt2massive} has been given in (\ref{eq_4ptT}), which shows that all we need to do is to take the following folded limit of the three-vertex seed integral (\ref{eq_seedIntResult}):
\bge
  \lim_{r_1,r_4\to 1}\mathcal{I}_{\aa_1\aa_2\aa_3}^{-2,0,-2}(r_1,r_2,r_3,r_4).
\ede 
In the folded limit $r_1\to 1$ and $r_4\to 1$, we expect that various individual terms in (\ref{eq_seedIntResult}) diverge, but the divergence must cancel out in the full result, as a consequence of choosing the Bunch-Davies initial condition.\footnote{Choosing the Bunch-Davies initial condition is implicit in our choice for all the mode functions.} 

Specifically, all the hypergeometric functions in the $\mathbf{F}$ factors in both (\ref{eq_boldA}) and (\ref{eq_boldB}) could develop divergent terms when we take their arguments to 1. With the knowledge that these divergent terms must cancel among themselves, we can directly throw them away when evaluating the function $\mathbf{F}(a,b;z)$ at $z=1$. Using the expansion of hypergeometric function at argument unity, we get the following finite result for $\mathbf{F}(a,b;z)$ as $z\to 1$ \cite{Qin:2023ejc}:
\begin{align}
  \operatorname{Fin}\left\{\lim_{z\to 1}\mathbf{F}(a,b;z)\right\} = \Gamma\left[\bgm a,b,1-b,1/2-a-b\\ 1-b-a/2,1/2-b-a/2\edm\right], ~~~~~~ (1/2-a-b \notin \mathbb{Z}),
\end{align}
where $\operatorname{Fin}\{\}$ means the finite part of the expression within. The case of  $1/2-a-b \in \mathbb{Z}$ can be computed by taking the limit. For example, the limit of a term in $\mathcal{I}^\text{FF}$ when $p\to-2$ can be computed as:
\begin{align}
  \operatorname{Fin} \Bigg\{ \lim_{\substack{p\to-2 \\ r \to 1}} &
  \bigg[\Big(\FR{r}{2}\Big)^{\ii \wt\nu}\mathbf{F}(p+\ii\wt\nu+\fr52,-\ii\wt\nu;r^2)+(\wt\nu\to-\wt\nu)\bigg]\Bigg\} =\sqrt{2}\pi^{3/2}\operatorname{sech}(\pi\wt\nu).
\end{align}
With all limits of all $\mb{F}$ factors properly taken as above, we find a finite result for (\ref{eq_4ptT}), which can again be separated into several pieces according to their analytical properties at $k_1\to 0$ or $k_2\to 0$:
\begin{align}
\label{eq_4ptResult}
  &\sum_{\aa_1,\aa_2,\aa_3=\pm} \mathcal{I}_{\aa_1\aa_2\aa_3}^{-2,-2,-2}\big(1,r_2,r_3,1\big)\n\\
  =&~\Big\{\Big[\mathcal{T}_\text{SS}(r_2,r_3)+\mathcal{T}_\text{SB}(r_2,r_3)+\mathcal{T}_\text{BS}(r_2,r_3)+\mathcal{T}_\text{SS}(r_2,r_3)\Big]\n\\
  &+(\wt\nu_1\to-\wt\nu_1)+(\wt\nu_2\to-\wt\nu_2)+(\wt\nu_1\to-\wt\nu_1,\wt\nu_2\to-\wt\nu_2)\Big\}+\text{c.c.}.
\end{align}
Here $r_2\equiv k_1/k_{34}$ and $r_3\equiv k_2/k_{34}$. The four pieces $\{\mathcal{T}_\text{SS},\mathcal{T}_\text{SB},\mathcal{T}_\text{BS},\mathcal{T}_\text{BB}\}$ are defined in a similar way as before, according to whether the expression is analytic in $r_2\to 0$ or $r_3\to 0$ limit. In particular, the signal-signal piece is nonanalytic when $r_2\to 0$ and when $r_3\to 0$, whose explicit expression is:
\begin{align}
\label{eq_tSS}
  \mathcal{T}_\text{SS}(r_2,r_3)
  =&~ -\FR{4\pi^{5/2}(e^{\pi\wt\nu_1}-\ii)(e^{\pi\wt\nu_2}-\ii)}{\sin(2\pi\ii\wt\nu_1)\sin(2\pi\ii\wt\nu_2)}r_2^{\ii \wt\nu_1+3/2}r_3^{\ii\wt\nu_2+3/2} 
  \mathcal{F}_4 \left[ \bgm \fr{\ii\wt\nu_1+\ii\wt\nu_2+4}2,\fr{\ii\wt\nu_1+\ii\wt\nu_2+5}2\\1+\ii\wt\nu_1,1+\ii\wt\nu_2\edm \middle| r_2^2,r_3^2\right],
\end{align}
where $\mathcal{F}_4$ is again the dressed Appell $F_4$ function defined in (\ref{eq_dressedF4}). Next, the signal-background piece $\mathcal{T}_\text{SB}$ is nonanalytic in $r_2\to 0$ but analytic in $r_3\to 0$:
\begin{align}
\label{eq_tSB}
  \mathcal{T}_\text{SB}(r_2,r_3)
=&  ~\FR{\sin(\ii\pi\wt\nu_2)}{\sqrt{\pi}(e^{\pi\wt\nu_1}-\ii)}\Big(\FR{r_2}2\Big)^{3/2-\ii\wt\nu_1} \sum_{n_1,n_2,n_3=1}^\infty \FR{(-1)^{n_{123}}}{n_1!n_2!n_3!}r_3^{2+n_3}\Big(\FR{r_3}2\Big)^{2n_{12}} \n\\
&\times  \FR{\Gamma[-n_1-\ii\wt\nu_2,-n_2+\ii\wt\nu_2]}{n_3+2n_2+\ii \wt\nu_2+\fr12}\mathbf{F}(n_3+2n_{12}-\ii\wt\nu_1+\fr92,\ii\wt\nu_1;r_2^2). 
\end{align}

The background-signal piece $\mathcal{T}_\text{BS}$ is obtained from $\mathcal{T}_\text{SB}$ by switching $\wt\nu_1\leftrightarrow\wt\nu_2$ as well as $r_2\leftrightarrow r_3$. Finally, the background-background piece is analytic in both $r_2\to 0$ and $r_3\to 0$ limit, and its expression is:  
\begin{align}
\label{eq_tBB}
\mathcal{T}_\text{BB}(r_2,r_3)
=& ~\FR{\sin(\ii\pi\wt\nu_1)\sin(\ii\pi\wt\nu_2)r_2^2 r_3^2}{4 \pi^2}  
 \sum_{n_1,\cdots,n_4=0}^{\infty}\FR{(-1)^{n_{1234}}}{n_1!n_2!n_3!n_4!} \Big(\FR{r_2}{2}\Big)^{2n_{12}}\Big(\FR{r_3}{2}\Big)^{2n_{34}}\n\\
 &\times \Gamma\Big[-n_1-\ii \wt\nu_1,-n_2+\ii\wt\nu_1,-n_3+\ii\wt\nu_2,-n_4-\ii\wt\nu_2\Big] \n\\
 &\times  \mathcal{F}_2 \left[ 2n_{1234}+5 \middle|\bgm 2n_1+\ii\wt\nu_1+\fr12,2n_4+\ii\wt\nu_2+\fr12 \\ 2n_1+\ii\wt\nu_1+\fr32 , 2n_4+\ii\wt\nu_2+\fr32 \edm \middle|-r_2,-r_3\right].
\end{align}
Again, we have checked that our analytical result (\ref{eq_4ptResult}) for the 4-point graph in Fig.\ \ref{fig_4pt2massive} agrees well with a direct numerical integration.

\begin{figure}
\centering  
\includegraphics[width=0.48\textwidth]{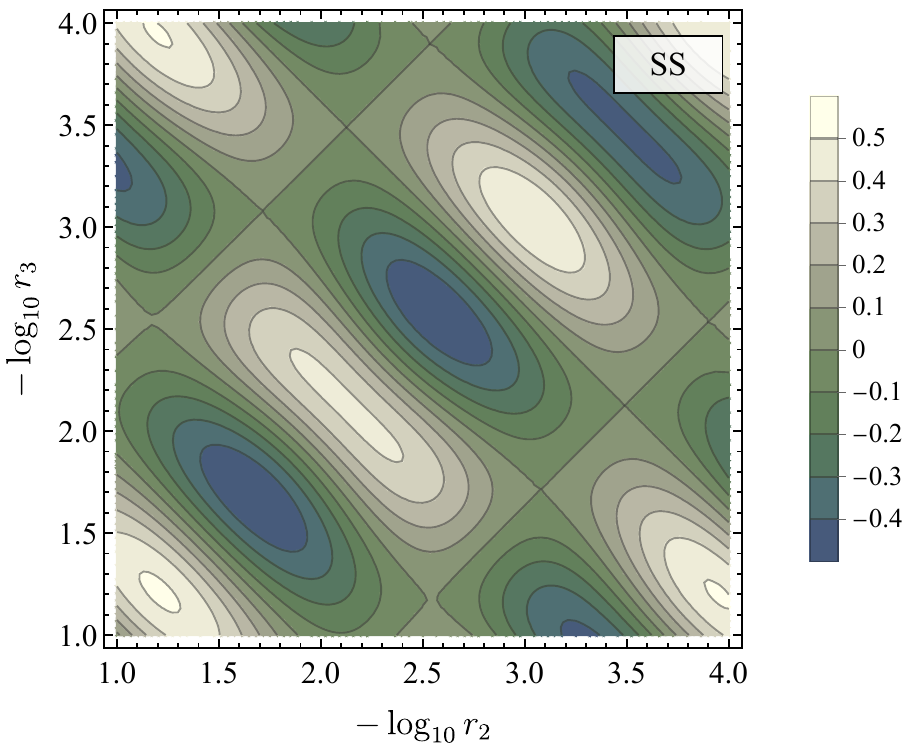}~~~
\includegraphics[width=0.48\textwidth]{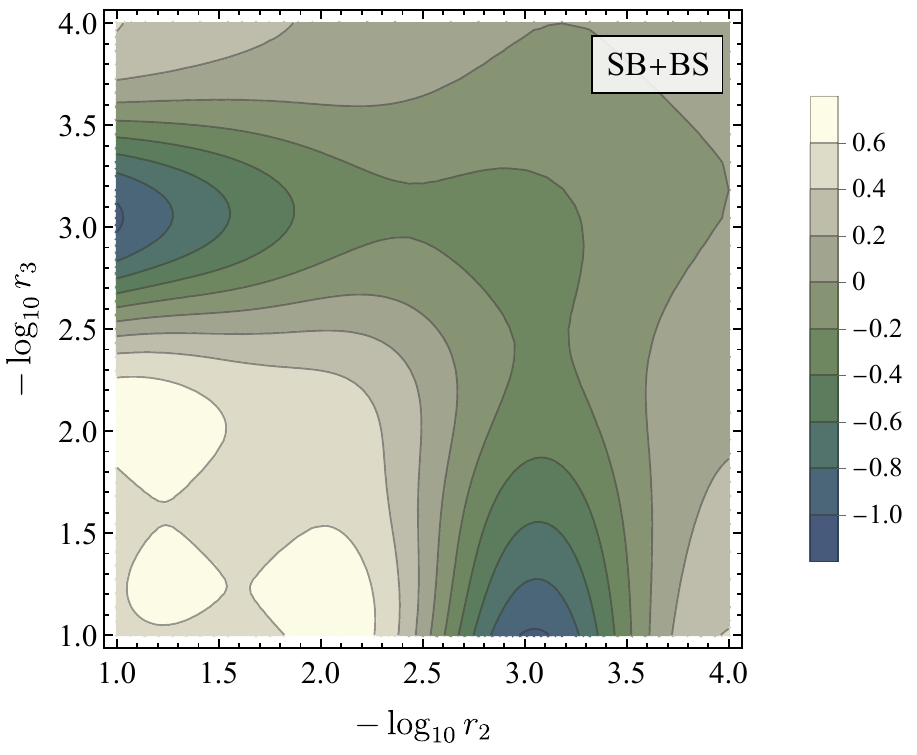}
\includegraphics[width=0.48\textwidth]{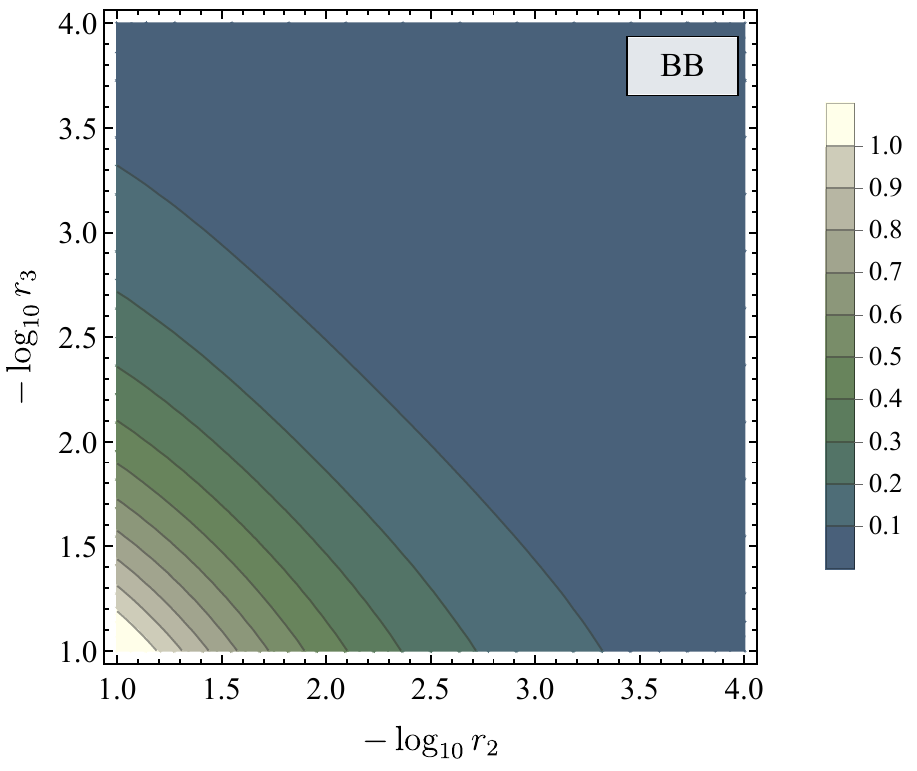}~~~
\includegraphics[width=0.48\textwidth]{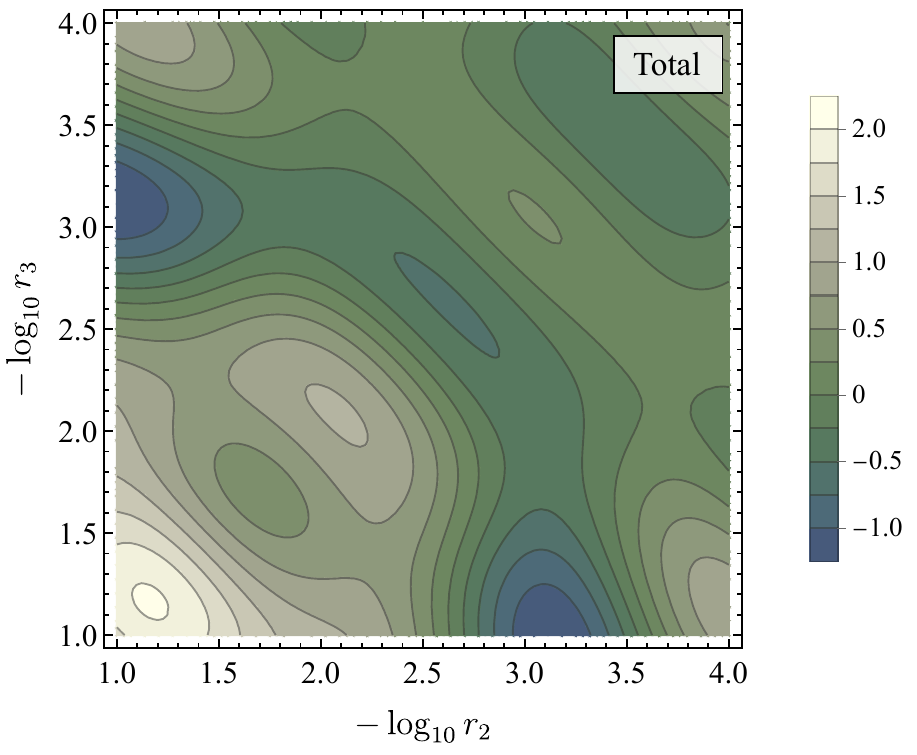}
\caption{The rescaled amplitudes $\wh{\mathcal{T}}_{i}(r_2,r_3)$, defined in (\ref{eq_That}), for the 4-point correlator in Fig.\;\ref{fig_4pt2massive} as functions of momentum ratio $r_2=k_1/k_{34}$ and $r_2=k_2/k_{34}$. The four panels correspond respectively to the signal-signal (SS), signal-background (SB+BS), the background-background (BB), and the total result (SS+SB+BS+BB). In these plots, we take $\wt\nu_1=1$ and $\wt\nu_2=2$. 
} 
\label{fig_4ptPlot}
\end{figure}

\begin{figure}
\centering  
\includegraphics[width=0.58\textwidth]{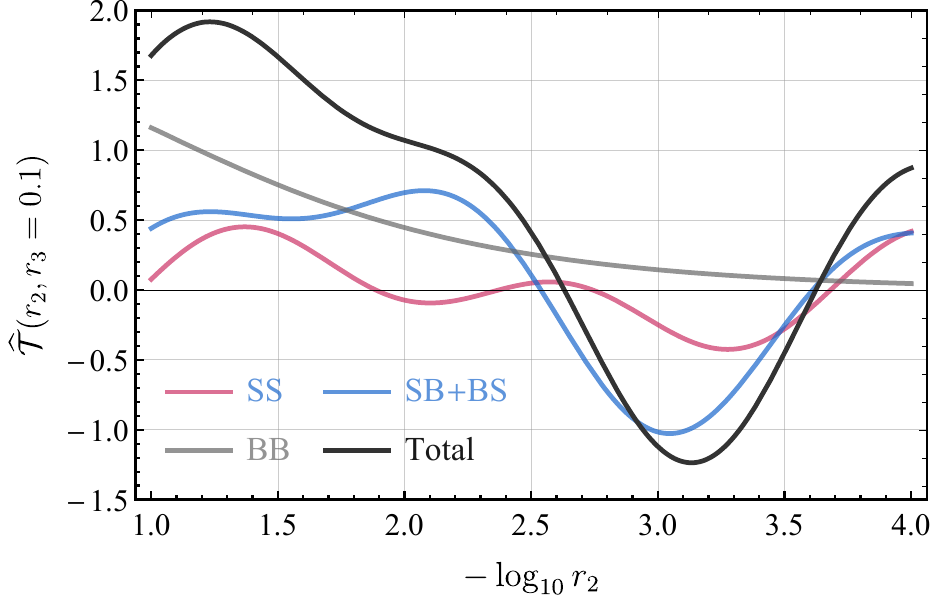} 
\caption{The same with Fig.\;\ref{fig_4ptPlot}, but with $r_3=0.1$ fixed.} 
\label{fig_tr2}
\end{figure}

Already from (\ref{eq_tSS})-(\ref{eq_tSB}) we can observe oscillatory dependences on $r_2$ and $r_3$ with two different frequencies $\wt\nu_1$ and $\wt\nu_2$, which manifest themselves through the complex powers $r_2^{\pm\ii\wt\nu_1}$ and $r_3^{\pm\ii\wt\nu_2}$. However, when we combine (\ref{eq_tSS}) and (\ref{eq_tSB}) with $\wt\nu_{1,2}\to\pm\wt\nu_{1,2}$ terms as well as complex conjugations in (\ref{eq_4ptResult}), we also see that the two oscillatory signals are mixed together in a complicated way. It would be helpful to illustrate these oscillatory signals with plots. In Fig.\;\ref{fig_4ptPlot}, we take a special example with $\wt\nu_1=1$ and $\wt\nu_2=2$, and plot the amplitudes $\wh{\mathcal{T}}_i$ as functions of $-\log_{10}r_2$ and $-\log_{10}r_3$. Here we define the amplitudes $\wh{\mathcal{T}}_i$ as:
\begin{align}
\label{eq_That}
  \wh{\mathcal{T}}_i(r_2,r_3)
  =&~\bigg\{(r_2r_3)^{-3/2}\Big[\mathcal{T}_i(r_2,r_3)+\mathcal{T}_i(r_3,r_2)\Big]+(\wt\nu_1\to-\wt\nu_1)+(\wt\nu_2\to-\wt\nu_2)\n\\
  &~+(\wt\nu_1\to-\wt\nu_1,\wt\nu_2\to-\wt\nu_2)\Big\}+\text{c.c.},
\end{align}
where $i=\text{SS}$, $\text{SB}+\text{BS},$ $\text{BB}$, and the total result $\text{SS}+\text{SB}+\text{BS}+\text{BB}$. By a summation of labels like $\text{SB}+\text{BS}$, we mean the summation of corresponding amplitudes $\wh{\mathcal{T}}_\text{SB}+\wh{\mathcal{T}}_\text{BS}$. Here we choose to multiply the original amplitudes (\ref{eq_tSS})-(\ref{eq_tBB}) by $(r_2r_3)^{-3/2}$ to cancel the ``comoving dilution'' factor $(r_2r_3)^{3/2}$ in the oscillatory signals. Also, in (\ref{eq_That}), we have symmetrized the amplitude with respect to $r_2\leftrightarrow r_3$. Given the definition $r_2\equiv k_1/k_{34}$ and $r_3=k_2/k_{34}$, this symmetrization amounts to the exchange $k_1\leftrightarrow k_2$. More realistically, to compute the full trispectrum, we should also symmetrize the amplitude over all four external momenta. However, from the Feynman diagram in Fig.\;\ref{fig_4pt2massive}, it is easy to see that the replacements $k_{1,2}\to k_{3,4}$ do not lead to any nontrivial signals in the limit $k_{1,2}\to 0$. Thus we do not include them in the plots of Fig.\;\ref{fig_4ptPlot}. 

From Fig.\;\ref{fig_4ptPlot}, we see that, while the two oscillatory signals are easily recognizable in $\wh{\mathcal{T}}_\text{SS}$, there are more nontrival patterns in $\wh{\mathcal{T}}_\text{SB}+\wh{\mathcal{T}}_\text{SB}$. As a result, when we add all pieces up, we get quite  complicated a shape, as shown in the lower-right panel of Fig.\;\ref{fig_4ptPlot}. To make this point more transparent, we also show, in Fig.\;\ref{fig_tr2}, the cross sections of Fig.\;\ref{fig_4ptPlot} with $r_3=0.1$ fixed, so that one can see more clearly how different components contribute to the total amplitude. The complicated pattern revealed in this example shows that we probably need more delicate templates than simple sinusoidal shapes in order to dig out signals with multiple massive exchanges. An interesting question is how to design  templates with appropriate simplicity but with essential information of the full amplitudes kept as much as possible. We leave this interesting question for a future study.

\section{Conclusion and Outlooks}
\label{sec_concl}

Inflation correlators are important theoretical data for QFTs in dS spacetime, and are promising targets for current and future cosmological observations. Inflation correlators mediated by massive fields are central objects in Cosmological Collider physics. Thus, the analytical computation of inflation correlators deserves a systematic investigation. 

Very often in weakly coupled theories, inflation correlators are dominated by tree-level exchanges. However, analytical computation of general tree graphs remains challenging in dS, due to the multi-layer integrals with time orderings in Schwinger-Keldysh formalism. In previous works, it has been shown that the partial Mellin-Barnes representation is useful in the analytical computation of inflation correlators. Even using this method, the complete analytical evaluation of tree graphs is still hampered by the complication of nested time integrals. 

In this work, we computed arbitrarily nested time integrals in PMB representation. The result is in general a multi-variable hypergeometric series. With our family-tree decomposition procedure, we can find series representation in terms of the inverse of any desired energy variable, or even the sum of several energy variables. This result largely solves the problem of analytical continuation of the nested time integrals in most physical regions. 

With our results, the analytical computation of inflation correlators with arbitrary massive exchanges at the tree level is reduced to a work of pole collecting in the Mellin integrand, which is largely trivial. Thus, barring possible issues with analytical continuation in special kinematics to be commented below, we can say that the problem of analytical computation of tree-level inflation correlators is solved. 

At this point, we want to comment on the meaning of analytical computation. As we have seen, most tree-level inflation correlators with massive exchanges have to be expressed in terms of gigantic hypergeometric series which are not yet named. One may say that we can systematically classify hypergeometric functions with increasing number of variables and parameters and give each of them a name. However, given that we know so little about these series in general, this is not super meaningful, and also is not really different from directly giving names to inflation correlators. The meaning of analytical calculation is thus somewhat obscure. Normally, when we say that we obtain an analytical answer, what we really mean is that we have good understanding of this answer in at least two ways: First, we know its analytical properties. This includes how the answer changes with parameters, and where it blows up or shows other singular behavior. Second, we know how to find numerical values of this answer for any choices of parameters with reasonable precision and computation time. Therefore, we can claim that we get an analytical answer only when we have gained sufficient knowledge about this answer. Having an answer is not enough; we need to understand it.

From this viewpoint, it seems to us that our result for the nested time integrals can be called an analytical answer: We know how to write down this answer as Taylor series for most kinematics. As long as there is a largest energy variable and we can use it to form small energy ratios, we can always express our answer as power series of these given small numbers. This means, on the one hand, we know the analytical properties of the answer at any soft energy limit, and also know how to take analytical continuations to different parameter spaces. On the other hand, having a convergent series often means that we can do fast numerical evaluation of the answer. This is proved true in our examples for two massive exchanges: In many cases, the numerical evaluation of our series solution is way faster than direct numerical integration of the original graph. 

Our results have opened new possibilities in the analytical study of inflation correlators. Many interesting problems can be pursued along this direction. We mention some of them as below.

First, a main result of this work is a simple procedure to write down the analytical answer for arbitrary nested time integrals in PMB representation. Since the same time integral also appears in the loop computation, the result here could be useful for the computation of loop correlators as well. Thus it would be interesting to apply our method to make more complete loop computation in PMB representation. 

Second, a nice feature of PMB representation is that the energy dependence and the momentum dependence of a graph are separated: The energy dependence is fully in the time integral, while the momentum dependence is fully in the loop momentum integral (or trivially factored out in a tree graph). Thus, our result on the nested time integral will be useful for studying energy dependence of a graph. This is particularly relevant to digging out the local CC signal in a graph, since, by definition, the local signal is a nonanalytic power in the energy ratios. We leave a more systematic study of local signals to a future work.

Third, it is important that the PMB representation does not assume full dS isometries of the problem. Therefore, it is straightforward to apply our results here to fields with dS-breaking dispersions such as non-unit sound speed, helical chemical potential, or even more exotic dispersion relations. Our method is also applicable to correlation functions in more general FRW background. It would also be interesting to investigate the analytical structure of general nested time integrals and correlation functions, including the location of poles and branch points. We leave these generalizations to future studies. 

Finally, it remains challenging to take analytical continuation of our series expressions to the parameter region where no small energy ratios exist. A pragmatic solution is using numerical interpolation to bridge different parameter regions in which various series expression converge. While this can indeed be implemented in some cases, it is not clear to us if this method works for all possible kinematics. Analytically, we may need more sophisticated methods to take analytical continuation for multi-layer series, which sounds like a nontrivial mathematical problem. We leave these more mathematically oriented problems for future studies as well.

\paragraph{Acknowledgments.} We thank Zhehan Qin for useful discussions. We also thank the anonymous referee for suggesting the discussion of phenomenology of multiple massive exchanges. This work is supported by NSFC under Grant No.\ 12275146, the National Key R\&D Program of China (2021YFC2203100), and the Dushi Program of Tsinghua University. 

\newpage
\begin{appendix}

\section{Mathematical Appendix}
\label{app_math}

\subsection{Mellin-Barnes representation}
We use MB representation for quite a few special functions in the main text, which we collect here. All expressions here can be found in standard mathematical handbooks such as \cite{nist:dlmf}.

 First, the Hankel functions $\text{H}_{\nu}^{(j)}(az)$ of $j$'th kind ($j=1,2$) frequently appear. Their MB representations are given by:
\bge
\label{eq_HankelMB}
  \text{H}_{\nu}^{(j)}(az)=\int_{-\ii\infty}^{\ii\infty}\FR{\di s}{2\pi\ii}\FR{(az/2)^{-2s}}{\pi}e^{(-1)^{j+1}(2s-\nu-1)\pi\ii/2}\Gamma\Big[s-\FR{\nu}{2},s+\FR{\nu}{2}\Big].~~~~(j=1,2)
\ede
Next, we use an exponential integral $\text{E}_p(z)$ defined in the following way:
\bge
  \label{eq_EInt} 
  \mathrm{E}_p(z)=\int_1^{\infty}\FR{e^{-zt}}{t^p}\di t. 
\ede
This exponential integral is related to a confluent hypergeometric function $\text{U}(a,b;z)$ via
\bge
\label{eq_EinU}
  \mathrm{E}_p(z)=z^{p-1}e^{-z}\mathrm{U}(p,p;z),
\ede
The confluent hypergeometric function $U(a,b;z)$ has the following MB representations: 
\bge
  \mathrm{U}(a,b;z)= \int_{-\ii\infty}^{+\ii\infty}\FR{\di s}{2\pi\ii}\Gamma\bgb a+s,1+a-b+s,-s \\ a, 1+a-b \edb z^{-a-s},
\ede
\bge
\label{eq_UMB2}
  \mathrm{U}(a,b,z)=z^{1-b}e^z\int_{-\ii\infty}^{+\ii\infty}\FR{\di s}{2\pi\ii}\Gamma\bgb b-1+s, s\\ a+s\edb z^{-s}.
\ede
The validities of these expressions put constraints on the range of $z$, which are always satisfied in the cases studied in this work, and thus we do not spell them out. With these expressions, we can get two different MB representations for the exponential integral $\text{E}_p(z)$. First, we have a partially resolved representation:
\begin{align}
  \mathrm{E}_p(z)=e^{-z}\int_{-\ii\infty}^{+\ii\infty}\FR{\di s}{2\pi\ii}\Gamma\bgb p+s,1+s,-s \\ p \edb z^{-s-1}.
\end{align}
Second, we have the following completely resolved representation:
\bge
\label{eq_EpMB2}
  \mathrm{E}_p(z) =\int_{-\ii\infty}^{+\ii\infty}\FR{\di s}{2\pi\ii}\FR{\Gamma(s)z^{-s}}{s+p-1} .
\ede
We note that the denominator $1/(s+p-1)$ in the integrand of the last expression comes from the $\Gamma$ factors $\Gamma(b-1+s)/\Gamma(a+s)$ in (\ref{eq_UMB2}) with $a=b=p$ as required by (\ref{eq_EinU}). After taking $a=b=p$, most left poles of $\Gamma(b-1+s)$ are canceled by the zeros of $1/\Gamma(a+s)$, with only one pole left, which is exactly the denominator $1/(s+p-1)$ in (\ref{eq_EpMB2}). Thus we see that the pole from $1/(s+p-1)$ should be treated as a left pole.

\subsection{Special functions}

Following our previous works on similar topics, we use shorthand notations for the products and fractions of Euler $\Gamma$ functions:
\begin{align}
  \Gamma\left[ z_1,\cdots,z_m \right]
  \equiv&~ \Gamma(z_1)\cdots \Gamma(z_m) ,\\
  \Gamma\left[\bgm z_1,\cdots,z_m \\w_1,\cdots, w_n\edm\right]
  \equiv&~\FR{\Gamma(z_1)\cdots \Gamma(z_m)}{\Gamma(w_1)\cdots \Gamma(w_n)}.
\end{align}
A number of hypergeometric series have been well studied and designated with special names. Several of these hypergeometric functions are used in the main text, and we collect their definitions here. More details about these functions can be found in \cite{Slater:1966}. First, the (generalized) hypergeometric function ${}_pF_q$ is defined by the following way when the series converges: 
\begin{align}
\label{eq_pFq}
  {}_pF_q\left[\bgm a_1,\cdots,a_p \\ b_1,\cdots,b_q \edm  \middle| z \right]
  =\sum_{n=0}^\infty
  \FR{(a_1)_n\cdots (a_p)_n}{(b_1)_n\cdots(b_q)_n}\FR{z^n}{n!},
\end{align}
where $(a)_n\equiv\Gamma(a+n)/\Gamma(a)$ is the Pochhammer symbol. In most cases, it turns out simpler to use the following dressed version of hypergeometric function: 
\begin{align}
\label{eq_dressedF}
  {}_p\mathcal{F}_q\left[\bgm a_1,\cdots,a_p \\ b_1,\cdots,b_q \edm  \middle| z \right]
  =&~\Gamma\bgb a_1,\cdots,a_p \\ b_1,\cdots,b_q\edb
  {}_pF_q\left[\bgm a_1,\cdots,a_p \\ b_1,\cdots,b_q \edm  \middle| z \right]\n\\
  =&~\sum_{n=0}^\infty\Gamma\left[\begin{matrix}
        a_1+n, \cdots, a_p+n \\
        b_1+n, \cdots, b_q+n
    \end{matrix}\right]\FR{z^n}{n!}.
\end{align}
A special case of Gauss hypergeometric function is frequently used in the main text, and thus we give a particular symbol to it:
\begin{align}
\label{eq_boldF}
    \mb{F}(a,b;z)\equiv \Gamma[a,b]\times {}_2 F_1\left[\bgm a/2,(1+a)/2 \\ 1-b \edm \middle| z \right].
\end{align}

Next we come to hypergeometric functions of two variables. First, there are four Appell functions $F_1,\cdots,F_4$. Two of them are used in the main text. We only present the definition of their dressed versions:
\begin{align}
\label{eq_dressedF2}
  \mathcal{F}_2\left[a \middle| \bgm b_1,b_2\\ c_1,c_2 \edm\middle| x,y\right]= \sum_{m,n=0}^\infty\Gamma\bgb a+m+n,b_1+m,b_2+n \\ c_1+m, c_2+n \edb\FR{x^my^n}{m!n!}.
\end{align}
\begin{align}
\label{eq_dressedF4}
  \mathcal{F}_4\left[\bgm a,b\\ c_1,c_2 \edm\middle| x,y\right]= \sum_{m,n=0}^\infty\Gamma\bgb a+m+n,b+m+n \\ c_1+m, c_2+n \edb\FR{x^my^n}{m!n!}.
\end{align}
Second, a more general class of two-variable hypergeometric functions are called Kampé de Fériet function in the literature, whose definition is: 
\begin{align}
  &{}^{p+q}F_{r+s}\left[\bgm a_1,\cdots,a_p\\ c_1,\cdots,c_r\edm \middle| \bgm b_1,b_1';\cdots;b_q,b_q'\\ d_1,d_1';\cdots;d_s,d_s'\edm\middle|x,y \right]\n\\
  =&\sum_{m,n=0}^\infty\FR{(a_1)_{m+n}\cdots(a_p)_{m+n}}{(c_1)_{m+n}\cdots(c_r)_{m+n}}\FR{(b_1)_{m}(b_1')_{n}\cdots(b_q)_{m}(b_q')_{n}}{(d_1)_{m}(d_1')_{n}\cdots(d_s)_{m}(d_s')_{n}}\FR{x^my^n}{m!n!}.
\end{align}
Again, we use the dressed version in the main text:
\begin{align}
  &{}^{p+q}\mathcal{F}_{r+s}\left[\bgm a_1,\cdots,a_p\\ c_1,\cdots,c_r\edm \middle| \bgm b_1,b_1';\cdots;b_q,b_q'\\ d_1,d_1';\cdots;d_s,d_s'\edm\middle|x,y \right]\n\\
  =&\sum_{m,n=0}^\infty 
  \Gamma\bgb a_1+m+n,\cdots,a_p+m+n\\ c_1+m+n,\cdots,c_r+m+n\edb
  \Gamma\bgb b_1+m,\cdots,b_q+m\\ d_1+m,\cdots,d_s+m\edb
  \Gamma\bgb b_1'+n,\cdots,b_q'+n\\ d_1'+n,\cdots,d_s'+n\edb
   \FR{x^my^n}{m!n!}.
\end{align}
Finally, there is a particular $n$-variable hypergeometric function that appears in the main text, called Lauricella's $F_A$ function:
\begin{align}
  F_A\left[a\middle|\bgm b_1,\cdots,b_N \\ c_1,\cdots,c_N \edm\middle|z_1,\cdots,z_N\right]=\sum_{m_1,\cdots,m_N=0}^\infty\FR{(a)_{m_1+\cdots+m_N}(b_1)_{m_1}\cdots (b_N)_{m_N}}{(c_1)_{m_1}\cdots(c_N)_{m_N}}\FR{z_1^{m_1}\cdots z_N^{m_N}}{m_1!\cdots m_N!}.
\end{align}
Again, we use this function in its dressed form:
\begin{align}
  &\mathcal{F}_A\left[a\middle|\bgm b_1,\cdots,b_N \\ c_1,\cdots,c_N \edm\middle|z_1,\cdots,z_N\right]\n\\
  =&\sum_{m_1,\cdots,m_N=0}^\infty
  \Gamma\bgb a+m_1+\cdots+m_N,b_1+m_1,\cdots,b_N+m_N \\ c_1+m_1,\cdots,c_N+m_N \edb \FR{z_1^{m_1}\cdots z_N^{m_N}}{m_1!\cdots m_N!}.
\end{align}

\section{Details of Computing the Three-Vertex Seed Integral}
\label{app_detail}

In this appendix we collect some intermediate steps in the computation of the three-vertex seed integral (\ref{eq_6ptseed}) in Sec.\ \ref{sec_2mass}. First, there are four independent branches in the seed integrals (\ref{eq_6ptseed}), shown below. The other four can be found by taking complex conjugation.  
\begin{align}
\label{eq_Ippp}
  \mathcal{I}_{+++}
  =& -\ii E_1^{p_1+1}E_2^{p_2+1}E_3^{p_3+1}\ell_1^3\ell_2^3\int_{-\infty}^0\di\tau_1\di\tau_2\di\tau_3(-\tau_1)^{p_1}(-\tau_2)^{p_2}(-\tau_3)^{p_3}e^{\ii (E_1\tau_1+ E_2\tau_2+ E_3\tau_3)}\n\\
   &\times \Big[D_{+-}^{(\wt\nu_1)}(\ell_1;\tau_1,\tau_2)+\Big(D_{-+}^{(\wt\nu_1)}(\ell_1;\tau_1,\tau_2)- D_{+-}^{(\wt\nu_1)}(\ell_1;\tau_1,\tau_2)\Big)\theta(\tau_1-\tau_2)\Big]\n\\
   &\times \Big[D_{-+}^{(\wt\nu_2)}(\ell_2;\tau_2,\tau_3)+\Big(D_{+-}^{(\wt\nu_2)}(\ell_2;\tau_2,\tau_3)- D_{-+}^{(\wt\nu_2)}(\ell_2;\tau_2,\tau_3)\Big)\theta(\tau_3-\tau_2)\Big]
\end{align}
\begin{align}
\label{eq_Ippm}
  &\mathcal{I}_{++-}
  = +\ii E_1^{p_1+1}E_2^{p_2+1}E_3^{p_3+1}\ell_1^3\ell_2^3\int_{-\infty}^0\di\tau_1\di\tau_2\di\tau_3(-\tau_1)^{p_1}(-\tau_2)^{p_2}(-\tau_3)^{p_3}e^{\ii (E_1\tau_1+ E_2\tau_2- E_3\tau_3)}\n\\
   &\times \Big[D_{+-}^{(\wt\nu_1)}(\ell_1;\tau_1,\tau_2)+\Big(D_{-+}^{(\wt\nu_1)}(\ell_1;\tau_1,\tau_2)- D_{+-}^{(\wt\nu_1)}(\ell_1;\tau_1,\tau_2)\Big)\theta(\tau_1-\tau_2)\Big]D_{+-}^{(\wt\nu_2)}(\ell_2;\tau_2,\tau_3).
\end{align}
\begin{align}
\label{eq_Impp}
  &\mathcal{I}_{-++}
  = +\ii E_1^{p_1+1}E_2^{p_2+1}E_3^{p_3+1}\ell_1^3\ell_2^3\int_{-\infty}^0\di\tau_1\di\tau_2\di\tau_3(-\tau_1)^{p_1}(-\tau_2)^{p_2}(-\tau_3)^{p_3}e^{\ii (-E_1\tau_1+ E_2\tau_2+ E_3\tau_3)} 
     \n\\
   &\times D_{-+}^{(\wt\nu_1)}(\ell_1;\tau_1,\tau_2)\Big[D_{-+}^{(\wt\nu_2)}(\ell_2;\tau_2,\tau_3)+\Big(D_{+-}^{(\wt\nu_2)}(\ell_2;\tau_2,\tau_3)- D_{-+}^{(\wt\nu_2)}(\ell_2;\tau_2,\tau_3)\Big)\theta(\tau_3-\tau_2)\Big].
\end{align}
\begin{align}
\label{eq_Ipmp}
  \mathcal{I}_{+-+}
  =& +\ii E_1^{p_1+1}E_2^{p_2+1}E_3^{p_3+1}\ell_1^3\ell_2^3\int_{-\infty}^0\di\tau_1\di\tau_2\di\tau_3(-\tau_1)^{p_1}(-\tau_2)^{p_2}(-\tau_3)^{p_3}e^{\ii (E_1\tau_1- E_2\tau_2+ E_3\tau_3)} \n\\
   &\times D_{+-}^{(\wt\nu_1)}(\ell_1;\tau_1,\tau_2)D_{-+}^{(\wt\nu_2)}(\ell_2;\tau_2,\tau_3).
\end{align}
Then, we classify the terms in the above four integrals according to whether the adjacent two time variables are time-ordered (T) or factorized (F). Thus, we get the following eight different terms:
\begin{align}
\label{eq_IFF}
  \mathcal{I}_{+++}^\text{(FF)}
  =& -\ii E_1^{p_1+1}E_2^{p_2+1}E_3^{p_3+1}\ell_1^3\ell_2^3\int_{-\infty}^0\di\tau_1\di\tau_2\di\tau_3(-\tau_1)^{p_1}(-\tau_2)^{p_2}(-\tau_3)^{p_3}e^{\ii (E_1\tau_1+ E_2\tau_2+ E_3\tau_3)}\n\\
   &\times D_{+-}^{(\wt\nu_1)}(\ell_1;\tau_1,\tau_2) D_{-+}^{(\wt\nu_2)}(\ell_2;\tau_2,\tau_3),
\end{align}
\begin{align}
  \mathcal{I}_{+++}^\text{(TF)}
  =& -\ii E_1^{p_1+1}E_2^{p_2+1}E_3^{p_3+1}\ell_1^3\ell_2^3\int_{-\infty}^0\di\tau_1\di\tau_2\di\tau_3(-\tau_1)^{p_1}(-\tau_2)^{p_2}(-\tau_3)^{p_3}e^{\ii (E_1\tau_1+ E_2\tau_2+ E_3\tau_3)}\n\\
   &\times \Big(D_{-+}^{(\wt\nu_1)}(\ell_1;\tau_1,\tau_2)- D_{+-}^{(\wt\nu_1)}(\ell_1;\tau_1,\tau_2)\Big)\theta(\tau_1-\tau_2) D_{-+}^{(\wt\nu_2)}(\ell_2;\tau_2,\tau_3), 
\end{align}
\begin{align}
  \mathcal{I}_{+++}^\text{(FT)}
  =& -\ii E_1^{p_1+1}E_2^{p_2+1}E_3^{p_3+1}\ell_1^3\ell_2^3\int_{-\infty}^0\di\tau_1\di\tau_2\di\tau_3(-\tau_1)^{p_1}(-\tau_2)^{p_2}(-\tau_3)^{p_3}e^{\ii (E_1\tau_1+ E_2\tau_2+ E_3\tau_3)}\n\\
   &\times D_{+-}^{(\wt\nu_1)}(\ell_1;\tau_1,\tau_2) \Big(D_{+-}^{(\wt\nu_2)}(\ell_2;\tau_2,\tau_3)- D_{-+}^{(\wt\nu_2)}(\ell_2;\tau_2,\tau_3)\Big)\theta(\tau_3-\tau_2),
\end{align}
\begin{align}
  \mathcal{I}_{+++}^\text{(TT)}
  =& -\ii E_1^{p_1+1}E_2^{p_2+1}E_3^{p_3+1}\ell_1^3\ell_2^3\int_{-\infty}^0\di\tau_1\di\tau_2\di\tau_3(-\tau_1)^{p_1}(-\tau_2)^{p_2}(-\tau_3)^{p_3}e^{\ii (E_1\tau_1+ E_2\tau_2+ E_3\tau_3)}\n\\
   &\times \Big(D_{-+}^{(\wt\nu_1)}(\ell_1;\tau_1,\tau_2)- D_{+-}^{(\wt\nu_1)}(\ell_1;\tau_1,\tau_2)\Big)\theta(\tau_1-\tau_2) \n\\
   &\times \Big(D_{+-}^{(\wt\nu_2)}(\ell_2;\tau_2,\tau_3)- D_{-+}^{(\wt\nu_2)}(\ell_2;\tau_2,\tau_3)\Big)\theta(\tau_3-\tau_2) ,
\end{align}
\begin{align}
  \mathcal{I}_{++-}^\text{(F)}
  =& +\ii E_1^{p_1+1}E_2^{p_2+1}E_3^{p_3+1}\ell_1^3\ell_2^3\int_{-\infty}^0\di\tau_1\di\tau_2\di\tau_3(-\tau_1)^{p_1}(-\tau_2)^{p_2}(-\tau_3)^{p_3}e^{\ii (E_1\tau_1+ E_2\tau_2- E_3\tau_3)}\n\\
   &\times  D_{+-}^{(\wt\nu_1)}(\ell_1;\tau_1,\tau_2) D_{+-}^{(\wt\nu_2)}(\ell_2;\tau_2,\tau_3),
\end{align}
\begin{align}
  \mathcal{I}_{++-}^\text{(T)}
  =& +\ii E_1^{p_1+1}E_2^{p_2+1}E_3^{p_3+1}\ell_1^3\ell_2^3\int_{-\infty}^0\di\tau_1\di\tau_2\di\tau_3(-\tau_1)^{p_1}(-\tau_2)^{p_2}(-\tau_3)^{p_3}e^{\ii (E_1\tau_1+ E_2\tau_2- E_3\tau_3)}\n\\
   &\times \Big(D_{-+}^{(\wt\nu_1)}(\ell_1;\tau_1,\tau_2)- D_{+-}^{(\wt\nu_1)}(\ell_1;\tau_1,\tau_2)\Big)\theta(\tau_1-\tau_2) D_{+-}^{(\wt\nu_2)}(\ell_2;\tau_2,\tau_3),
\end{align}
\begin{align}
  \mathcal{I}_{-++}^\text{(F)}
  =& +\ii E_1^{p_1+1}E_2^{p_2+1}E_3^{p_3+1}\ell_1^3\ell_2^3\int_{-\infty}^0\di\tau_1\di\tau_2\di\tau_3(-\tau_1)^{p_1}(-\tau_2)^{p_2}(-\tau_3)^{p_3}e^{\ii (-E_1\tau_1+ E_2\tau_2+ E_3\tau_3)} 
     \n\\
   &\times D_{-+}^{(\wt\nu_1)}(\ell_1;\tau_1,\tau_2) D_{-+}^{(\wt\nu_2)}(\ell_2;\tau_2,\tau_3),
\end{align}
\begin{align}
\label{eq_ImppN}
  \mathcal{I}_{-++}^\text{(T)}
  =& +\ii E_1^{p_1+1}E_2^{p_2+1}E_3^{p_3+1}\ell_1^3\ell_2^3\int_{-\infty}^0\di\tau_1\di\tau_2\di\tau_3(-\tau_1)^{p_1}(-\tau_2)^{p_2}(-\tau_3)^{p_3}e^{\ii (-E_1\tau_1+ E_2\tau_2+ E_3\tau_3)} \n\\
   &\times D_{-+}^{(\wt\nu_1)}(\ell_1;\tau_1,\tau_2) \Big(D_{+-}^{(\wt\nu_2)}(\ell_2;\tau_2,\tau_3)- D_{-+}^{(\wt\nu_2)}(\ell_2;\tau_2,\tau_3)\Big)\theta(\tau_3-\tau_2). 
\end{align}
We then take the PMB representation for all these terms and regroup them according to (\ref{eq_IFFdef})-(\ref{eq_ITTdef}). The time integrals can then be carried out using the method in Sec.\ \ref{sec_timeint}. To write down the result after the time integral, we note that there is a common factor in all terms after taking the PMB representation: 
\begin{align}
  \mathbb{H}(\{s\})
  \equiv&~\FR{1}{(4\pi)^2} \Gamma\Big[s_1-\FR{\ii\wt\nu_1}2,s_1+\FR{\ii\wt\nu_1}2,s_2-\FR{\ii\wt\nu_1}2,s_2+\FR{\ii\wt\nu_1}2\Big]\n\\
  &\times\Gamma\Big[s_3-\FR{\ii\wt\nu_2}2,s_3+\FR{\ii\wt\nu_2}2,s_4-\FR{\ii\wt\nu_2}2,s_4+\FR{\ii\wt\nu_2}2\Big].
\end{align}
Then, we can write down the results of all terms with time integrals finished:
\begin{align}
    \mathcal{I}_{+++}^\text{(FF)}&+\mathcal{I}_{++-}^\text{(F)}+\mathcal{I}_{-++}^\text{(F)}+\mathcal{I}_{+-+}\n\\
    &= -\ii \int_{s_1,\cdots,s_4} \, \Big[-\ii e^{\ii\pi(2s_{23}-p_{123}/2)}-e^{\ii\pi(2s_2-p_{12}/2+p_3/2)}-e^{\ii\pi(2s_3+p_1/2-p_{23}/2)}+\ii e^{\ii\pi(p_2/2-p_{13}/2)}\Big] \n\\
    &\times \mathbb{H}(\{s\}) (r_1 r_2 r_3 r_4)^{3/2}\Big(\FR{r_1}2\Big)^{-2s_{1}}\Big(\FR{r_2}2\Big)^{-2s_{2}}\Big(\FR{r_3}2\Big)^{-2s_{3}} \Big(\FR{r_4}2\Big)^{-2s_{4}}\n\\
    &\times \Gamma\Big[p_1-2s_1+\fr52,p_2-2s_{23}+4,p_3-2s_4+\fr52\Big].
\end{align}
\begin{align}
    \mathcal{I}_{+++}^\text{(FT)}+\mathcal{I}_{-++}^\text{(T)} 
  =&-4\ii \int_{s_1,\cdots,s_4}\sin[\pi(s_3-s_4)]\sin[\pi(s_2-p_1/2+3/4)]e^{-\ii(p_{23}-2s_{234}+13/2)\pi/2}\mathbb{H}(\{s\}) \n\\
  &\times (r_1 r_2 r_3^2)^{3/2} \Big(\FR{r_1}2\Big)^{-2s_{1}}\Big(\FR{r_2}2\Big)^{-2s_{2}}\Big(\FR{r_3}2\Big)^{-2s_{34}} \Big(\FR{r_3}{r_4}\Big)^{p_3+1}\n\\
  &\times  \Gamma(p_1-2s_1+5/2) {}_2\mathcal{F}_1\left[\bgm p_3-2s_4+5/2,p_{23}-2s_{234}+13/2 \\ p_3-2s_4+7/2 \edm \middle|-\FR{r_3}{r_4}\right].
\end{align}
\begin{align}
    \mathcal{I}_{+++}^\text{(TF)}+\mathcal{I}_{++-}^\text{(T)}
    =&-4\ii \int_{s_1,\cdots,s_4}\sin[\pi(s_2-s_1)]\sin[\pi(s_3-p_3/2+3/4)]e^{-\ii(p_{12}-2s_{123}+13/2)\pi/2}\mathbb{H}(\{s\}) \n\\
    &\times (r_2^2 r_3 r_4)^{3/2}\Big(\FR{r_4}2\Big)^{-2s_{4}}\Big(\FR{r_3}2\Big)^{-2s_{3}}\Big(\FR{r_2}2\Big)^{-2s_{12}} \Big(\FR{r_2}{r_1}\Big)^{p_1+1}\n\\
    &\times  \Gamma(p_3-2s_4+5/2) {}_2\mathcal{F}_1\left[\bgm p_1-2s_1+5/2,p_{12}-2s_{123}+13/2 \\ p_1-2s_1+7/2 \edm \middle|-\FR{r_2}{r_1}\right].
\end{align}
\begin{align}
    \mathcal{I}_{+++}^\text{(TT)}
  =&  -4 \int_{s_1,\cdots,s_4}\sin[\pi(s_1-s_2)]\sin[\pi(s_3-s_4)]\mathbb{H}(\{s\}) e^{\ii\pi(s_{1234}-p_{123}/2)}\n\\
   &\times  r_2^3 r_3^3 \Big(\FR{r_2}{r_1}\Big)^{p_1+1}\Big(\FR{r_3}{r_4}\Big)^{p_3+1} \Big(\FR{r_2}{2}\Big)^{-2s_{12}}\Big(\FR{r_3}{2}\Big)^{-2s_{34}}\n\\
   &\times  \mathcal{F}_2 \left[ p_{123}-2s_{1234}+9 \middle|\bgm p_1-2s_1+5/2,p_3-2s_4+5/2 \\ p_1-2s_1+7/2 , p_3-2s_4+7/2 \edm \middle|-\FR{r_2}{r_1},-\FR{r_3}{r_4}\right].
\end{align}
It then remains to finish the Mellin integrals. As indicated in the main text, we pick up all left poles given in (\ref{eq_leftPoles}). The residues of $\mathbb{H}(\{s\})$ are given by
\begin{align}
    &\underset{\{s\}=\{-n+\ii\aa\wt\nu_1/2\}}{\operatorname{Res}}  \mathbb{H}(\{s\})\n\\
    &=\FR{1}{(4\pi)^2} \FR{(-1)^{n_1+n_2+n_3+n_4}}{n_1!n_2!n_3!n_4!}\Gamma\Big[-n_1+\ii \aa_1 \wt\nu_1,-n_2+\ii \aa_2 \wt\nu_1,-n_3+\ii \aa_3 \wt\nu_2,-n_4+\ii \aa_4 \wt\nu_2\Big].
\end{align}
Summing up all the left poles, we can finish the Mellin integral:
\begin{align}
\mathcal{I}_{+++}^\text{(FF)}&+\mathcal{I}_{++-}^\text{(F)}+\mathcal{I}_{-++}^\text{(F)}+\mathcal{I}_{+-+}\n\\
    &= \sum_{\{n\},\{\aa\}} \FR{1}{(4\pi)^2}\Big[-e^{\ii\pi(\ii\aa_2\wt\nu_1+\ii\aa_3\wt\nu_2-p_{123}/2)}+\ii e^{\ii\pi(\ii\aa_2\wt\nu_1-p_{12}/2+p_3/2)}+\ii e^{\ii\pi(\ii \aa_3\wt\nu_2+p_1/2-p_{23}/2)}+ e^{\ii\pi(p_2/2-p_{13}/2)}\Big]\n\\
    &\times \FR{(-1)^{n_1+n_2+n_3+n_4}}{n_1!n_2!n_3!n_4!}(r_1 r_2 r_3 r_4)^{3/2}\Big(\FR{r_1}2\Big)^{2n_{1}-\ii \aa_1\wt\nu_1}\Big(\FR{r_2}2\Big)^{2n_{2}-\ii \aa_2\wt\nu_1}\Big(\FR{r_3}2\Big)^{2n_{3}-\ii \aa_3\wt\nu_2} \Big(\FR{r_4}2\Big)^{2n_{4}-\ii \aa_4\wt\nu_2}\n\\
    &\times \Gamma\Big[p_1+2n_1-\ii\aa_1\wt\nu_1+{5}/{2},p_2+2n_{23}-\ii\aa_2\wt\nu_1-\ii\aa_3\wt\nu_2+4,p_3+2n_4-\ii\aa_4\wt\nu_2+{5}/{2}\Big] \n\\
    &\times \Gamma\Big[-n_1+\ii \aa_1 \wt\nu_1,-n_2+\ii \aa_2 \wt\nu_1,-n_3+\ii \aa_3 \wt\nu_2,-n_4+\ii \aa_4 \wt\nu_2\Big]\n\\
    &= \sum_{n_3,\{\aa\}} \FR{1}{(4\pi)^2}\Big[-e^{\ii\pi(\ii\aa_2\wt\nu_1+\ii\aa_3\wt\nu_2-p_{123}/2)}+\ii e^{\ii\pi(\ii\aa_2\wt\nu_1-p_{12}/2+p_3/2)}+\ii e^{\ii\pi(\ii \aa_3\wt\nu_2+p_1/2-p_{23}/2)}+ e^{\ii\pi(p_2/2-p_{13}/2)}\Big]\n\\
    &\times  \mb{F}(p_1-\ii\aa_1\wt\nu_1+{5}/{2},\ii\aa_1\wt\nu_1,r_1^2)\mb{F}(p_3-\ii\aa_4\wt\nu_2+{5}/{2},\ii\aa_4\wt\nu_2,r_4^2)\n\\
    &\times \FR{(-1)^{n_3}}{n_3!}(r_1 r_2 r_3 r_4)^{3/2}\Big(\FR{r_1}2\Big)^{-\ii \aa_1\wt\nu_1}\Big(\FR{r_2}2\Big)^{-\ii \aa_2\wt\nu_1}\Big(\FR{r_4}2\Big)^{-\ii \aa_4\wt\nu_2}\Big(\FR{r_3}2\Big)^{2n_{3}-\ii \aa_3\wt\nu_2}\n\\
    &\times \Gamma(-n_3+\ii \aa_3 \wt\nu_2)\mb{F}(p_2+2n_3-\ii\aa_2\wt\nu_1-\ii\aa_3\wt\nu_2+4,\ii\aa_2\wt\nu_1,r_2^2)\n\\
    =& \sum_{\{\aa\}}\FR{-2^{p_2-1}}{\pi^{1/2}} \Big[-e^{\ii\pi(\ii\aa_2\wt\nu_1+\ii\aa_3\wt\nu_2-p_{123}/2)}+\ii e^{\ii\pi(\ii\aa_2\wt\nu_1-p_{12}/2+p_3/2)}+\ii e^{\ii\pi(\ii \aa_3\wt\nu_2+p_1/2-p_{23}/2)}+ e^{\ii\pi(p_2/2-p_{13}/2)}\Big]\n\\ 
    &\times \operatorname{csch}(\aa_2\pi\wt\nu_1)\operatorname{csch}(\aa_3\pi\wt\nu_2)(r_1 r_2 r_3 r_4)^{3/2}\Big(\FR{r_1}2\Big)^{-\ii \aa_1\wt\nu_1}r_2^{-\ii \aa_2\wt\nu_1}r_3^{-\ii \aa_3\wt\nu_2}\Big(\FR{r_4}2\Big)^{-\ii \aa_4\wt\nu_2}\n\\
    &\times \mb{F}(p_1-\ii\aa_1\wt\nu_1+{5}/{2},\ii\aa_1\wt\nu_1,r_1^2)\mb{F}(p_3-\ii\aa_4\wt\nu_2+{5}/{2},\ii\aa_4\wt\nu_2,r_4^2)\n\\
    &\times \mathcal{F}_4 \left[ \bgm-\ii\aa_2\wt\nu_1/2-\ii\aa_3\wt\nu_2/2+2+p_2/2,-\ii\aa_2\wt\nu_1/2-\ii\aa_3\wt\nu_2/2+5/2+p_2/2\\1-\ii\aa_2\wt\nu_1,1-\ii\aa_3\wt\nu_2\edm \middle| r_2^2,r_3^2\right],
\end{align}
where
\begin{align}
    \mb{F}(a,b,c)\equiv \Gamma[a,b] {}_2 F_1\left[\bgm a/2,(1+a)/2 \\ 1-b \edm \middle| c \right].
\end{align}
Similarly,
\begin{align}
    \mathcal{I}_{+++}^\text{(FT)}+&\mathcal{I}_{-++}^\text{(T)}\n\\
    =&\sum_{\{n\},\{\aa\}} \FR{-4\ii}{(4\pi)^2} \sin[\pi(\ii \aa_3 \wt\nu_2/2-\ii\aa_4\wt\nu_2/2)]\sin[\pi(\ii\aa_2\wt\nu_1/2-p_1/2+3/4)]e^{-\ii(p_{23}-\ii (\aa\wt\nu)_{234}+13/2)\pi/2} \n\\
  &\times \FR{(-1)^{n_1+n_2+n_3+n_4}}{n_1!n_2!n_3!n_4!}(r_1 r_2r_3^2)^{3/2}\Big(\FR{r_1}2\Big)^{2n_{1}-\ii \aa_1\wt\nu_1}\Big(\FR{r_2}2\Big)^{2n_{2}-\ii \aa_2\wt\nu_1}\Big(\FR{r_3}2\Big)^{2n_{3}-\ii \aa_3\wt\nu_2} \Big(\FR{r_3}{r_4}\Big)^{p_3+1}\n\\
  &\times  \Gamma(p_1+2n_1-\ii\aa_1\wt\nu_1+5/2) \Gamma\Big[-n_1+\ii \aa_1 \wt\nu_1,-n_2+\ii \aa_2 \wt\nu_1,-n_3+\ii \aa_3 \wt\nu_2,-n_4+\ii \aa_4 \wt\nu_2\Big] \n\\
  &\times {}_2\mathcal{F}_1\left[\bgm p_3+2n_4-\ii\aa_4\wt\nu_2+5/2,p_{23}+2n_{234}-\ii(\aa\wt\nu)_{234}+13/2 \\ p_3+2n_4-\ii\aa_4\wt\nu_2+7/2 \edm \middle|-\FR{r_3}{r_4}\right]\n\\
  =& \sum_{m,n_3,n_4,\{\aa\}}  \FR{-4\ii}{(4\pi)^2} \sin[\pi(\ii \aa_3 \wt\nu_2/2-\ii\aa_4\wt\nu_2/2)]\sin[\pi(\ii\aa_2\wt\nu_1/2-p_1/2+3/4)]e^{-\ii(p_{23}-\ii (\aa\wt\nu)_{234}+13/2)\pi/2} \n\\
  &\times \FR{(-1)^{n_3+n_4}}{n_3!n_4!m!}(r_1 r_2r_3^2)^{3/2}\Big(\FR{r_1}2\Big)^{-\ii \aa_1\wt\nu_1}\Big(\FR{r_2}2\Big)^{-\ii \aa_2\wt\nu_1}\Big(\FR{r_3}2\Big)^{2n_{34} -\ii \aa_{34}\wt\nu_2} \Big(-\FR{r_3}{r_4}\Big)^{m}\Big(\FR{r_3}{r_4}\Big)^{p_3+1}\n\\
  &\times \FR{\Gamma[-n_3+\ii \aa_3 \wt\nu_2,-n_4+\ii \aa_4 \wt\nu_2]}{m+p_3+2n_4-\ii\aa_4\wt\nu_2+5/2} \mb{F}(p_1-\ii\aa_1\wt\nu_1+{5}/{2},\ii\aa_1\wt\nu_1,r_1^2)\n\\
  &\times  \mb{F}(m+p_{23}+2n_{34}-\ii(\aa\wt\nu)_{234}+13/2,\ii\aa_2\wt\nu_1,r_2^2); 
\end{align}
Here in the last line we have used a shorthand notation $(\aa\wt\nu)_{234}\equiv\aa_2\wt\nu_1+\aa_3\wt\nu_2+\aa_4\wt\nu_2$.
\begin{align}
\label{eq_INF}
    \mathcal{I}_{+++}^\text{(TF)}&+\mathcal{I}_{++-}^\text{(T)}\n\\
    =& \sum_{m,n_1,n_2,\{\aa\}} \FR{-4\ii}{(4\pi)^2}\sin[\pi(\ii \aa_2 \wt\nu_1/2-\ii\aa_1\wt\nu_1/2)]\sin[\pi(\ii\aa_3\wt\nu_2/2-p_3/2+3/4)]e^{-\ii(p_{12}-\ii (\aa\wt\nu)_{123}+13/2)\pi/2} \n\\
  &\times \FR{(-1)^{n_1+n_2}}{n_1!n_2!m!}(r_2^2 r_3 r_4)^{3/2}\Big(\FR{r_4}2\Big)^{-\ii \aa_4\wt\nu_2}\Big(\FR{r_3}2\Big)^{-\ii \aa_3\wt\nu_2}\Big(\FR{r_2}2\Big)^{2n_{12}-\ii \aa_{12}\wt\nu_1} \Big(-\FR{r_2}{r_1}\Big)^{m}\Big(\FR{r_2}{r_1}\Big)^{p_1+1}\n\\
  &\times \FR{\Gamma[-n_2+\ii \aa_2 \wt\nu_1,-n_1+\ii \aa_1 \wt\nu_1]}{m+p_1+2n_1-\ii\aa_1\wt\nu_1+5/2} \mb{F}(p_3-\ii\aa_4\wt\nu_2+{5}/{2},\ii\aa_4\wt\nu_2,r_4^2)\n\\
  &\times  \mb{F}(m+p_{12}+2n_{12}-\ii(\aa\wt\nu)_{123}+13/2,\ii\aa_3\wt\nu_2,r_3^2).
\end{align}
Here in the last line we have used a shorthand notation $(\aa\wt\nu)_{123}\equiv\aa_1\wt\nu_1+\aa_2\wt\nu_1+\aa_3\wt\nu_2$.
\begin{align}
    \mathcal{I}_{+++}^\text{(TT)}
  =& \sum_{\{n\},\{\aa\}}\FR{-4}{(4\pi)^2}\sin[\pi(\ii\aa_1\wt\nu_1/2-\ii\aa_2\wt\nu_1/2)]\sin[\pi(\ii\aa_3\wt\nu_2/2-\ii\aa_4\wt\nu_2/2)]e^{\ii\pi(\ii(\aa\wt\nu)_{1234}/2-p_{123}/2)}\n\\
  &\times \FR{(-1)^{n_1+n_2+n_3+n_4}}{n_1!n_2!n_3!n_4!}r_2^3 r_3^3 \Big(\FR{r_2}{r_1}\Big)^{p_1+1}\Big(\FR{r_3}{r_4}\Big)^{p_3+1} \Big(\FR{r_2}{2}\Big)^{2n_{12}-\ii(\aa\wt\nu)_{12}}\Big(\FR{r_3}{2}\Big)^{2n_{34}-\ii(\aa\wt\nu)_{34}} \n\\
   &\times  \Gamma\Big[-n_1+\ii \aa_1 \wt\nu_1,-n_2+\ii \aa_2 \wt\nu_1,-n_3+\ii \aa_3 \wt\nu_2,-n_4+\ii \aa_4 \wt\nu_2\Big]\n\\
   &\times  \mathcal{F}_2 \left[ p_{123}+2n_{1234}-\ii(\aa\wt\nu)_{1234}+9 \middle|\bgm p_1+2n_1-\ii\aa_1\wt\nu_1+\fr52,p_3+2n_4-\ii\aa_4\wt\nu_2+\fr52 \\ p_1+2n_1-\ii\aa_1\wt\nu_1+\fr72 , p_3+2n_4-\ii\aa_4\wt\nu_2+\fr72 \edm \middle|-\FR{r_2}{r_1},-\FR{r_3}{r_4}\right].
\end{align}
Here in the last line, we used another shorthand notation $(\aa\wt\nu)_{1234}\equiv\aa_1\wt\nu_1+\aa_2\wt\nu_1+\aa_3\wt\nu_2+\aa_4\wt\nu_2$.
Then, we finish the summation over $\{\aa\}=\{\aa_1,\cdots,\aa_4\}$, regroup terms according to their analytical properties at $r_i\to 0$, and find the final result as summarized in (\ref{eq_seedIntResult}).

\end{appendix}

 \newpage
\bibliography{CosmoCollider} 
\bibliographystyle{utphys}

\end{document}